\documentclass[aps,prd,superscriptaddress,nofootinbib,amsmath,amsfonts,preprintnumbers,groupedaddress,showpacs,10pt,english]{revtex4-1}
\usepackage{amsmath}
\usepackage{amssymb}
\usepackage{babel}
\usepackage{wrapfig}
\usepackage{cancel}

\makeatletter

\usepackage{array,multirow,graphicx}
\usepackage{dcolumn}
\usepackage{newlfont}
\usepackage{bm}
\usepackage[colorlinks,citecolor=blue,urlcolor=blue,linkcolor=blue]{hyperref}
\usepackage[figtopcap]{subfigure}
\usepackage{color}

\begin{document}

\title{ Non-flat and non-Extensive Thermodynamic Effects of M\o ller tetradic theory of gravitation on cosmology}

\author{G.~G.~L.~Nashed}
\email{nashed@bue.edu.eg}
\affiliation {Centre for Theoretical Physics, The British University in Egypt, P.O. Box
43, El Sherouk City, Cairo 11837, Egyp.}

\author{A. T. Shafeek}
\email{mrmrshafeek13@gmail.com}
\affiliation{Department of Mathematics, Faculty of Education,\\
\small Ain Shams University, Roxy, Cairo, Egypt. }

\date{\today}

\begin{abstract}
We derive non-flat cosmological models for two cases (i.e., dust and radiation) in the context of M\o ller's tetradic theory (MTT) of gravitation  using the tetrad that creates the non-flat Friedmann-Robertson-Walker (FRW) metric. These two models are affected by the free dimensional parameter, $\lambda$, that characterized MTT, which approaches zero in the flat case for both models.  Using standard definitions of thermodynamics, we calculate the radius horizon, Hawking temperature, and entropy of our non-flat models in the framework of cosmology and show the effect of $\lambda$ on open and closed universes. We then use the first law of thermodynamics to construct non-flat cosmological models via the non-extensive thermodynamic approach. The resulting models are affected by $\lambda$  and the extensive parameter, $\delta$, which quantifies the effect of non-extensive thermodynamics. When we set, $\lambda=0$ and $\delta=1$, we return to Einstein's general relativity models. We study the evolution of our models in the presence of collisionless non-relativistic matter and describe precise forms of the dark energy density and equation-of-state parameter constraining the non-extensive thermodynamic parameter. We show that insertion of the non-extensive thermodynamic parameter affects the non-flat FRW universe in a manner that noticeably differs from that observed under normal thermodynamics. We also show that the deceleration of the open universe behaves as dark energy in a future epoch, i.e., when the redshift approaches -1, i.e., $z\approx$-1.
%{ Finally, we calculate the distance modulus and show that our models are consistent with the observational data when the redshift exceeds four, i.e., $z>0.4$.}
\end{abstract}


%\begin{center}
\maketitle
%%%%%%%%%%%%%%%%%%%%%%%%%%%%%%%%%%% Section 1 %%%%%%%%%%%%%%%%%%%%%%%%%%%%%%%%%%%%%%%%
\section{Introduction}\label{S1}
%%%%%%%%%%%%%%%%%%%%%%%%%%%%%%%%%%%%%%%%%%%%%%%%%%%%%%%%%%%%%%%%%%%%%%%%%%%%%%%%%%%%%%

The recent observation of a distant type Ia supernova reveals that our universe is entering an accelerating epoch \cite{Perlmutter:1997zf,Dalal:2000xw,Riess:1998cb,Riess:2004nr}. This epoch of cosmic acceleration is due to the existence  of unknown
energy matter that is able to break the strong-energy condition $\rho_{_{_{DE}}}+3p_{_{_{DE}}}>0$, where $\rho_{_{_{DE}}}$ and $p_{_{_{DE}}}$
 are the energy-density and pressure of dark energy, respectively. Many approaches to study dark energy have been proposed  to
investigate accelerating cosmologies at late time epochs \cite{Sahni:2002kh,Eidelman:2004wy,Kamenshchik:2001cp,Zhu:2003sq,Sen:2002ss,Godlowski:2003pd,Godlowski:2004pt,Godlowski:2004gh,Puetzfeld:2004df,Biesiada:2004td}.

 Several methods \cite{Carroll:2003wy,Capozziello:2003tk,Deffayet:2001pu,freese2002cardassian,Nojiri:2003wx,ArkaniHamed:2003uy} have been developed to modify gravity and describe accelerating cosmologies at late time epochs. In the present study, we take advantage of modifying gravity constructed on the absolute parallelism space which is known in the literature by M\o ller tetradic theory of gravitation. The concepts of absolute parallelism were given first in physics by Einstein \cite{einstein1928riemann}, who attempted to unify the gravitational and electromagnetic fields by using several degrees of freedom characterized by tetrads. This trial
failed,  because a solution for the Schwarzschild black hole could not be obtained.

M\o ller considered tetrad theory as purely gravitational theory and found that the treatment of  the energy-momentum complex in the frame of this geometry is more suitable than that in Einstein's theory of  general relativity (GR). He then assumed a restriction on the relevant Lagrangians so that the metric tensor becomes identical to that in Einstein's field equations. Afterward, he \cite{moller1978crisis} put aside this restriction and searched for a
 larger class of Lagrangians, with the aim of possibly shifting from Einstein's GR field equations to stronger gravitational fields. M\o ller's theory was extended into scalar tetradic theory, as introduced by
S\'aez \cite{Saez:1983tap}. Meyer investigated  M\o ller's theory as a special case of  Poinc\'are gauge theory \cite{meyer1982moller}.
  Hayashi and Nakano  \cite{hayashi1967extended} constructed a tetrad theory of gravitation
as a gauge theory of the space-time translation group. Hayashi and Shirafuji \cite{Hayashi:1979qx} subsequently studied the
geometric and observational grounds of tetrad theory, eventually finding the Lagrangian consists of a quadratic form of the torsion tensor; the group then constructed ``new GR theory."
 Assuming invariance under parity operations,  Hayashi and Shirafuji \cite{Hayashi:1979qx} used
the most general Lagrangian to obtain three expressions involving three free parameters to be constrained from an experiment. The group showed that two of these three parameters could be determined from  solar-system
experiments, while the third only has an upper boundary  \cite{Hayashi:1979qx, miyamoto1971possible}. Moreover, the numerical values of the first two parameters are very small, and consistent with a zero value. If we set these two parameters equal to zero,  the new GR coincides with that proposed
by Hayashi and Nakano and M\o ller; thus, we refer to this concept as HNM theory. If the torsion tensor has a non-vanishing axial-vector component, then HNM theory will differ from GR theory.

Many applications based on the HNM framework have been developed in recent years. Some cosmological applications \cite{saez1984mo11er}, for example, include investigations of gravitational radiation \cite{schweizer1979poincare,schweizer1980post}, studies on the energy-momentum
complex \cite{Mikhail:1994bj,Mikhail:1994rj}, derivations of  a general solution with spherical symmetry \cite{Shirafuji:1995xc}, and derivations of solutions with
axial symmetry   \cite{saez1984stationary}.

Many amended  gravitational theories present a different demonstration of the two stages of accelerated expansion of our universe \cite{Capozziello:2011et, Nojiri:2010wj}, and they can  improve the renormalizability of GR and therefore may present gravitational theory real to a quantum formulation \cite{Weinberg:1995mt}. The standard method in the direction of constructing   modified gravitational theories  is to include  extra terms in the Einstein-Hilbert action   \cite{Clifton:2011jh, DeFelice:2010aj}. Nevertheless, we can begin from the equivalent  construction of gravity using torsion, specifically, we can use  the teleparallel equivalent of general relativity (TEGR) \cite{Unzicker:2005in, Aldrovandi:2013wha, Maluf:2013gaa}, and modify the  corresponding Lagrangian, i.e.,  the torsion scalar $T$, in different methods, giving  $f(T)$ gravity \cite{Cai:2015emx, Bengochea:2008gz, Linder:2010py},  $f(T,T_G)$ gravity \cite{Kofinas:2014owa, Kofinas:2014daa}, $f(T,B)$ gravity \cite{Bahamonde:2015zma,Nashed:2014sea,Karpathopoulos:2017arc,Ren:2021uqb, Bohmer:2021eoo}, scalar-torsion theories \cite{Geng:2011aj, Hohmann:2018rwf, Bahamonde:2019shr},  and so on.  $f(T)$ gravitational theories have shown many interesting results,  in cosmology \cite{Chen:2010va, Zheng:2010am, Bamba:2010wb, Cai:2011tc, Capozziello:2011hj, Wei:2011aa, Amoros:2013nxa, Otalora:2013dsa, Bamba:2013jqa, Li:2013xea, Nashed20071047,Paliathanasis:2014iva, Malekjani:2016mtm, Farrugia:2016qqe, Qi:2017xzl, Cai:2018rzd,Nashed20062241, Anagnostopoulos:2019miu, Nunes:2019bjq, Cai:2019bdh, Yan:2019gbw, ElHanafy:2019zhr, Saridakis:2019qwt,Awad2017, Wang:2020zfv, Ren:2021tfi}  as well as   spherically symmetric spacetime \cite{Boehmer:2011gw, Gonzalez:2011dr, Ferraro:2011ks,Nashed2010, Wang:2011xf, Atazadeh:2012am, Rodrigues:2013ifa, Nashed:2013bfa, Nashed:2014iua, Junior:2015fya,Shirafuji19971355, Kofinas:2015hla,Shirafuji:1997wy, Das:2015gwa,Nashed2002521, Rani:2016gnl, Rodrigues:2016uor, Mai:2017riq, Singh:2019ykp, Nashed:2020kjh, Elizalde:2020icc,Bhatti:2018fsc, Ashraf:2020yyo, ElHanafy:2015egm,Ditta:2021wfl}.
The aim of the present study to derive cosmological models for non-flat space and study their thermodynamic properties using standard thermodynamic and Tsallis definitions.
The rest of this paper is organized as follows: In Section \ref{MTT}, the basic tools of MTT are presented. In Section  \ref{FRW}, previous trials of cosmological studies in the MTT framework are presented. In Section  \ref{3s}, we derive a cosmological solution of  MTT for non-flat space and obtain its corresponding cosmological quantities, including the  Hubble parameter, and deceleration. In Section  \ref{4p},  we study the thermodynamics of the models derived in  Section  \ref{3s} and evaluate their radius horizon, entropy, and Hawking temperature.  In Section  \ref{tas}, we study , in detail, the cosmological evolution of the non-flat models of MTT in the presence of
collisionless non-relativistic matter; we also give exact formulas for the  dark energy density
and equation-of-state (EoS) parameters.
%{ In subsection \ref{obs}, we compare our models with Type Ia supernovae observations for $z>0.4$ and find an agreement with the observational data.} The final section summarizes the results of the present study.
%\let\thefootnote\relax\footnotetext{\small $^{*}$ Corresponding author: e-mail address: mrmrshafeek13@gmail.com}
\noindent
\small
\section{M\o ller's Tetrad theory of gravitation}\label{MTT}
 M\o ller  modified Einstein's GR theory by constructing a new theory of gravitation based on a four-dimensional
Weitzenb\"ock geometry \cite{moller1978crisis,Hayashi:1967se,1984GReGr..16..501S,robertson1932groups}. This theory is known in the literature  as  MTT of gravitation. The main motivation of M\o ller for this modification was to obtain a gravitational theory free from singularities while retaining the main merits of GR.

In MTT,  the field variables include 16-tetrad covariant components $\lambda^{i}_{{\ \nu}}$ where $i=1\cdots 4$ are the mesh indices that are raised and lowered by  the Minkowski metric, which has the form $\eta_{i j}=diag(+1,\,-1,\,-1,\,-1)$, and the Greek indices are coordinate indices that are lowered and raised by the metric tensor. In the framework of Weitzenb\"ock geometry,  we can define the metric tensor as follows:
\begin{equation}\label{1}
  g_{\alpha\beta}=\eta_{i j} \lambda^{i}_{{\ \alpha}}\lambda^{j}_{{\ \beta}}\,.
\end{equation}

The contorsion tensor $\gamma_{\mu\nu\alpha}$ plays a key role in M\o ller's theory; this tensor is defined as:
\begin{equation}\label{2}
  \gamma_{\mu\nu\alpha}=\eta_{i j}\lambda^i{}_\mu\lambda^j{}_{\nu;\alpha}\,,
\end{equation}
where the semicolon denotes the covariant differentiation with respect to the  Christoffel symbols.

M\o ller derived the field equations of his theory by applying a variational principle to the Lagrangian density
%\footnote{ In this study we use the  reduced Planck units in which ($\hbar = c = k_B = 1$).},
\begin{equation}\label{3}
  {\cal L}=(-g)^{1/2}[\alpha_{1}c^{\mu}c_{\mu}+\alpha_{2}\gamma^{\mu\nu\rho}\gamma_{\mu\nu\rho}+\alpha_{3}\gamma^{\mu\nu\rho}\gamma_{\rho\nu\mu}]\,,
\end{equation}
where $\alpha_{1}, \alpha_{2}$ and  $\alpha_{3}$ are constants, $g=det(g_{\mu\nu})$ and $c_{\mu}$ is a vector field defined by:
\[c_{\mu}=\gamma^{\rho}_{\ \mu\, \rho}\,.\]
M\o ller indicated that his  theory coincides with GR under weak fields, which leads to the constants  $\alpha_{1}, \alpha_{2}$  and  $\alpha_{3}$ taking the forms:
\begin{equation}\label{4}
  \alpha_{1}=-1/\chi, \qquad \qquad  \alpha_{2}=\lambda/\chi, \qquad \qquad \alpha_{3}=(1-2\lambda)/\chi,
\end{equation}
where $\lambda$ is a free dimensionless parameter and $\chi=8\pi$, in the relativistic units where the speed of light and Newtonian gravitational constant have unit values. This same choice of parameters was obtained by Hayashi and Nakano \cite{Hayashi:1967se}. M\o ller presented his field equations in the following form:
\begin{equation}\label{5}
  G_{\mu\nu} + H_{\mu \nu}= -\chi T_{\mu \nu},\qquad \qquad  M_{\mu\nu}=0\,,
\end{equation}
where $G_{\mu\nu}$ is the Einstein tensor and defined as:
\begin{equation}\label{6}
  G_{\mu\nu} = R_{\mu\nu}-\frac{1}{2} R g_{\mu \nu}\,,
\end{equation}
and $T_{\mu\nu}$ is the energy--momentum tensor and defined in standard form as:
\begin{equation}\label{7}
 T_{\mu\nu}=\rho_{0}\frac{dx^{\mu}}{ds}\frac{dx^{\nu}}{ds}\,,
\end{equation}
where $\rho_{0}$ is the proper density.  The tensors $H_{\mu\nu}$ and $F_{\mu\nu}$ are defined as
\begin{equation}\label{8}
  H_{\mu\nu}=\lambda\left[\gamma_{\alpha\sigma\mu} \gamma^{\alpha\sigma}_{\ \ \ \nu} +\gamma_{\alpha\sigma\mu}\gamma_{\nu}^{\ \ \alpha\sigma}+\gamma_{\alpha\sigma\nu}\gamma_{\mu}^{\ \ \alpha\sigma}+ g_{\mu\nu}\left(\gamma_{\alpha\sigma\lambda}\gamma^{\lambda\sigma\alpha} -\frac{1}{2}\gamma_{\alpha\sigma\lambda}\gamma^{\alpha\sigma\lambda}\right)\right],
\end{equation}
and
\begin{equation}\label{9}
  F_{\mu\nu}=\lambda\left[c_{\mu,\nu}-c_{\nu,\mu}-c_{\alpha}(\gamma^{\alpha}_{\ \ \mu\nu}-\gamma^{\alpha}_{\ \ \nu\mu})+\gamma_{\mu \nu\  \ \ ; \ \alpha}^{\ \ \ \alpha}\right]\,,
\end{equation}
where the comma and semi colon respectively, refer to ordinary and covariant derivatives.  Equations (\ref{8}) and (\ref{9})  are symmetric and skewed symmetric tensors, respectively.

{ Equations (\ref{8}) and (\ref{9}) clearly show that M\o ller's theory is reduced to standard Einstein GR theory when $\lambda=0$ otherwise it does not coincide and in that case, the field equations are not invariant under local Lorentz transformation. The main reason for the broken of the local Lorentz transformation is the axial vector \cite{Hayashi:1979qx,Hohmann:2020dgy}.  In this sense, many tetrads that reproduce the same metric can reproduce different physics.}
 MTT has many applications in the astrophysics and cosmological domains \cite{Mikhail:1994rj,Mikhail:1994bj,1984GReGr..16..501S} . Given our interest  in cosmology, we provide  a brief summary of the results obtained  by S\'aez and De Juan, who carried out studies in the framework of cosmology using a flat space \cite{1984GReGr..16..501S}.

\section{Friedmann--Lemaitre--Robertson--Walker (FLRW) in MTT}\label{FRW}
In this section, we apply the field equations of MTT to FRW. First, we present a brief summary of the results obtained in this domain by S\'aez and De Juan.  The tetrad field featuring  homogeneity and isotropy given by Robertson \cite{robertson1932groups} in the spherical polar coordinates $[t, r, \theta, \phi]$ has the form:
\begin{equation}\label{11}
  \lambda_{i}^{^{\ \mu}}=\begin{pmatrix}
  1& 0& 0& 0\hspace{0.cm}\\
  0& \displaystyle\frac{l_1\sin\theta\cos\phi}{4a(t)}& \displaystyle\frac{l_2\cos\theta\cos\phi-4r\sqrt{k}\sin\phi}{4ra(t)}& -\displaystyle\frac{l_2\sin\phi+4r\sqrt{k}\cos\theta\cos\phi}{4ra(t)\sin\theta} \hspace{0.cm}\\
  0& \displaystyle\frac{l_1\sin\theta\sin\phi}{4a(t)}& \displaystyle\frac{l_2\cos\theta\sin\phi+4r\sqrt{k}\cos\phi}{4ra(t)}& \displaystyle\frac{l_2\cos\phi-4r\sqrt{k}\cos\theta\sin\phi}{4ra(t)\sin\theta}\hspace{0.cm}\\
  0& \displaystyle\frac{l_1\cos\theta}{4a(t)}& -\displaystyle\frac{l_2\sin\theta}{4ra(t)}& \displaystyle\frac{\sqrt{k}}{a(t)}
  \end{pmatrix}\,,
\end{equation}
where $l_1=4+ kr^{2}$, $l_2=4- kr^{2}$, $a(t)$ is the cosmic scale factor and $k=0,1,-1$ are the spatial curvature indices corresponding to spatially flat, closed and open universe, respectively.\\
The Friedmann Robertson-Walker metric (FRW) is given by:\\
\begin{equation}\label{12}
 ds^{2}=dt^{2}-\frac{a^{2}(t)}{\left(1+\frac{1}{4}kr^{2}\right)^2}\left\{dr^{2}+r^2d\theta^{2}+r^2\sin^2\theta d\phi^{2}\right\}.
\end{equation}

By applying M\o ller's field equations, i.e., Eqs. (\ref{5}) and (\ref{6}) to the tetrad defined in Eq.(\ref{11}), S\'aez and De Juan obtained the following set of differential equations:\\
\begin{equation}\label{13}
\frac{\dot a^2}{a^{2}}+\frac{h}{a^{2}}=\frac{\chi}{3}T^{^{0}}_{_{\ 0}}\,,
\end{equation}
{
\begin{equation}\label{14}
\frac{2\ddot a}{a}+\frac{\dot a^2}{a^{2}}+\frac{h}{a^{2}}=\chi T^{^{1}}_{_{\ 1}}=\chi T^{^{2}}_{_{\ 2}}=\chi T^{^{3}}_{_{\ 3}}\,,
\end{equation}}
where $h=k(1-3\lambda)$. We use relativistic units in which $c=G=1$ in this study,the energy-momentum tensor for a perfect fluid is given by:
\begin{equation}\label{15}
T^{^{0}}_{_{\ 0}}=\rho, \qquad \qquad \qquad
T^{^{1}}_{_{\ 1}}=T^{^{2}}_{_{\ 2}}=T^{^{3}}_{_{\ 3}}=-p.
\end{equation}
Equations (\ref{13}) and (\ref{14}) give rise to the following:\\
\begin{equation}\label{16}
\frac{\dot a^2}{a^{2}}+\frac{h}{a^{2}}=\frac{\chi}{3}\rho,
\end{equation}
\begin{equation}\label{17}
\frac{2\ddot a}{a}+\frac{\dot a^2}{a^{2}}+\frac{h}{a^{2}}=-\chi p\,,
\end{equation}
where  $p$ is the pressure of the fluid and $\rho$ is the energy density.\\
From the conservation law, $T^{\mu \nu}_{ \ \ \  ; \nu}=0$, we obtain:
\begin{equation}\label{19}
a\frac{d\rho}{da}=-3(\rho+p)\,.
\end{equation}
In the framework of cosmology, $EoS$ can be assumed to take the form\footnote{{ In this study we assume specific values of the EoS by assuming the dust $\omega=0$, radiation,  $\omega=1/3$,  and dark energy,  $\omega=-1$, cases. We cannot study the case when  $\omega$ is arbitrary because the field equations (\ref{13}) and (\ref{14})  cannot have exact solution.}}:
\begin{equation}\label{EoS}
p=w\rho\,,
\end{equation}
where $w$ is a parameter of the EoS.
We now consider  the dust case, $p=0$, i.e., $w=0$ in Eq. (\ref{19})  and get: \\
\begin{equation}\label{20}
\rho a^{3}=N\,,
\end{equation}
where $N$ is a constant of integration. Using Eqs. (\ref{16}), (\ref{17}) and (\ref{20}) S\'aez and De Juan obtained the following relations:\\
for $h>0$:\\
{
\begin{eqnarray}\label{21}
&&a(\beta)=b(1-\cos\beta), \qquad  h^{1/2}t(\beta)=b(\beta-\sin\beta)\Longrightarrow a(t)\approx\frac {\sqrt [3]{6\,h\,t^2} \left( 2
\sqrt [3]{6\,b^2}-\sqrt [3]{h\,t^2} \right) }{4\sqrt [3]{b}}\,,\nonumber\\
&& \Longrightarrow t(z)=\frac {2\sqrt [4]{b}\sqrt { \left( 6\,b-1+6\,b\,z \right) \sqrt {9\,
 \left( 1+z \right) b-6}+9\,\sqrt { \left( 1+z \right) b} \left[ 2\,
 \left( 1+z \right) b-1 \right] }}{\sqrt[4] {9h^2 \left( 1+z \right) }}\,,
\end{eqnarray}}
for $h<0$:\\
{
\begin{eqnarray}\label{22}
&&a(\beta)=b(\cosh\beta-1), \qquad \mid h\mid^{1/2}t(\beta)=b(\sinh\beta-\beta)\Longrightarrow
 a(t)\approx\frac {\sqrt [3]{6 \mid h\mid t^2} \left( 2\sqrt [3]{6b^2}+\sqrt [3]{ \mid h\mid t^2} \right) }{4\sqrt [3]{b}}
\,,\nonumber\\
&& \Longrightarrow t(z)=\frac {2\sqrt [4]{b}\sqrt { \left( 6\,b+1+6\,b\,z \right) \sqrt {9\,
 \left( 1+z \right) b+6}-9\,\sqrt { \left( 1+z \right) b} \left[ 2\,
 \left( 1+z \right) b+1 \right] }}{\sqrt[4] {9h^2 \left( 1+z \right) }}\,,
\end{eqnarray}}
{ where $b=\frac{\chi N}{6\mid h\mid}$ and $\beta$ is the  time parameter}. From Eqs. (\ref{21}) and (\ref{22}), the Hubble parameter  takes the form:
\begin{equation}\label{23}
H(\beta)=\frac{\frac{da(\beta)}{d\beta}}{a(\beta)}=\begin{cases}
\displaystyle\frac{\sqrt{h}\sin\beta}{b(1-\cos\beta)^{2}}\,,     &  \textrm{for}\quad   h>0\,,\\
\\
\displaystyle\frac{\sqrt{h}\sinh\beta}{b(\cosh\beta-1)^{2}}\,,   &  \textrm{for} \quad   h<0\,.
\end{cases}
\end{equation}
{
\begin{equation}\label{23t}
H(t)=\displaystyle\frac{\dot a(t)}{a(t)}\approx\begin{cases}\displaystyle\frac {4\sqrt [3]{6\,b^2}-\sqrt [3]{h\,t^2}}{3t \left( 2\,
\sqrt [3]{6\,b^2}-\sqrt [3]{h\,t^2} \right) }
\,,     &  \textrm{for}\quad   h>0\,,\\
\\
\displaystyle\frac {4\sqrt [3]{6\,b^2}+\sqrt [3]{ \mid h\mid\,t^2}}{3t \left( 2\,
\sqrt [3]{6\,b^2}+\sqrt [3]{ \mid h\mid\,t^2} \right) }

\,,   &  \textrm{for} \quad   h<0\,.
\end{cases}
\end{equation}}
Because the deceleration parameter is given by \cite{Weinberg:1972kfs}:
\begin{equation}\label{24}
q=\frac{dH^{-1}}{dt}-1=\frac{-\ddot a}{aH^{2}}\,,
\end{equation}
\begin{equation}\label{25}
q(\beta)=\begin{cases}
\displaystyle\frac{2\sin^{2}\beta-\cos\beta(1-\cos\beta)}{\sin^{2}\beta}-1\,,       &  \textrm{for} \quad  h>0,\\
\\
\displaystyle\frac{2\sinh^{2}\beta-\cosh\beta(\cosh\beta-1)}{\sinh^{2}\beta}-1\,,   &  \textrm{for} \quad   h<0\,.
\end{cases}
\end{equation}
{
\begin{equation}\label{25r}
q(t)\approx\begin{cases}
\displaystyle\frac {\sqrt[3]{6h\,t^2\,b^2}-\sqrt[3]{h^2\,t^4}+2\sqrt[3]{36\,b^4}}{4 \left( \sqrt [3]{6\,b^2}-\sqrt [3]{h\,t^2} \right) ^{2}}
\,,       &  \textrm{for}\,\,\,  h>0,\\
\\
-\displaystyle\frac {\sqrt [3]{6\,h\,t^2\,b^2}+\sqrt [3]{h^2\,t^4}-2\,\sqrt [3]{36\,b^4}}{4 \left(\sqrt [3]{6\,b^2}+\sqrt [3]{h\,t^2
} \right) ^{2}}
\,,   &  \textrm{for}\,\,\,   h<0\,.
\end{cases}
\end{equation}}
According to Eqs. (\ref{16}) and (\ref{17}), when $p=0$, we obtain the following energy density equation:\\
\begin{equation}\label{26}
\rho=\frac{-3\ddot a}{4\pi a}\,.
\end{equation}
Substituting Eq. (\ref{24}) into Eq. (\ref{26}) we obtain:\\
\begin{equation}\label{27}
\rho=\frac{-3qH^{2}}{4\pi}\,.
\end{equation}
Substituting  Eqs. (\ref{23}) and (\ref{25}) into  Eq.(\ref{27}) we obtain:
\begin{equation}\label{28}
\rho(\beta)=\begin{cases}
\displaystyle\frac{3h[\sin^{2}\beta-\cos\beta(1-\cos\beta)]}{4\pi b^{2}(1-\cos\beta)^{4}}\,,       &  \textrm{for}\,\,\,   h>0\,,\\
\\
\displaystyle\frac{3h[\sinh^{2}\beta-\cosh\beta(\cosh\beta-1)}{4\pi b^{2}(\cosh\beta-1)^{4}}\,,    &  \textrm{for}\,\,\,   h<0\,.
\end{cases}
\end{equation}
{
\begin{equation}\label{28r}
\rho(t)\approx\begin{cases}
\displaystyle\frac {\sqrt [3]{6\,b^2}+\sqrt [3]{h\,t^2}}{3\pi \,{t}^{2
} \left( \sqrt [3]{h\,t^2}-2\sqrt [3]{6\,b^2}\right) }
\,,       &  \textrm{for}  \,\,\, h>0\,,\\
\\
\displaystyle\frac{\sqrt [3]{h\,t^2}-\sqrt [3]{6\,b^2}}{3\pi \,{t}^{
2} \left( 2\,\sqrt [3]{6\,b^2}+\sqrt [3]{h\,t^2} \right) }
\,,    &  \textrm{for}  \,\,\, h<0\,.
\end{cases}
\end{equation}}
The relation between the scale factor and redshift is given as follows \cite{Capozziello:2014zda,Awad:2017yod}\\
 \begin{equation}\label{29}
1+z=\frac{a_{_{z=0}}}{a}\,,
\end{equation}
From Eqs.(\ref{21}) and (\ref{22}),  we obtain:
\begin{equation}\label{30}
\beta=\begin{cases}
\cos^{-1}[1-1/b(1+z)]\,,       &  \textrm{for} \quad  h>0\,,\\
\\
\cosh^{-1}[1+1/b(1+z)]      &  \textrm{for} \quad  h<0\,,
\end{cases}
\end{equation}
where,  $a_{_{z=0}}=1$, which is the value of the scale factor at a preset time.  In the following section, a set of graphs is plotted for the case of $h>0$ (i.e., $\lambda>1/3$ for the open universe and $\lambda<1/3$ for the closed universe).
\begin{figure}
\centering
 \subfigure[\, Displays the scale factor vs. the cosmic time  for the closed universe.]{\label{fig:1a}\includegraphics[scale=0.25]{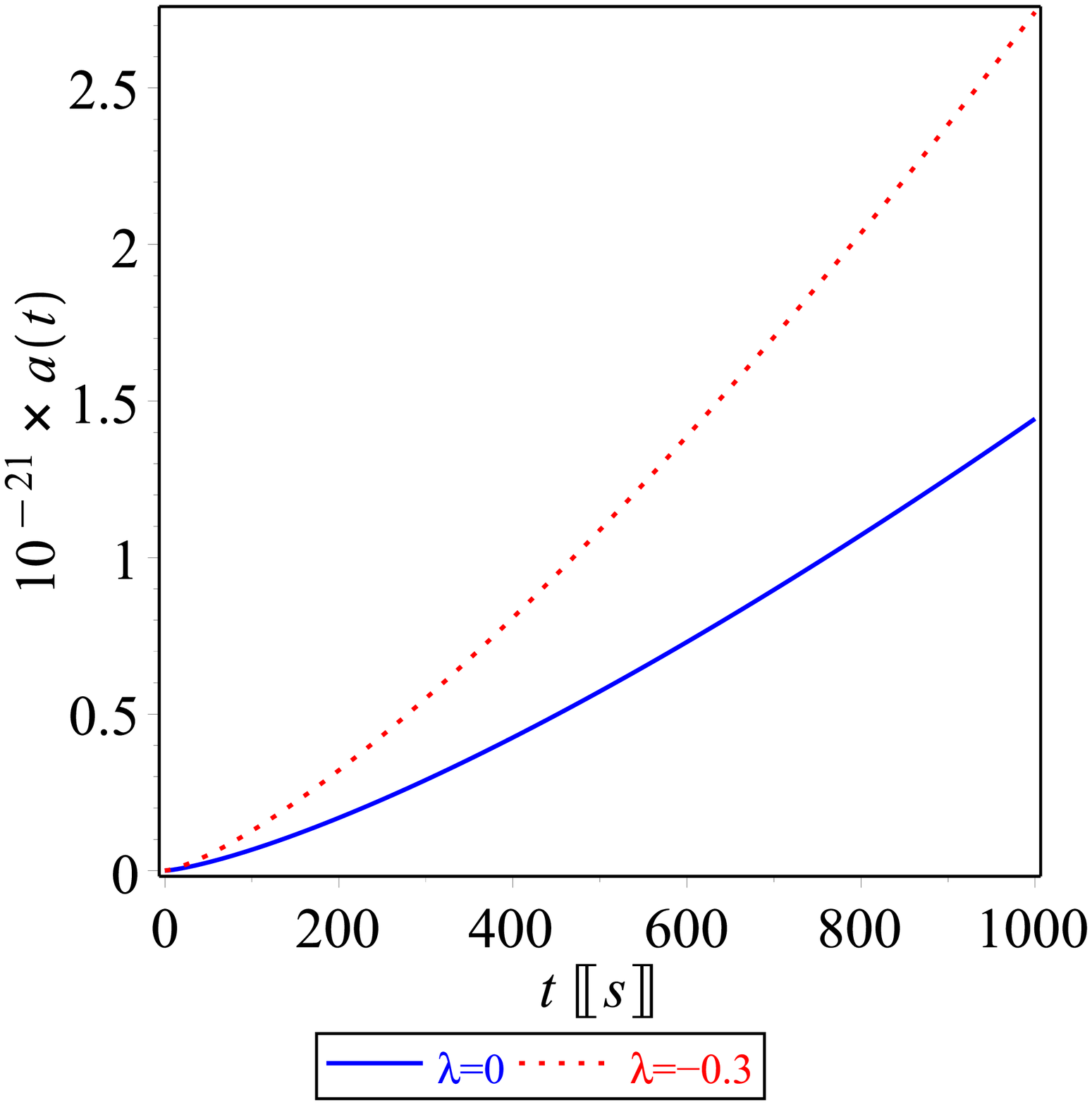}}\hspace{0.2cm}
\subfigure[\, Displays the Hubble parameter vs. the cosmic time for the closed universe.]{\label{fig:1b}\includegraphics[scale=0.25]{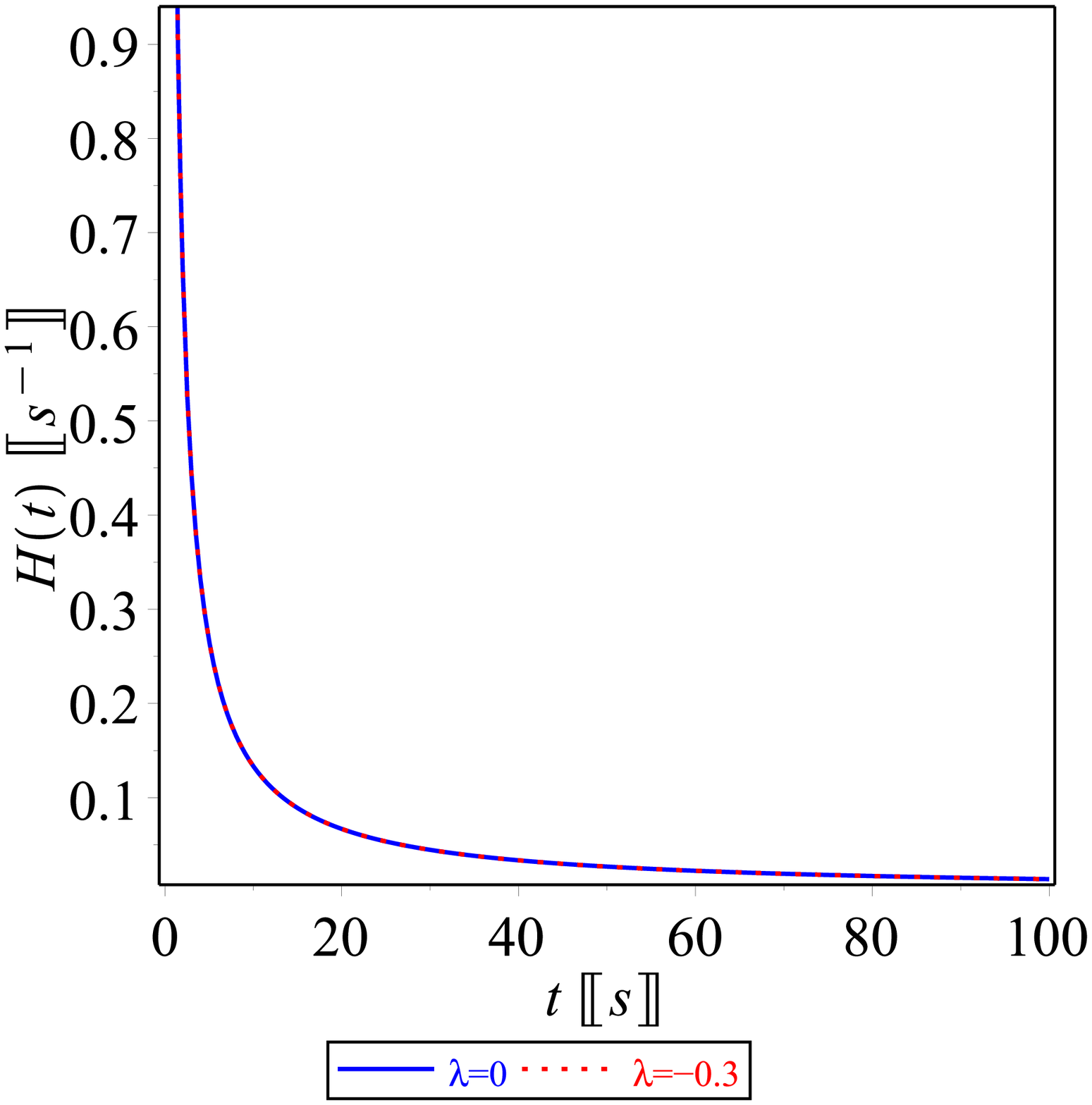}}\hspace{0.2cm}
\subfigure[\, Displays the deceleration  parameter vs. the cosmic time for the closed universe.]{\label{fig:1c}\includegraphics[scale=0.25]{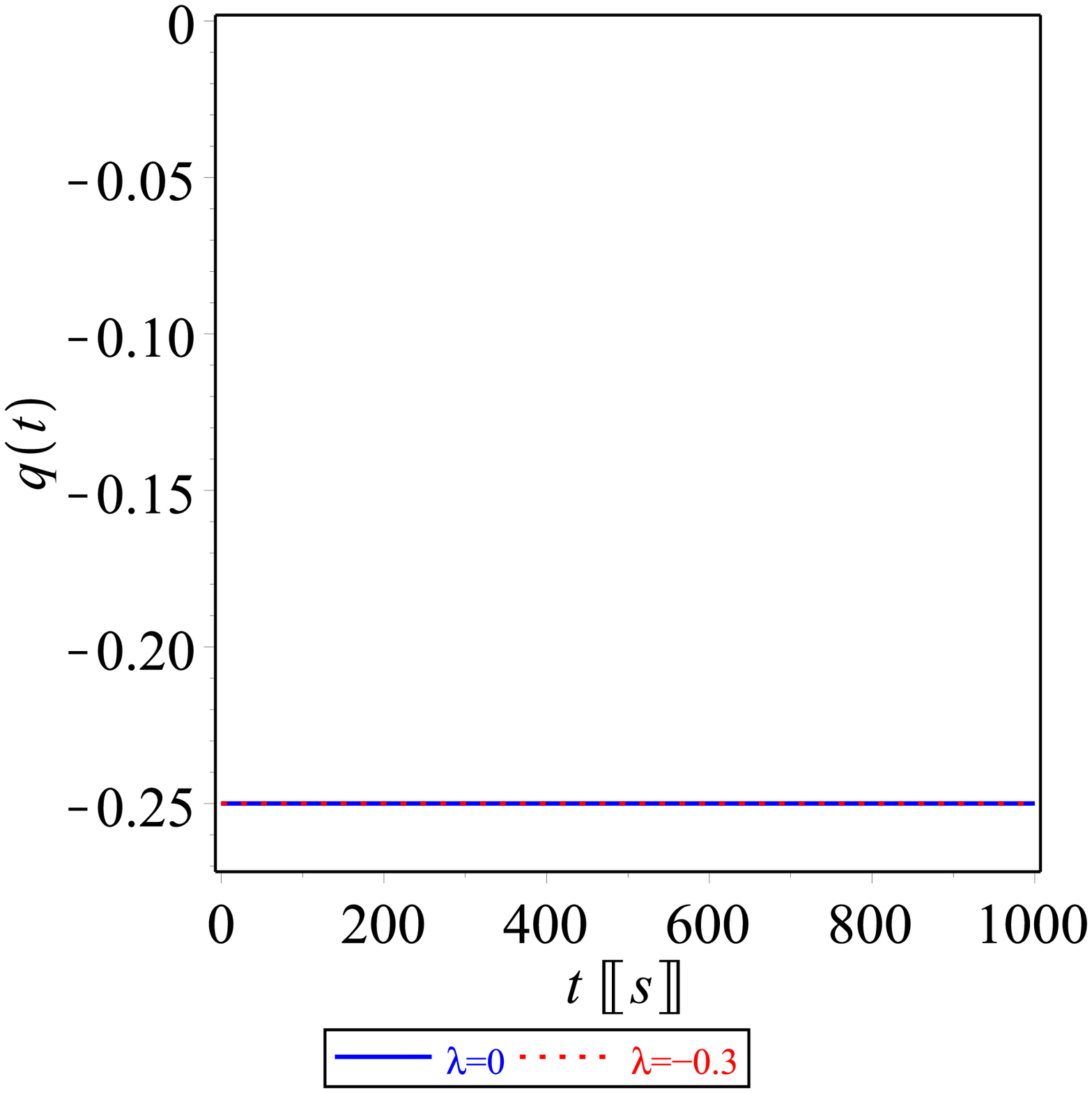}}\hspace{0.2cm}
%\subfigure[\, Displays the density  parameter vs. the cosmic time for the closed universe.]{\label{fig:1d}\includegraphics[scale=0.3]{Den}}
\caption{Plots of the scale factor $a(t)$, Hubble parameter $H(t)$ and deceleration parameter $q(t)$
%and density parameter $\rho(t)$
in the dust case.}
\label{Fig:1}
\end{figure}

 {Figure \ref{Fig:1} \subref{fig:1a}: presents the closed universe behavior of the scale factor versus the cosmic time $t$,  taking into account the effect of  $h$. The behavior of the scale in the present case is different from that in GR. { For $h>0$, the scale factor in the case of the open universe and under the effect  of $h$ begins with zero and then increases,  in  GR this factor takes negative values. Thus, we exclude this case from our consideration.}
 Figure  \ref{Fig:1}\subref{fig:1b}: shows that  the Hubble  parameter decreases with the cosmic time and that its behavior under the effect of  $h$ coincides with that in GR. Figure  \ref{Fig:1}\subref{fig:1c}: shows that  the deceleration   parameter has a constant negative value with the cosmic time and that its behavior under the effect of  $h$ coincides  from that in GR. { The case  $h<0$ is not physically because either $k>0$ or $k<0$ we always has an imaginary quantity in the scale factor.}
 %Finally, figure  \ref{Fig:1}\subref{fig:1d}: shows that  the density  parameter decreases with the cosmic time and that its behavior under the effect of  $h$ coincides  from that in GR.
 }

We studied the dust case ($w=0$) in the present section, in the following sections we discuss other cases of the EoS parameter.
\section{ Solution of M\o ller's field equations using different forms of EoS}\label{3s}
The most common examples of cosmological fluids with a constant $\omega$ are the dust ($\omega=0$), radiation ($\omega=1/3$) and dark energy ($\omega=-1$) cases \cite{Akarsu:2011zd}. Substituting Eq. (\ref{EoS}) into Eqs. (\ref{16}) and (\ref{17}) we obtain
\begin{equation}\label{32}
\frac{\dot a^2}{a^{2}}+\frac{h}{a^{2}}=\frac{\chi}{3}\rho,
\end{equation}
\begin{equation}\label{33}
\frac{2\ddot a}{a}+\frac{\dot a^2}{a^{2}}+\frac{h}{a^{2}}=-\chi \omega\rho\,.
\end{equation}
We now solve   Eqs. (\ref{32}) and (\ref{33}) together to obtain  $a(t)$ and $\rho(t)$ for different values of $\omega$.
\begin{itemize}
	\item[(i)]\underline{ The radiation case, i.e., $\omega=1/3$}:\\
Substituting   $\omega=1/3$, into Eqs.  (\ref{32}) and (\ref{33}) we obtain the following energy density equation:
\begin{equation}\label{34}
\rho(t)=\frac{2304c_{1}^{2}}{(c_{1}^{2}c_{2}^{2}+c_{1}^{2}t^{2}+2c_{2}c_{1}^{2}t-768\chi)^{2}}\,,
\end{equation}
{ where $c_1$ and $c_2$ are constants of integration.}
The cosmic scale factor takes the form:
\begin{equation}\label{35}
a(t)=\frac{1}{c_{1}}\sqrt{h(768\chi-c_{1}^{2}c_{2}^{2}-c_{1}^{2}t^{2}-2c_{2}c_{1}^{2}t)}\,.
\end{equation}
Because $p=\omega\rho$, the pressure of a fluid is given as follows:\\
\begin{equation}\label{36}
p(t)=\frac{768c_{1}^{2}}{(c_{1}^{2}c_{2}^{2}+c_{1}^{2}t^{2}+2c_{2}c_{1}^{2}t-768\chi)^{2}}\,.
\end{equation}
Using Eq. (\ref{35}), we obtain the Hubble parameter in the following form:
\begin{equation}\label{37}
H(t)=\frac{c_{1}^{2}t+c_{2}c_{1}^{2}}{(c_{1}^{2}c_{2}^{2}+c_{1}^{2}t^{2}+2c_{2}c_{1}^{2}t-768\chi)}\,.
\end{equation}
Using Eqs. (\ref{24}) and (\ref{37}), we obtain a formula for the deceleration as follows:
\begin{equation}\label{38}
q(t)=\frac{768\chi}{c_{1}^{2}(t+c_{2})^{2}}.
\end{equation}
Using Eqs. (\ref{29}) and (\ref{35}), we  obatin the relation between the cosmic time and  redshift as follows:
\begin{equation}\label{39}
t=\frac{1}{c_{1}}\bigg({\sqrt{-\frac{c_{1}^{2}}{h(1+z)^{2}}+768\chi}-c_{1}c_{2}}\bigg)\,,
\end{equation}
\end{itemize}
where $c_1$ and $c_2$ are constants of integration.
{ In this  study, we use the boundary condition $a(0)=0$ and $a(t_0)=1$ and get
 $c_1=\frac{32\sqrt{3\chi}t_0 h}{1+h t_0{}^2}$ and $c_2=\frac{h t_0{}^2+1}{h t_0}$.  The parameter $h$ in this case  should take a negative value  so that the value under the square root of Eqs. (\ref{35}) and (\ref{39}) become real. This  means  $\lambda$ must be  $\lambda>1/3$ if we deal with a closed universe and  $\lambda<1/3$ if we deal with open universe.}
\begin{figure}
\centering
 \subfigure[\, Displays the scale factor vs.  cosmic time $t$ for the open universe.]{\label{fig:2a}\includegraphics[scale=0.27]{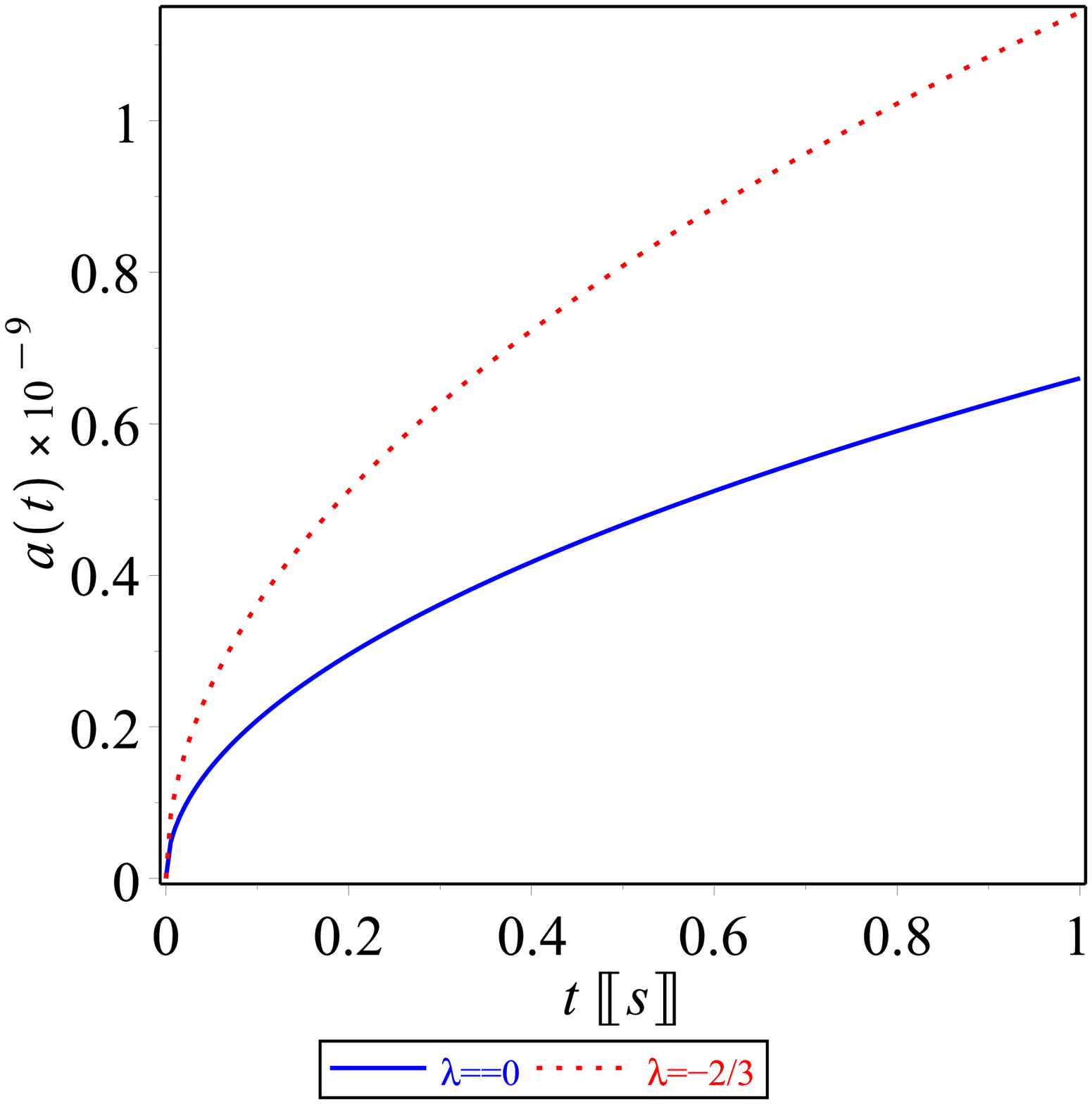}}\hspace{0.2cm}
\subfigure[\, Displays the Hubble parameter vs.  cosmic time $t$ for the open universe.]{\label{fig:2b}\includegraphics[scale=0.27]{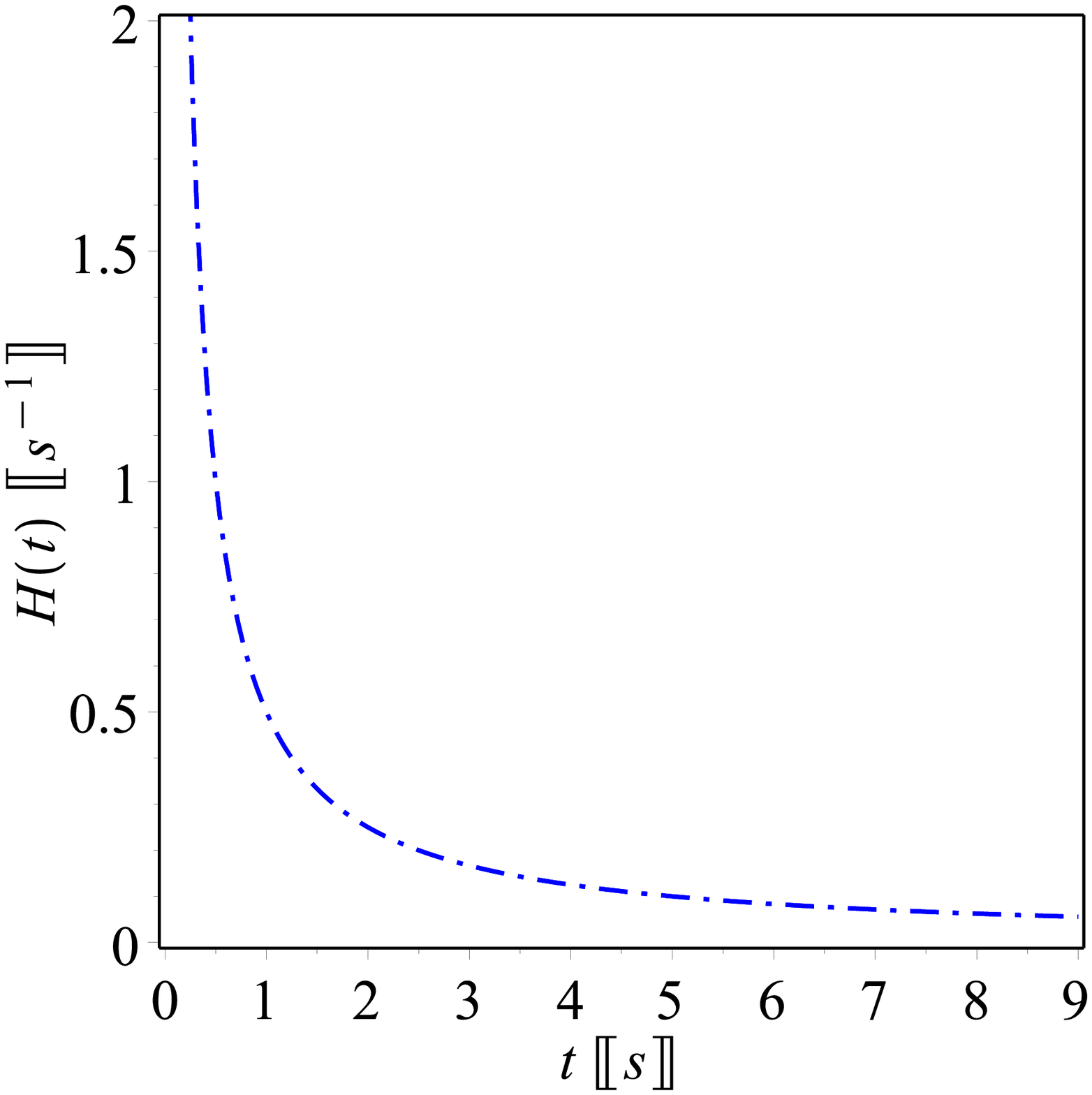}}
\subfigure[ \, Displays the deceleration parameter vs.  cosmic time $t$ for the open universe.]{\label{fig:2c}\includegraphics[scale=0.27]{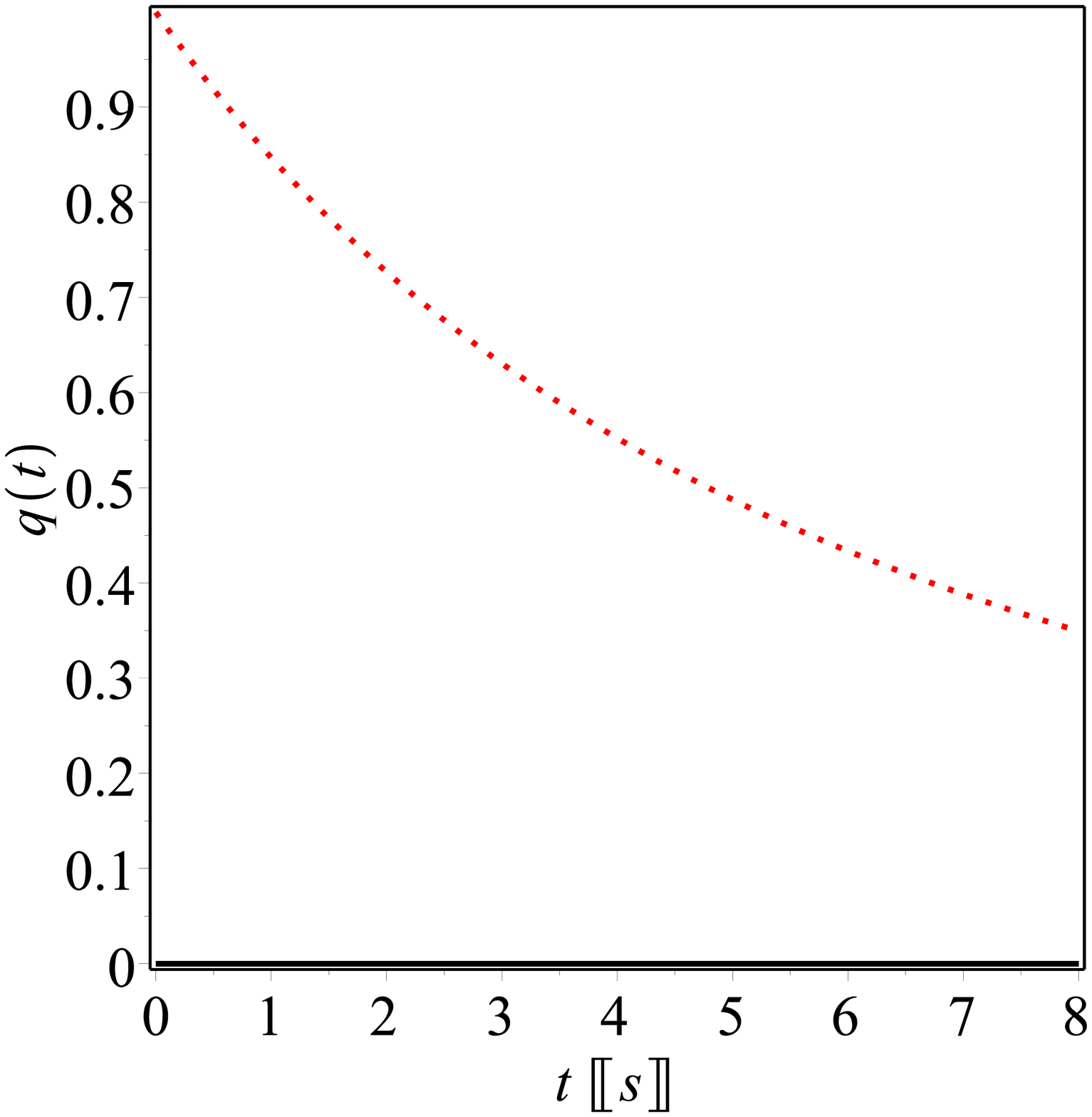}}
\caption{Plots of the scale factor $a(t)$,  Hubble parameter $H(t)$, and  deceleration parameter $q(t)$ in the radiation case.}
\label{Fig:2}
\end{figure}

 { In Figure \ref{Fig:2} \subref{fig:2a}}, we depict  the scale factor versus the cosmic time.  { The behavior of the scale factor in the open universe differs from that in the closed universe. In the closed universe,  the model begins with a big bang at $t=0$ and then evolves in a monotonically increasing manner for $\lambda\neq 0$. In the GR , $\lambda=0$, the scale factor has an oscillatory behavior with a future finite time Big Crunch singularity\footnote{{ It will be difficult to plot the scale factor of open universe and compare it with the plot of GR.}}.}  In the open universe, the scale factor increases with the cosmic time in both GR and MTT. We can thus say that the scale factor expands during a radiation-dominated epoch, which is consistent with observations \cite{Wetterich:2013aca}.
{  In Figure \ref{Fig:2} \subref{fig:2b}}, we depict the Hubble parameter versus the cosmic time $t$, this parameter takes positive values and decreases with the cosmic time.
 { In Figure \ref{Fig:2} \subref{fig:2c}},  we plot  the deceleration parameter versus the cosmic time $t$. This parameter has a positive value which means a decelerating epoch occurs during the expansion of the universe in a radiation dominated era ( i.e., for a decelerating universe, $q<0$ and vice versa \cite{Mamon:2016dlv}).

\begin{itemize}
	\item[(ii)] \underline{The case of dark energy, i.e., \ $\omega=-1$}:\\
In this case, the energy-density and pressure are given as follows:
\begin{equation}\label{45}
\rho(t)=c_{1}, \qquad \qquad  p(t)=-c_{1}\,.
\end{equation}
The scale factor, Hubble parameter, and deceleration parameter of this model take the following forms:
\begin{equation}\label{46}
a(t)=\frac{\sqrt{3}[9h+\exp(-2\sqrt{3\chi c_{1}}(-t+c_{2})/3)]}{[6\sqrt{\chi c_{1}}\exp(-\sqrt{3\chi c_{1}}(-t+c_{2})/3)]}\,,
\end{equation}
\begin{equation}\label{47}
H(t)=-\frac{\sqrt{3\chi c_{1}}[9h-\exp(-2\sqrt{3\chi c_{1}}(-t+c_{2})/3)]}{[27h+3\exp(-2\sqrt{3\chi c_{1}}(-t+c_{2})/3)]}\,,
\end{equation}
and
\begin{equation}\label{48}
q(t)=-\frac{81[h+\exp(-2\sqrt{3\chi c_{1}}(-t+c_{2})/3)/9]^{2}}{[9h-\exp(-2\sqrt{3\chi c_{1}}(-t+c_{2})/3)]^{2}}.
\end{equation}
Using the above equations, we obtain the cosmic time as a function of the redshift as follows:
\begin{equation}\label{49}
t=c_{2}+\frac{3\ln\bigg(9h\sqrt{3}(1+z)/[6\sqrt{\chi c_{1}}-2\sqrt{3}(1+z)]\bigg)}{\sqrt{3\chi c_{1}}}\,.
\end{equation}
\end{itemize}
In the case of $w=-1$, we obtain a vacuum solution, because the energy density and pressure take  constant values, which is similar to AdS. { Eqs. (\ref{46}) $\sim$ (\ref{48})  in the late
time limit, i.e.,  $t>>t_0$,  reduce to $a(t) \propto exp(H_0t)$; $H\rightarrow H_0 =\sqrt{\frac{\chi c_1}{3}}$; $q(t)\rightarrow -1$ which means that in late time dark energy domination is no longer influenced by the shape of the
universe.}
In the following section, we calculate a number of thermodynamic quantities for the dust and radiation cases.
\section{ Thermodynamic behavior of the dust and radiation cases}\label{4p}
The apparent horizon of the FRW model for the non-flat case takes the form \cite{Akbar:2006kj}:
\begin{equation}\label{50}
r_{\mathrm a}=\frac{1}{\sqrt{H^{2}+\frac{k}{a^{2}}}}\,.
\end{equation}
For the flat universe, the value of the apparent horizon is identical to the inverse of the Hubble horizon. The  Hawking temperature $T$ and Hawking  entropy $S$  of the FRW model take  the forms \cite{Cai:2008gw}:\\
\begin{equation}\label{51}
T=\frac{1}{2\pi r_{\mathrm  a}}\,,\qquad \qquad
S=\frac{A}{4}\,.
\end{equation}
 Equation (\ref{51}) shows that the entropy is  proportional to the area ($A=4\pi r_{a}^{2}$) of the apparent horizon. We calculate the apparent horizon, $T$ and $S$ for the dust and radiation cases discussed in Section \ref{FRW}.\\
 \begin{itemize}
	\item[(i)] \underline{The dust case}\\
Using Eqs. (\ref{21}), (\ref{22}), (\ref{23}) and (\ref{50}), we obtain:\\
%\begin{equation}\label{56}
%r_{\mathrm a}(\beta)=\begin{cases}
%b(1-\cos\beta)^{2}/\sqrt{h\sin^{2}\beta+k(1-\cos\beta)^{2}}\,,        &  \textrm{for} \quad   h>0\,,\\
%\\
%b(\cosh\beta-1)^{2}/\sqrt{h\sinh^{2}\beta+k(\cosh\beta-1)^{2}}\,,    &  \textrm{for} \quad    h<0\,,
%\end{cases}
%\end{equation}
{\begin{equation}\label{56r}
r_{\mathrm a}(t)\approx\begin{cases}
\displaystyle\frac {3\sqrt {2} \sqrt [3]{h}t \left( 2\,\sqrt [3]{6\,b^2}-\sqrt [3]{h\,t^2}\right)}{4\sqrt {2\,{h}^{2/3}{b}^{4/3}{6}^{2/3}-4
\,h\sqrt [3]{6\,t^2\,b^2}+2\,{h}^{4/3}{t}^{4/3}+3\,
\sqrt [3]{6\,b^2\,t^2}}}\,,        &  \textrm{for}\,   h>0\,,\\
\\
\displaystyle\frac {3\sqrt {2} \sqrt [3]{\mid h\mid}t \left( 2\,\sqrt [3]{6\,b^2}+\sqrt [3]{\mid h\mid\,t^2}\right)}{4\sqrt {2\,{h}^{2/3}{b}^{4/3}{6}^{2/3}+4
\,h\sqrt [3]{6\,t^2\,b^2}+2\,{h}^{4/3}{t}^{4/3}+3\,
\sqrt [3]{6\,b^2\,t^2}}}
\,,    &  \textrm{for} \quad    h<0\,,
\end{cases}
\end{equation}}
$T$ is given by:
%\begin{equation}\label{57}
%T(\beta)=\begin{cases}
%\sqrt{h\sin^{2}\beta+k(1-\cos\beta)^{2}}/2\pi b(1-\cos\beta)^{2}\,,        &  \textrm{for}   \quad h>0\,,\\
%\\
%\sqrt{h\sinh^{2}\beta+k(\cosh\beta-1)^{2}}/2\pi b(\cosh\beta-1)^{2}\,,     &  \textrm{for} \quad   h<0\,,
%\end{cases}
%\end{equation}
{
\begin{equation}\label{57r}
T(t)\approx\begin{cases}
\displaystyle\frac{2\sqrt {2\,{h}^{2/3}{b}^{4/3}{6}^{2/3}-4
\,h\sqrt [3]{3\,t^2\,b^2}+2\,{h}^{4/3}{t}^{4/3}+3\,
\sqrt [3]{6\,b^2\,t^2}}}{6\pi\sqrt {2} \sqrt [3]{h}t \left( 2\,\sqrt [3]{6\,b^2}-\sqrt [3]{h\,t^2}\right)}   \,,        &  \textrm{for}   \quad h>0\,,\\
\\
\displaystyle\frac {2\sqrt {2\,{h}^{2/3}{b}^{4/3}{6}^{2/3}+4
\,h\sqrt [3]{6\,t^2\,b^2}+2\,{h}^{4/3}{t}^{4/3}+3\,
\sqrt [3]{6\,b^2\,t^2}}}{3\pi\sqrt {2} \sqrt [3]{\mid h\mid}t \left( 2\,\sqrt [3]{6\,b^2}+\sqrt [3]{\mid h\mid\,t^2}\right)}\,,     &  \textrm{for} \quad   h<0\,,
\end{cases}
\end{equation}}
and  $S$ is given by:
%\begin{equation}\label{58}
%S(\beta)=\begin{cases}
%\pi b^{2}(1-\cos\beta)^{4}/[h\sin^{2}\beta+h(1-\cos\beta)^{2}]\,,       &  \textrm{for} \quad  h>0\,,\\
%\\
%\pi b^{2}(\cosh\beta-1)^{4}/[h\sinh^{2}\beta+k(\cosh\beta-1)^{2}]\,,     &  \textrm{for} \quad  h<0\,.
%\end{cases}
%\end{equation}
{
\begin{equation}\label{58r}
S(t)\approx\begin{cases}
\displaystyle\frac {9\pi\sqrt [3]{h^2}t^2 \left( 2\,\sqrt [3]{6\,b^2}-\sqrt [3]{h\,t^2}\right)^2}{8\left(2\,{h}^{2/3}{b}^{4/3}{6}^{2/3}-4
\,h\sqrt [3]{6\,t^2\,b^2}+2\,{h}^{4/3}{t}^{4/3}+3\,
\sqrt [3]{6\,b^2\,t^2}\right)}\,,     &  \textrm{for} \quad  h>0\,,\\
\\
\displaystyle\frac {9\pi\sqrt [3]{h^2}t^2 \left( 2\,\sqrt [3]{6\,b^2}+\sqrt [3]{\mid h\mid\,t^2}\right)^2}{8\left(2\,{h}^{2/3}{b}^{4/3}{6}^{2/3}+4
\,h\sqrt [3]{6\,t^2\,b^2}+2\,{h}^{4/3}{t}^{4/3}+3\,
\sqrt [3]{6\,b^2\,t^2}\right)}\,,     &  \textrm{for} \quad  h<0\,.
\end{cases}
\end{equation}}
\end{itemize}
\begin{figure}
\centering
 \subfigure[\,{ Displays the apparent horizon vs.  cosmic time  for the closed universe of Eq. (\ref{56r}) when $h<0$ and $\lambda=0.4$.}]{\label{fig:3a}\includegraphics[scale=0.3]{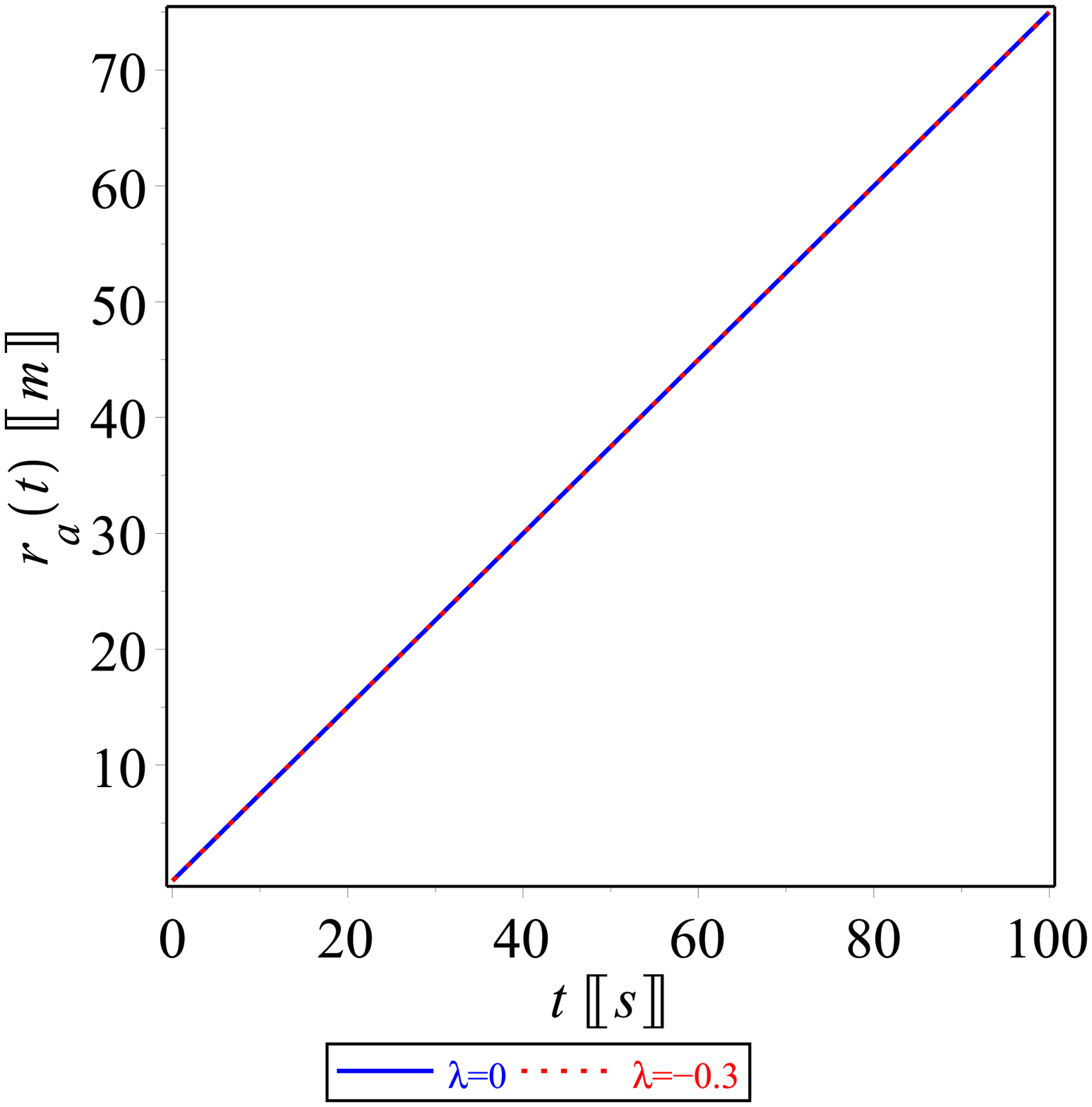}}\hspace{0.2cm}
\subfigure[\,{ Displays the Hawking temperature vs. cosmic time  for the closed universe of Eq. (\ref{56r}) when $h<0$ and $\lambda=0.4$.}]{\label{fig:3b}\includegraphics[scale=0.3]{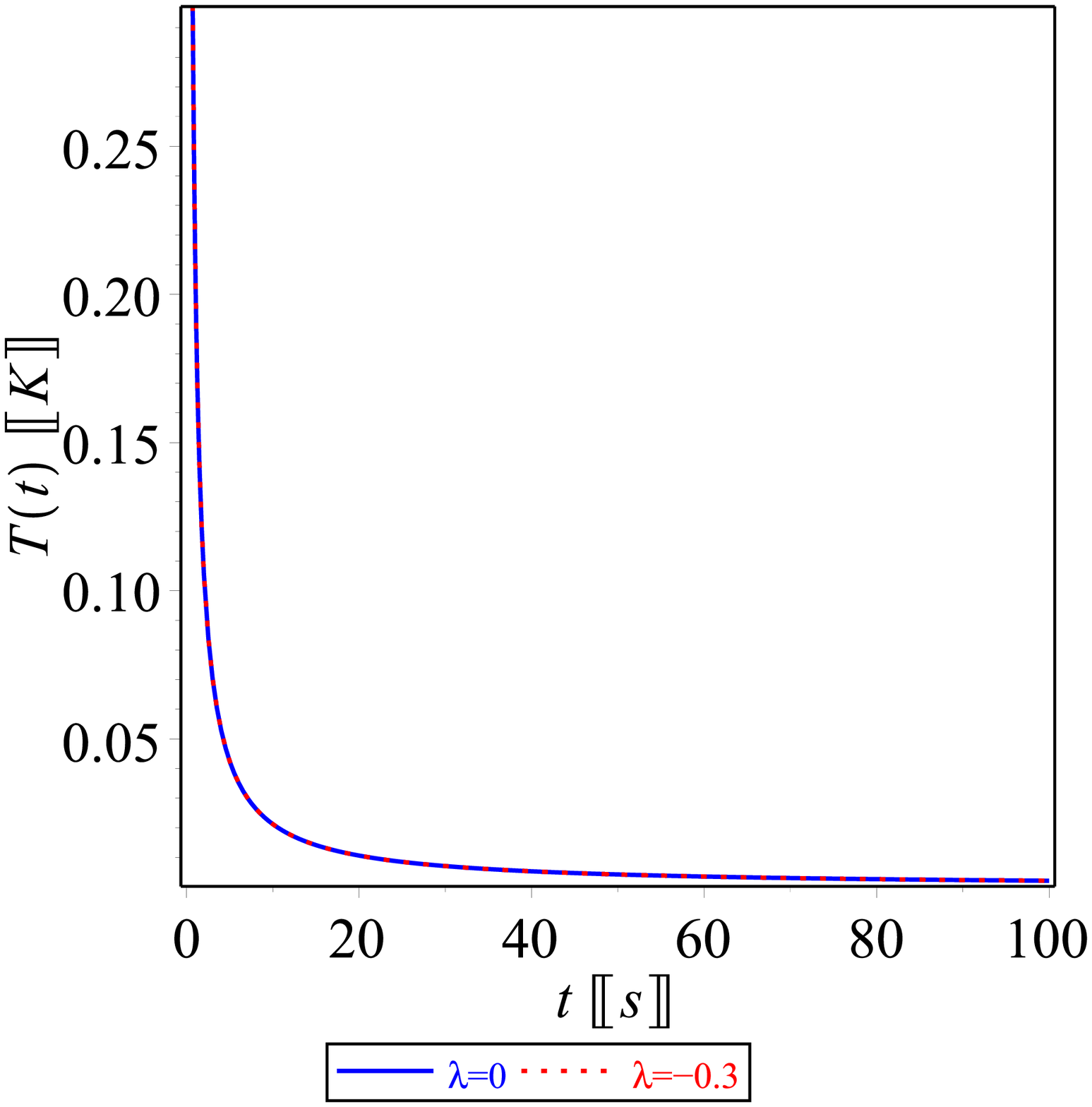}}
\caption{Plots of the apparent horizon and Hawking temperature  in the dust case.}
\label{Fig:3}
\end{figure}
  Figures \ref{Fig:3}\subref{fig:3a} and \ref{Fig:3}\subref{fig:3b} show that the apparent horizon increases whereas $T$ decreases with the cosmic time, and that both parameters have positive values. { The Figure \ref{Fig:3} considers} the closed universe of Eq. (\ref{56r}) when $h<0$ and $\lambda=0.4$. We do not consider the case of open universe because this case makes the radius horizon takes a negative value. 	
 \begin{itemize}
	\item[(ii)] \underline{The radiation case}\\
Substituting  Eqs. (\ref{35}) and (\ref{37}) into Eq.(\ref{50}), we obtain the apparent horizon as follows:
\begin{equation}\label{53}
r_{\mathrm a}=\frac{\sqrt{h}(c_{1}^{2}c_{2}^{2}+c_{1}^{2}t^{2}+2c_{2}c_{1}^{2}t-768\chi)}{\sqrt{h(c_{1}^{2}t+c_{2}c_{1}^{2})^{2}-kc_{1}^{2}(c_{1}^{2}c_{2}^{2}+
c_{1}^{2}t^{2}+2c_{2}c_{1}^{2}t-768\chi)}}\,.
\end{equation}
Substituting Eq. (\ref{53}) into Eq.(\ref{51}) we obtain the Hawking temperature and entropy in the following forms:\\
\begin{equation}\label{54}
T=\frac{\sqrt{h(c_{1}^{2}t+c_{2}c_{1}^{2})^{2}-kc_{1}^{2}(c_{1}^{2}c_{2}^{2}+c_{1}^{2}t^{2}+2c_{2}c_{1}^{2}t-768\chi)}}{2\pi \sqrt{h}(c_{1}^{2}c_{2}^{2}+c_{1}^{2}t^{2}+2c_{2}c_{1}^{2}t-768\chi)}
\end{equation}
\begin{equation}\label{55}
S=\frac{-\pi h[(t+c_{2})^{2}c_{1}^{2}-768\chi]^{2}}{c_{1}^{2}[(t+c_{2})^{2}(k-h)c_{1}^{2}-768k\chi}
\end{equation}
\end{itemize}
\begin{figure}
\centering
 \subfigure[\, {Displays the apparent horizon vs. cosmic time for the open universe of Eq. (\ref{56r}) when $h>0$ and $\lambda=-2/3$}.]{\label{fig:4a}\includegraphics[scale=0.3]{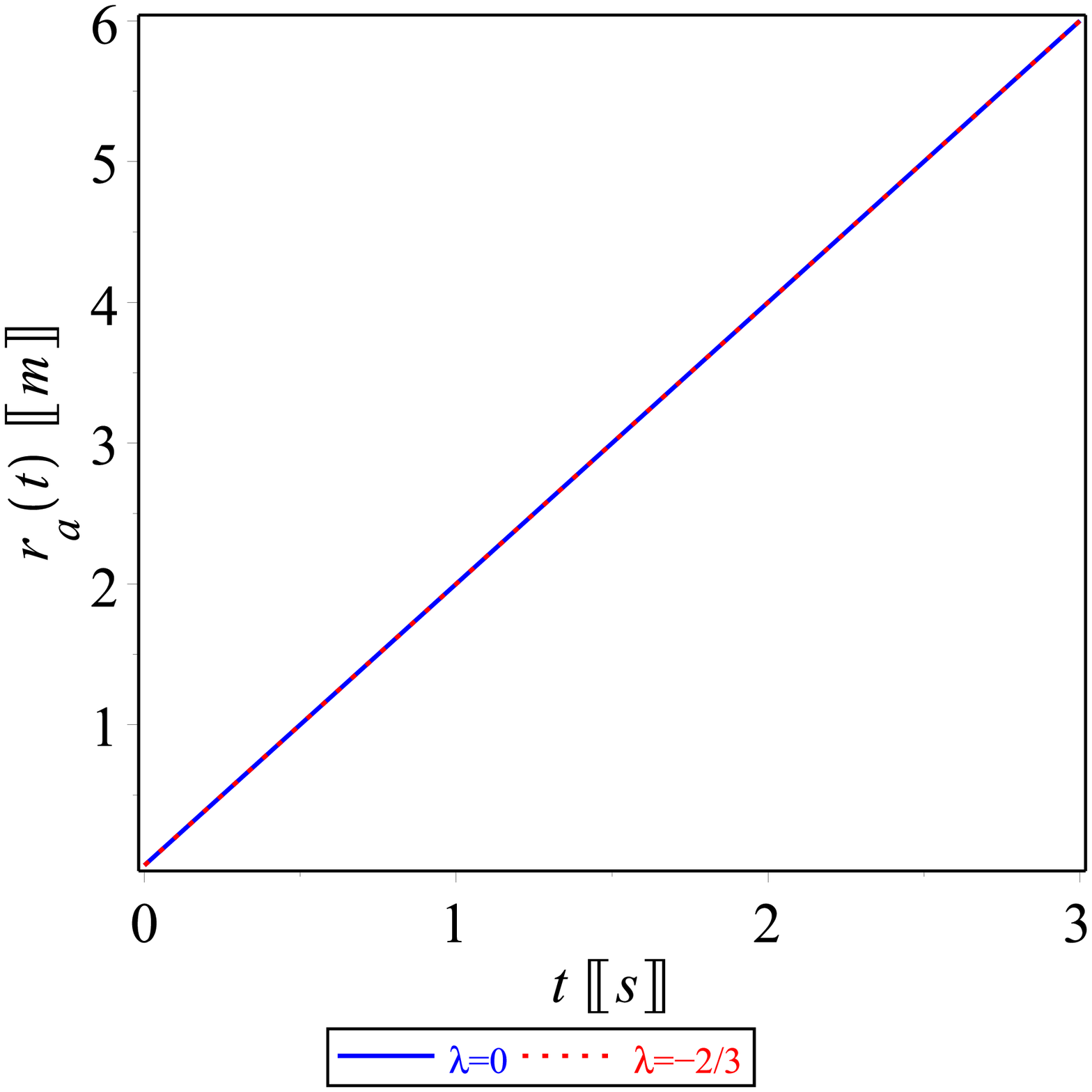}}\hspace{0.2cm}
\subfigure[\, { Displays the Hawking temperature vs. cosmic time for the open universe of Eq. (\ref{56r}) when $h>0$ and $\lambda=-2/3$}.]{\label{fig:4b}\includegraphics[scale=0.3]{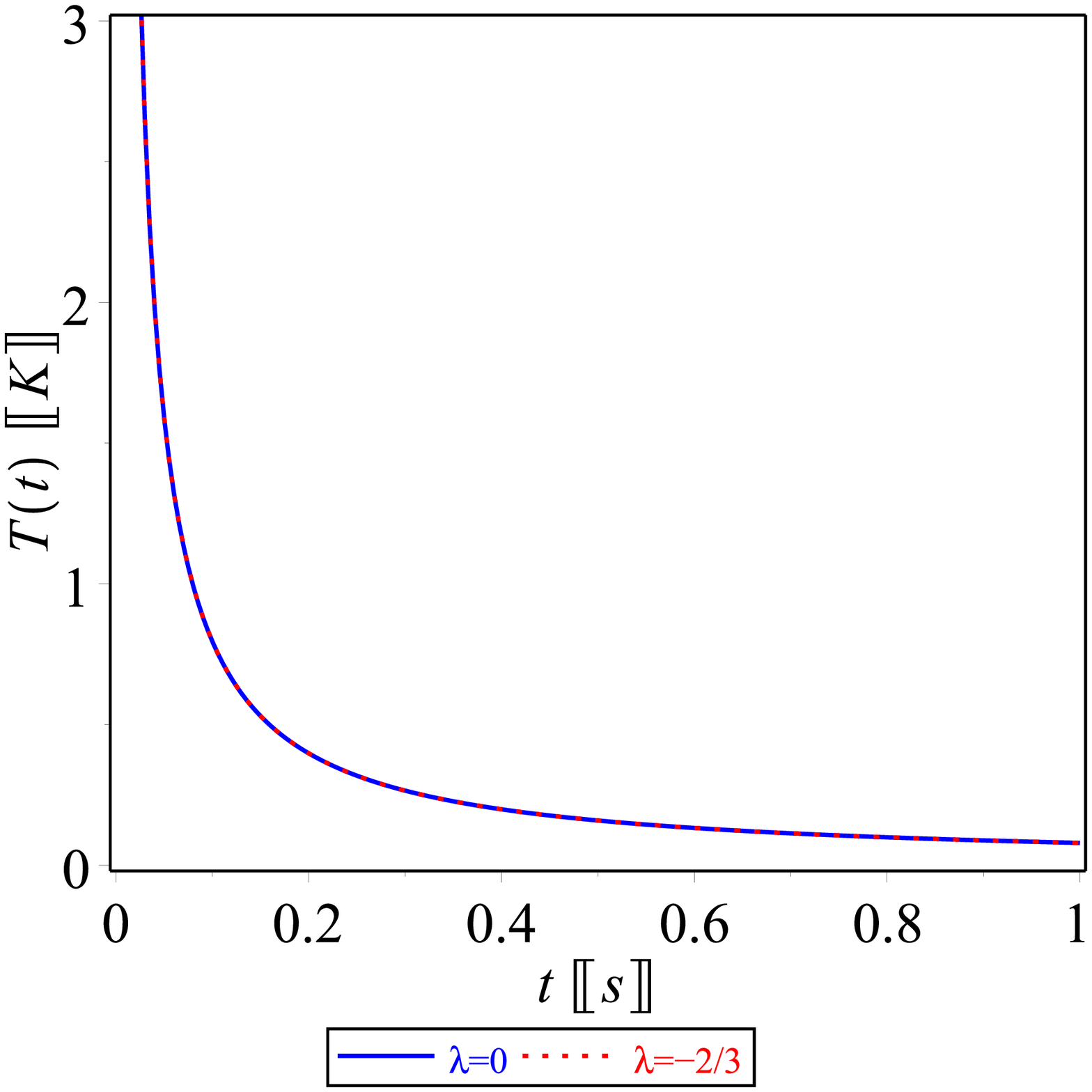}}
\caption{Plots of the apparent horizon and Hawking temperature  in the radiation case.}
\label{Fig:4}
\end{figure}

  { In Figure \ref{Fig:4}\subref{fig:4a}},  we plot the apparent horizon versus the cosmic time $t$, which shows that $r_a$ always has a positive value and increases with the cosmic time.  { The Figure \ref{Fig:4} considers  the open universe of Eq. (\ref{56r})} when $h>0$ and $\lambda=-2/3$. We do not consider the case of open universe because this case makes the radius horizon takes a negative value.
  %The behavior of $r_a$ under the effect of $h$ deviates slightly from its behavior in GR.
  In general, the apparent horizon is not constant but changes with time \cite{Akbar:2006mq}.
  Figure \ref{Fig:4}\subref{fig:4b}, shows that $T$ decreases with the cosmic time and is always positive. The behavior of $T$ under the influence of $h$ is not different from its behavior in GR. Therefore, the thermodynamic quantities in the case of radiation do not show a real difference from the GR framework because the effect of $h$ in the calculations is negligible.

In the following section, we study the effect of Tsallis entropy on  the dust and radiation cases.
\section{ Tsallis entropy}\label{tas}
In this section, we generalize the discussion presented in the previous section  by taking into account non-extensive effects. The generalization of entropy in non-extensive statistics, is described by the so-called Tsallis entropy \cite{Tsallis:2012js}, which is defined  as:\\
\begin{equation}\label{65}
S=\frac{\alpha}{4G}A^{\delta}\,,
\end{equation}
where $\alpha$ is a positive constant that has a dimension of $L^{2(1-\delta)}$, and $\delta$ is the non-extensive parameter. Equation (\ref{65}) shows that we recover GR when  $\delta=1$. From the first law of thermodynamics, we have:
\begin{equation}\label{66}
-dE=\delta Q,
\end{equation}
where $-dE=TdS$, is the internal energy, $T$ is the Hawking temperature, $S$ is the entropy,  and $\delta Q$ is the heat flow for an infinitesimal time interval $dt$ through the horizon, which is defined as \cite{Cai:2005ra,Lymperis:2018iuz}:\\
\begin{equation}\label{67}
\delta Q=A(\rho_{m}+p_{m})H r_{a}dt\,,
\end{equation}
where $\rho_{m}$ and $p_{m}$ are the matter energy density and pressure, respectively.
From Eq. (\ref{65}), we find that:\\
\begin{equation}\label{68}
dS=\frac{(4\pi)^{\delta}\, \delta\alpha \,  r_{\mathrm a}^{2\delta-1} \dot{r}_{\mathrm a} dt}{2G}\,,
\end{equation}
where $\dot{r}_{\mathrm a}$ is the derivative of the apparent  horizon with respect to the cosmic time and can easily be calculated from Eq.  (\ref{50}). Using Eqs. (\ref{50}), (\ref{51}), (\ref{66}) and (\ref{68}) we obtain:
\begin{equation}\label{69}
\frac{-(4\pi)^{2-\delta}G}{\alpha}(\rho_{m}+p_{m})=\delta \bigg(\dot H-\frac{k}{a^{2}}\bigg)\bigg(H^{2}+\frac{k}{a^{2}}\bigg)^{1-\delta},
\end{equation}
where $\dot H$ is the derivative of the Hubble parameter with respect to the cosmic time.
We can extract $\dot H$ from Eq.(\ref{69}) by using the conservation law of fluid matter, which takes the form:
\begin{equation}\label{70}
\dot{\rho}_{m}+3H(\rho_{m}+p_{m})=0\,.
\end{equation}
Substituting Eq.(\ref{70}) into Eq.(\ref{69}) and integrating both sides with respect to the cosmic time, we obtain:\\
\begin{equation}\label{71}
\frac{2(4\pi)^{2-\delta}}{3\alpha}G\rho_{m}=\frac{\delta}{2-\delta}\bigg(H^{2}+\frac{k}{a^{2}}\bigg)^{2-\delta}-\frac{f}{3\alpha},
\end{equation}
where $f$ is a constant of integration. Using Tsallis entropy,  we can obtain the modified Friedmann equations, as shown in Eqs. (\ref{69}) and (\ref{71}), as follows:
\begin{equation}\label{72}
H^{2}+\frac{h}{a^{2}}=\frac{8\pi G}{3}(\rho_{_{_{m}}}+\rho_{_{_{DE}}}),
\end{equation}
and,
\begin{equation}\label{73}
\dot H+H^{2}=-4\pi G\bigg[\frac{1}{3}(\rho_{_{_{m}}}+\rho_{_{_{DE}}})+p_{m}+p_{_{_{DE}}}\bigg],
\end{equation}
where $\rho_{_{_{DE}}}$ and $p_{_{_{DE}}}$ are the energy density and pressure of dark energy, respectively.\\
After rescaling $\tilde{\alpha}=\alpha(4\pi)^{1-\delta}$ and $\tilde{f}=f(4\pi)^{1-\delta}$ we obtain the matter and dark energy densities as follows:\\
\begin{equation}\label{74}
\rho_{_{_{m}}}=\frac{3\tilde{\alpha}}{8\pi G}\bigg[\frac{\delta}{2-\delta}\bigg(H^{2}+\frac{k}{a^{2}}\bigg)^{2-\delta}-\frac{\tilde{f}}{3\tilde{\alpha}}\bigg],
\end{equation}
\begin{equation}\label{75}
\rho_{_{_{DE}}}=\frac{3}{8\pi G}\bigg[H^{2}+\frac{h}{a^{2}}-\frac{\tilde{\alpha}\delta}{2-\delta}\bigg(H^{2}+\frac{k}{a^{2}}\bigg)^{2-\delta}+\frac{\tilde{f}}{3}\bigg].
\end{equation}
The pressures of matter and dark energy are given as:
\begin{equation}\label{76}
p_{_{_{m}}}=\frac{-\tilde{\alpha}}{4\pi G}\bigg[\delta\bigg(\dot H-\frac{k}{a^{2}}\bigg)\bigg(H^{2}+\frac{k}{a^{2}}\bigg)^{1-\delta}+\frac{3\delta}{2(2-\delta)}\bigg(H^{2}+\frac{k}{a^{2}}\bigg)^{2-\delta}-
\frac{\tilde{f}}{2\tilde{\alpha}}\bigg]\,,
\end{equation}
\begin{equation}\label{77}
 p_{_{_{DE}}}=\frac{-1}{4\pi G}\bigg[\dot H+\frac{3}{2}H^{2}+\frac{1}{2}\frac{h}{a^{2}}-\frac{3\tilde{\alpha}\delta}{2(2-\delta)}\bigg(H^{2}+\frac{k}{a^{2}}\bigg)^{2-\delta}-\tilde{\alpha}
\delta\bigg(\dot H-\frac{k}{a^{2}}\bigg)\bigg(H^{2}+\frac{k}{a^{2}}\bigg)^{1-\delta}+\frac{\tilde{f}}{2}\bigg].
\end{equation}
Using Eqs. (\ref{77}) and (\ref{75}), we can obtain the EoS parameter for dark energy as follows:
\begin{equation}\label{78}
\omega_{_{_{DE}}}=-\frac{1}{3}-\frac{2\bigg[\dot H+H^{2}-\tilde{\alpha}\delta\bigg(\dot H-\frac{k}{a^{2}}\bigg)\bigg(H^{2}+\frac{k}{a^{2}}\bigg)^{1-\delta}-\frac{\tilde{\alpha}\delta}{2-\delta}\bigg(H^{2}+\frac{k}{a^{2}}\bigg)^{2-\delta}
+\frac{\tilde{f}}{3}\bigg]}{3\bigg[H^{2}+\frac{h}{a^{2}}-\frac{\tilde{\alpha} \delta}{2-\delta}\bigg(H^{2}+\frac{k}{a^{2}}\bigg)^{2-\delta}+\frac{\tilde{f}}{3}\bigg]}.
\end{equation}
Equation (\ref{78}) shows that $\delta$ can take any value except 2, scholars generally believe that $\delta$ should be greater than $1$ \cite{Azmi:2015xqa,Cleymans:2013rfq}. If $\delta=1$, we obtain the standard extensive thermodynamic case of GR; if $\delta>1$ we obtain the non-extensive thermodynamic case. To investigate the pure effect of $\delta$, we set $\alpha$ to its standard value, i.e., $\alpha=1$ \cite{Lymperis:2018iuz}. We discuss the  results of Section \ref{tas} { through the following Figures.}
\begin{figure}
\centering
 \subfigure[\,Displays the energy density vs. the cosmic time for the open universe in the radiation case for Eqs. (\ref{74}) and (\ref{75}).]{\label{fig:5a}\includegraphics[scale=0.3]{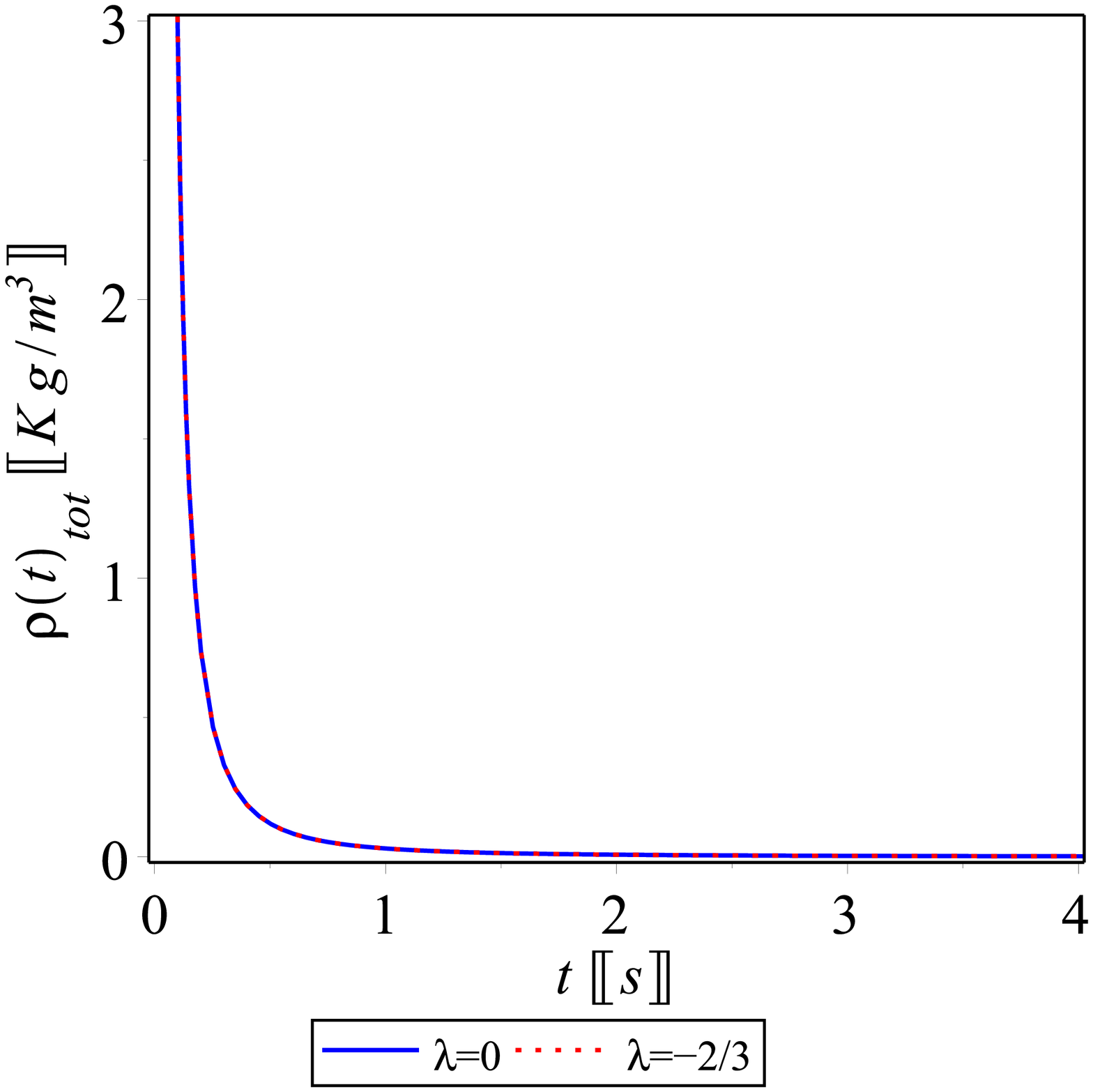}}\hspace{0.2cm}
\subfigure[\,Displays the energy density vs. the cosmic time for the closed universe in the dust case for Eqs. (\ref{74}) and (\ref{75}).]{\label{fig:5b}\includegraphics[scale=0.3]{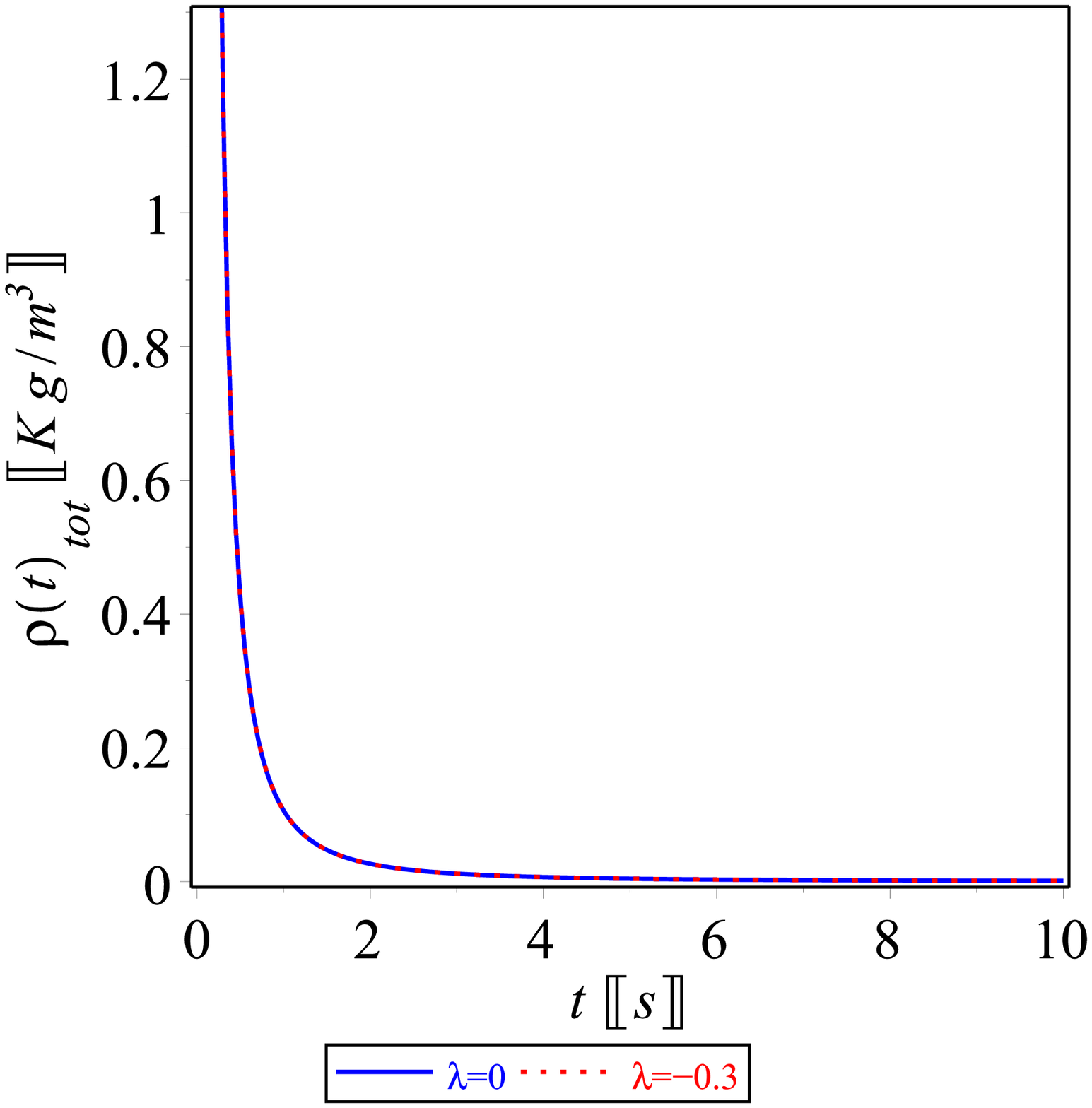}}
\caption{Plots of  energy density for the $ \ref{Fig:5}\, \subref{fig:5a}$ radiation  and $\ref{Fig:5}\, \subref{fig:5b}$ dust cases.}
\label{Fig:5}
\end{figure}
 { In Figure \ref{Fig:5}\subref{fig:5a}}, we plot the energy density $\rho(t)$ for the open universe in the radiation case. The energy density takes a positive value and decreases with the cosmic time. The Tsallis parameter has no effect on the energy density; $h$  exerts some effect on this parameter but is generally negligible. Thus, the results coincide with the GR case. The open cosmological model is acceptable, because the energy density condition ($\rho\geq 0$) is satisfied  \cite{1983NCimB..74..182B}.   {  In Figure \ref{Fig:5}\subref{fig:5b}}, we plot the energy density for the closed universe in the dust case. The behavior of the energy density when $\lambda=-0.3$  has no different from its behavior at $\lambda=0$. The energy density of the open universe, in the dust case, generally takes positive values; however, when $\lambda=0$, i.e., at the GR limit,  it takes negative values.  Therefore, we exclude the open universe from our consideration.
 \begin{figure}
\centering
 \subfigure[\,Displays the entropy vs.  $\beta$  for the closed universe in the dust case]{\label{fig:6a}\includegraphics[scale=0.3]{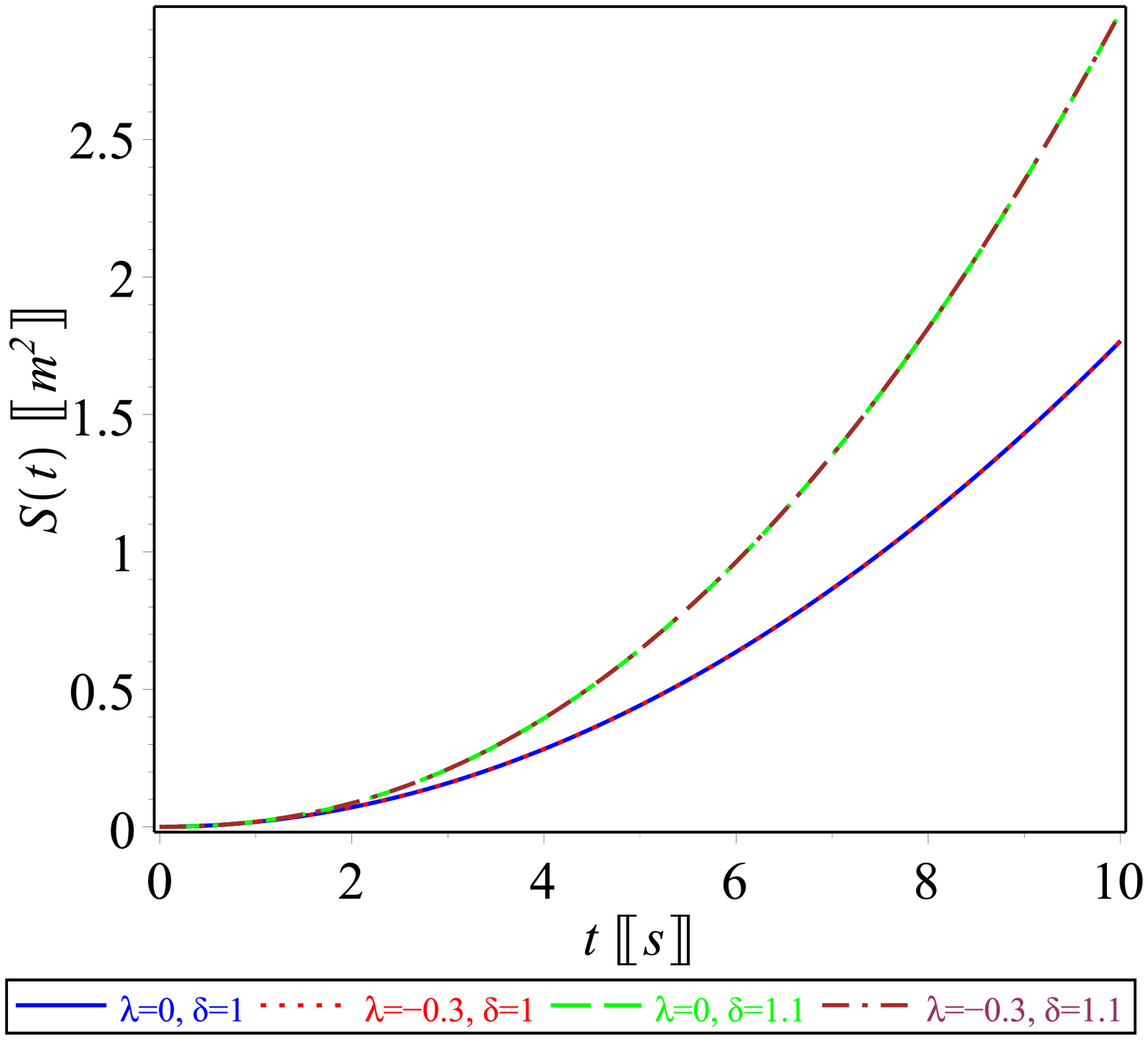}}\hspace{0.2cm}
\subfigure[\,Displays the entropy vs. the cosmic time for the open universe in the radiation case]{\label{fig:6b}\includegraphics[scale=0.3]{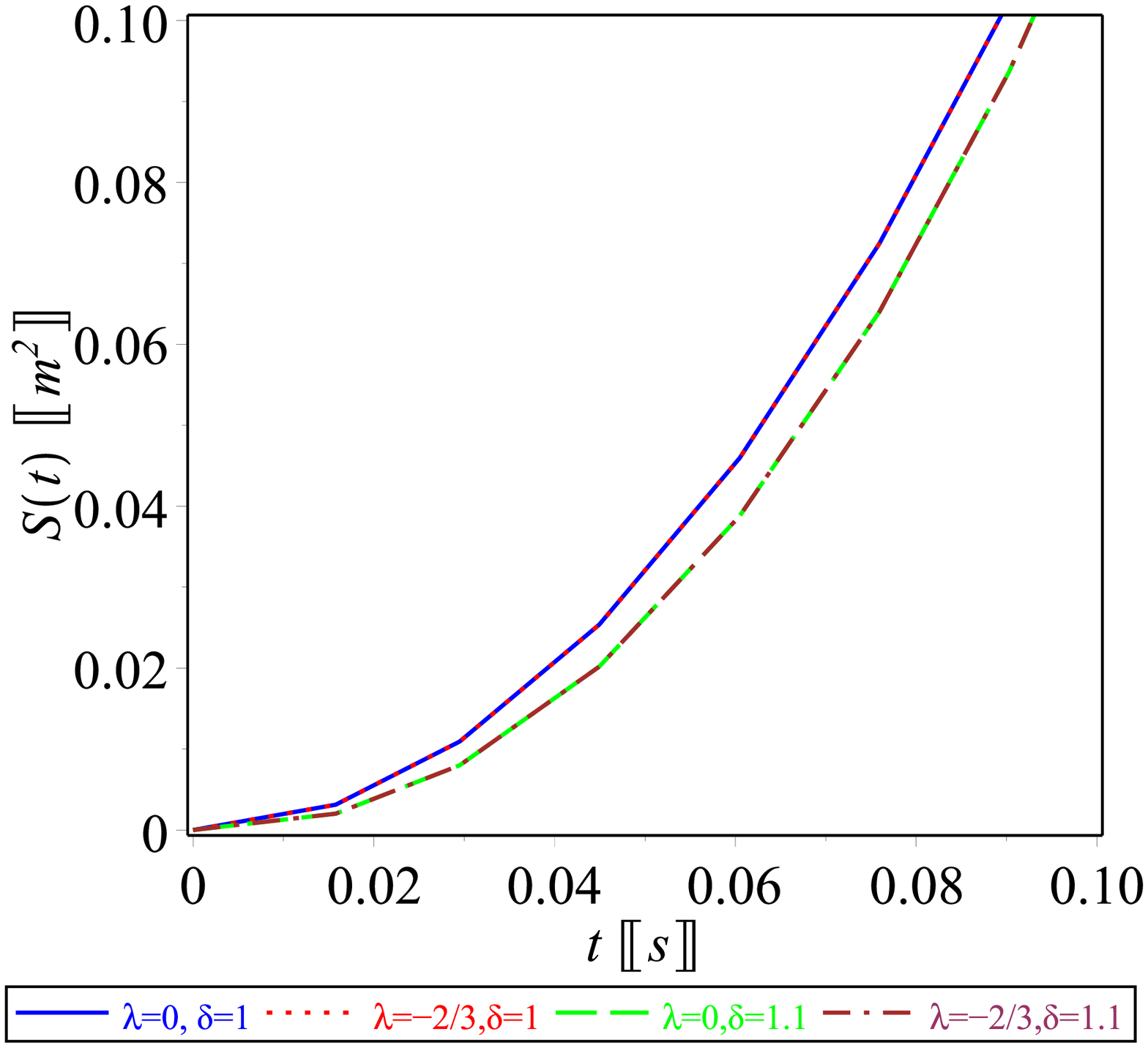}}
\caption{Plots of  entropy for the $\subref{fig:6a}$ dust case and $\subref{fig:6b}$ radiation case.}
\label{Fig:6}
\end{figure}

 { In Figure \ref{Fig:6}\subref{fig:6a}}, we plot the entropy in the dust case when $\lambda=-0.3$, and  $\lambda=0$. The behavior of entropy  when  $\lambda=-0.3$, is different from its behavior when $\lambda=0$. In both cases, we note that the standard entropy differs from the Tsallis entropy.  { In Figure \ref{Fig:6}\subref{fig:6b}}, we plot the standard and Tsallis entropies for the open universe in the radiation case  when $\lambda=-2/3$ and  $\lambda=0$. The behavior of the standard entropy differs from that of the Tsallis entropy, and both increase with the cosmic time. The  behavior of the entropy in the case of $\lambda=-2/3$ differs slightly from its behavior in the case of $\lambda=0$.
 \begin{figure}
\centering
 %\subfigure[\,Displays the EoS vs.  $t$  for the open universe in the radiation case]{\label{fig:7a}\includegraphics[scale=0.3]{20}}\hspace{0.2cm}
{\label{fig:7}\includegraphics[scale=0.4]{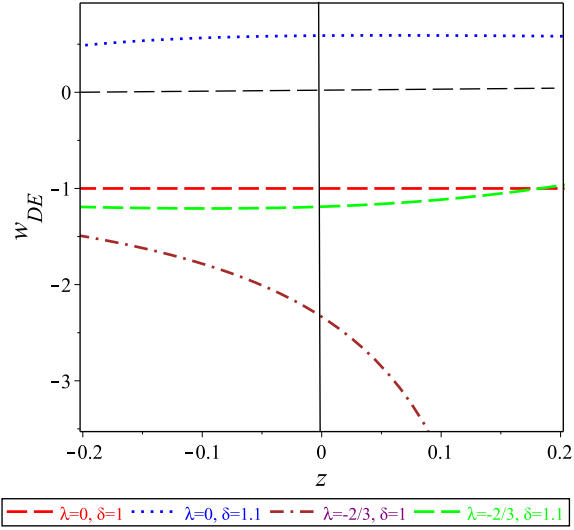}}
\caption{Behaviors of  the EoS of dark energy versus redshift, for the open universe in the radiation case using different values of
model dimensionless parameters $\lambda$  and $\delta$.}
\label{Fig:7}
\end{figure}
 % \begin{figure}
%\centering
% {\label{fig:7a}\includegraphics[scale=0.4]{20}}\hspace{0.2cm}
%\subfigure[Displays the entropy versus the cosmic time for the open universe in the radiation case.]{\label{fig:6b}\includegraphics[scale=0.4]{11}}
%\caption{Plot the EoS of dark energy for the radiation case}
%\label{Fig:7}
%\end{figure}

{ In Figure \ref{Fig:7}}, we plot the EoS of dark energy for the radiation case, and obtain $\omega_{DE}=-1$ (representing dark energy), when $\lambda=0$ and $\delta=1$. When $\lambda=0$ and $\delta=1.1$, $\omega_{DE}$ starts with a positive value and then decreases until it approaches $-1$, after which the universe enters a phantom regime. The parameter $h$ has a significant effect on the behavior of $\omega_{DE}$, because $\omega_{DE}$ starts with $-1$ when $\lambda=-2/3$ and $\delta=1$ . When  $\lambda=-2/3$ and $\delta=1.1$, $\omega_{DE}$ begins with a negative value, increases until it reaches a positive value, and then decreases until it reaches -1. The universe enters a phantom region once more (when $\omega_{DE}<-1$).

To study the behavior of the equation-of-state parameter, (\ref{78}),  and  to investigate how it is affected by $\delta$ and $h$  we plot it in
graph \ref{Fig:7}  and present $\omega_{_{_{DE}}}(z)$ for the case of open universe, and for different values of $\delta$ and $\lambda$. As we can see, for
increasing  the value of $\delta$ and decreasing the value of $\lambda$ the evolution of $\omega_{_{_{DE}}}(z)$ and its current value $z=0$ tend to obtain lower
values. In particular, while for $\delta=1$ and $\lambda= -2/3$ the dark-energy equation-of-state parameter lies completely in the
quintessence regime, $\delta=1.1$ and $\lambda=-2/3$ the universe will result in the phantom regime for $z>0.2$, and specifically
for $\delta > 1$ the phantom-divide crossing has been realized in the future. Hence, in the case of MTT
  we obtain the possibility to exhibit the crossing to the phantom regime, contrary to the case of GR.

The matter and dark energy density parameters are defined as follows \cite{Hobson:2006se}:
{ \begin{equation}\label{79}
\Omega_{m}=\frac{8\pi G}{3(H^{2}+\frac{h}{a^2})}\rho_{m}, \qquad \qquad \Omega_{_{DE}}=\frac{8\pi G}{{3(H^{2}+\frac{h}{a^2})}}\rho_{_{DE}}.
\end{equation}}
The energy density for the dust case is given by
\begin{equation}\label{80}
\rho_{m}=\frac{\rho_{m,0}}{a^{3}}\,, \qquad   \textrm{where} \qquad \quad \rho_{m,0}=\frac{3\Omega_{m,0}H_{0}{}^{2}}{8\pi G}\,, \qquad \textrm{and} \qquad \quad  \Omega_{m,0}=\Omega_{m}(z=0)=0.3\,,
\end{equation}
where $"0"$ refers to the  value of the present time. Substitution of Eq. (\ref{80}) in Eq. (\ref{79}) yields:
{ \begin{equation}\label{81}
H^2=\frac{H_{0}{}^2}{(1-\Omega_{DE})}\left(\frac{\Omega_{m,0}}{a^{3}}-\frac{h}{a^2H_{0}{}^2}\right).
\end{equation}}
Differentiation of Eq.(\ref{81}) yields:
{ \begin{equation}\label{82}
\dot H=-\frac{1}{2(1-\Omega_{DE})}\bigg(H^{2}\left[3(1-\Omega_{DE})+(1+z)\Omega_{DE}^{'}\right]+\frac{h}{a^2H_{0}{}^2}\bigg).
\end{equation}}
where $\Omega_{DE}^{'}=\frac{d\Omega_{DE}}{dz}$. Substituting  Eq.(\ref{75}) into Eq.(\ref{79}), we obtain the form of the dark energy density parameter with the help of Eq. (\ref{81}), as follows:
{ \begin{equation}\label{83}
\Omega_{DE}=1-\frac{\alpha\delta}{2-\delta}\Bigg[\frac{H_{0}{}^2}{(1-\Omega_{DE})}\left(\frac{\Omega_{m,0}}{a^{3}}-\frac{h}{a^2H_{0}{}^2}\right)+h(1+z)^2
\Bigg]^{-1}\Bigg[\frac{H_{0}{}^2}{(1-\Omega_{DE})}\left(\frac{\Omega_{m,0}}{a^{3}}-\frac{h}{a^2H_{0}{}^2}\right)+k(1+z)^2\Bigg]^{2-\delta}+\frac{\tilde{f}}{3}
%-H_{0}^{2}\Omega_{m,0}(1+z)^{3}\bigg[-k(1+z)^{2}+\bigg\{\frac{(2-\delta)\Omega_{m,0}H_{0}{}^{2}(1+z)^{3}}{\alpha\delta}\bigg(1+\frac{h}{\Omega_{m,0}H_{0}^{2}(1+z)}+\frac{\tilde{f}}{3\Omega_{m,0}
%H_{0}{}^{2}(1+z)^{3}}\bigg) \bigg\}^{\frac{1}{2-\delta}}\bigg]^{-1}.
\end{equation}}
The deceleration parameter is defined as follows:
\begin{equation}\label{84}
q=-1-\frac{\dot H}{H^{2}}.
\end{equation}
By substituting Eqs. (\ref{81}) and (\ref{82}) into Eq.(\ref{84}), we obtain $q(z)$ as follows \cite{Lymperis:2018iuz}:
{ \begin{equation}\label{85}
q(z)=-1+\frac{3 (1-\Omega_{DE})+(1+z)\Omega_{DE}^{'}}{2 (1-\Omega_{DE})}+\frac{h}{2H_{0}{}^2\left[H_{0}{}^2\Omega_{m,0}(1+z)-h\right]}\,.
\end{equation}}
For the radiation case ($p=\frac{1}{3}\rho$) the energy density is given as follows:
\begin{equation}\label{85}
\rho_{m}=\frac{\rho_{m,0}}{a^{4}}.
\end{equation}
Following the same procedure applied in the dust case  we obtain:
{ \begin{equation}\label{86}
H^2=\frac{H_{0}{}^2}{(1-\Omega_{DE})}\left(\frac{\Omega_{m,0}}{a^{4}}-\frac{h}{a^2H_{0}{}^2}\right)\,,
\end{equation}}
{ \begin{equation}\label{87}
\Omega_{DE}=1-\frac{\alpha\delta}{2-\delta}\Bigg[\frac{H_{0}{}^2}{(1-\Omega_{DE})}\left(\frac{\Omega_{m,0}}{a^{4}}-\frac{h}{a^2H_{0}{}^2}\right)+h(1+z)^2
\Bigg]^{-1}\Bigg[\frac{H_{0}{}^2}{(1-\Omega_{DE})}\left(\frac{\Omega_{m,0}}{a^{4}}-\frac{h}{a^2H_{0}{}^2}\right)+k(1+z)^2\Bigg]^{2-\delta}+\frac{\tilde{f}}{3},
\end{equation}}
The deceleration parameter as a function of the red shift is given by:
{\begin{equation}\label{88}
q(z)=-1+\frac{4 (1-\Omega_{DE})+(1+z)\Omega_{DE}^{'}}{2 (1-\Omega_{DE})}+\frac{h}{H_{0}{}^2\left[H_{0}{}^2\Omega_{m,0}(1+z)^2-h\right]}.
\end{equation}}
\begin{figure}
\centering
{\label{fig:8}\includegraphics[scale=0.4]{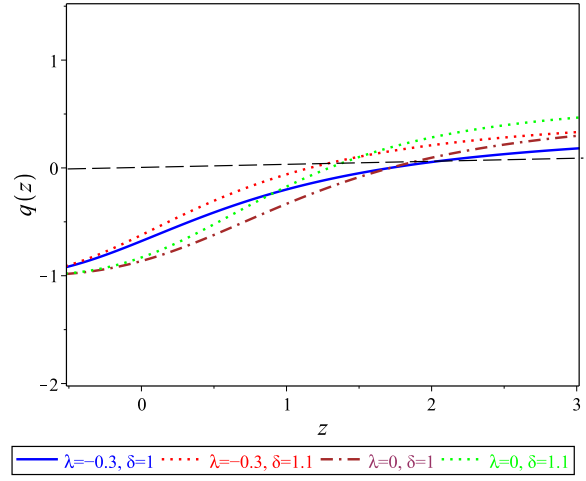}}\hspace{0.2cm}
%\subfigure[\,Displays the deceleration parameter $q(z)$ vs the red shift for the radiation case.]{\label{fig:8b}\includegraphics[scale=0.3]{22}}
\caption{Behaviors of  the deceleration parameter versus red shift, for the dust case with initial conditions using different values of
model dimensionless parameters $\lambda$  and $\delta$.}
\label{Fig:8}
\end{figure}
{  In Figure \ref{Fig:8}}  we depict the deceleration parameter for $\lambda=0$ and $\lambda\neq0$ using different values of $\delta$. { The Figure} demonstrates  that the deceleration parameter has positive values for positive values of the redshift and then takes negative values until it approaches AdS  which means the universe  enters an exponential expansion phase in the future.
  %In Fig.  \ref{Fig:8}\subref{fig:8b}, we plotted the deceleration parameter for the radiation case, we notice that it begins with a small negative value and then approaches to -1 in the GR and at both $\delta=1, 1.1$, while it  begins with a large negative value and then approaches to -1  at $\lambda=-2/3$.\\
%We will discuss the behavior of the matter and dark energy density parameters through the following diagrams.

\begin{figure}
\centering
 \subfigure[ Displays the energy density parameter versus the red shift at $\lambda=0.3$ and $\delta=1$ .]{\label{fig:9a}\includegraphics[scale=0.3]{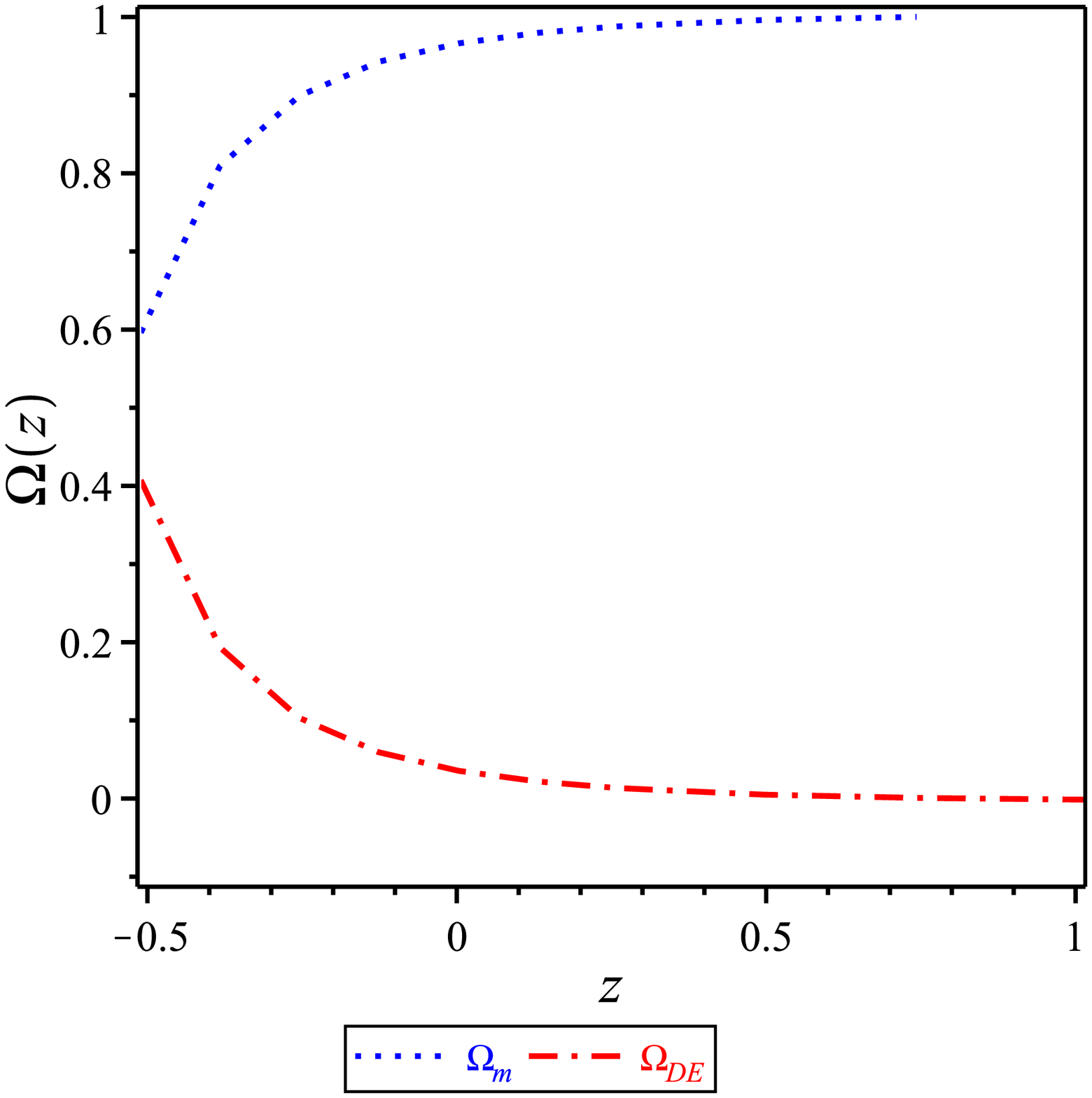}}\hspace{0.2cm}
\subfigure[ Displays the energy density parameter versus the red shift at $\lambda=0.3$ and $\delta=1.1$.]{\label{fig:9b}\includegraphics[scale=0.3]{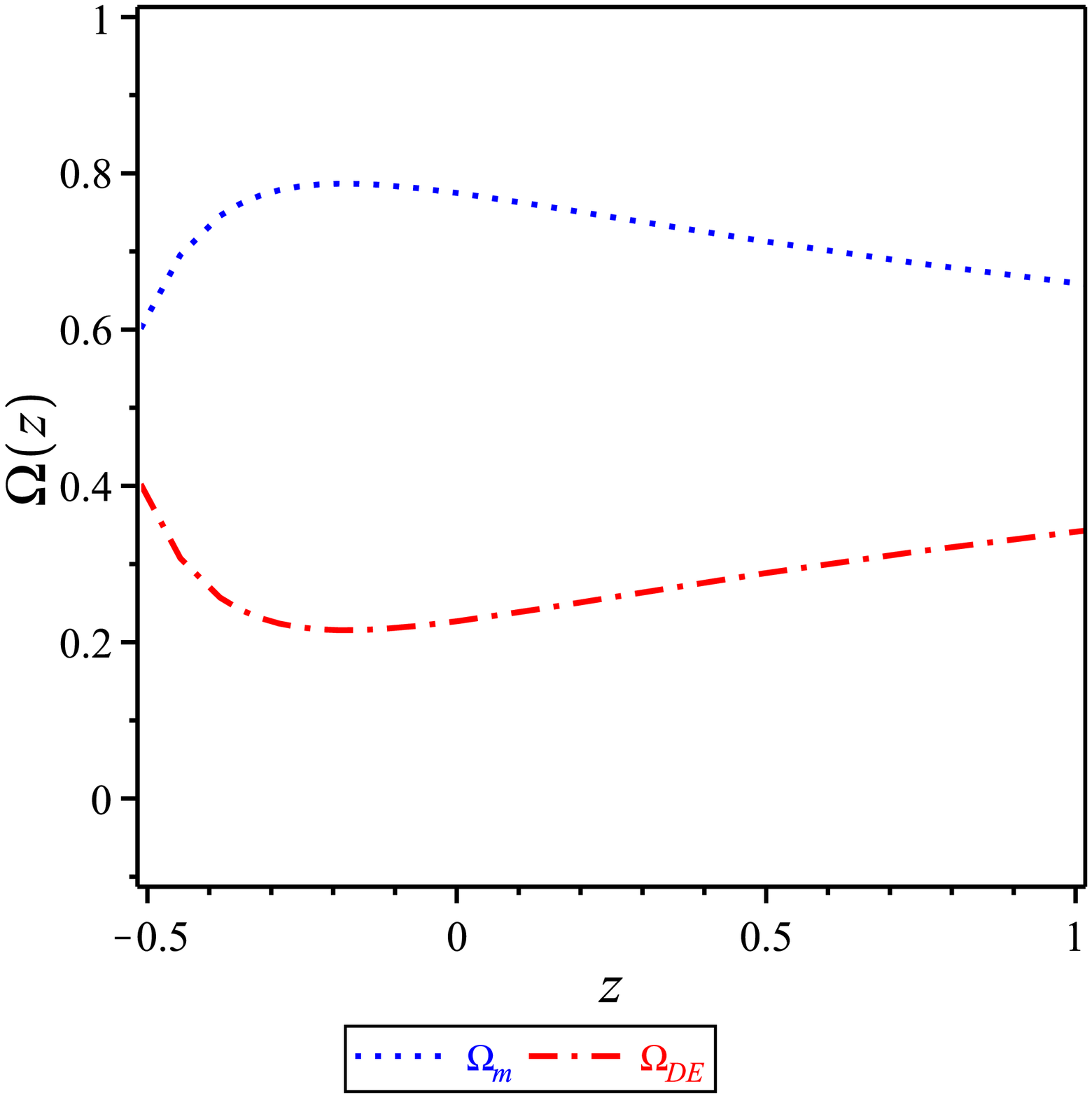}}
\caption{ Behaviors of  the energy density parameter versus redshift, for  the dust case with initial conditions\\ $H_0 = 67.9 (Kms^{-1}Mpc^{-1})$ and $\Omega_{DE_{0}}=0.7$, using different values of
model dimensionless parameters $\lambda$  and $\delta$.}
\label{Fig:9}
\end{figure}

 { In Figure \ref{Fig:9}\subref{fig:9a}}, we plot the matter and dark energy density parameters  for the closed universe in the dust case. { The Figure shows that $\Omega_{m}$} and $\Omega_{DE}$ have approximately similar effects, i.e., they diverge from each other and  $\Omega_{m}$ becomes dominant while $\Omega_{DE}$ is minimized for large positive redshift. When we take into account the effect of the Tsallis parameter, i.e., when $\delta\neq 1$ we obtain different behaviors of   $\Omega_{m}(z)$ and  $\Omega_{DE}(z)$, which converge with each other as shown {  in Figure \ref{Fig:9}\subref{fig:9b}}.

  Note that $\Omega_{m}+\Omega_{DE}=1$, and this relation corresponds to  actual observations because the standard inflationary models predict that $\Omega=1$,  \cite{Garriga:1998px}.
  When $z\longrightarrow-1$, $t\longrightarrow\infty$, represents the far future  \cite{Lymperis:2018iuz,ElHanafy:2019zhr}; $z=0$ represents the current value of the red shift.\\

\begin{figure}
\centering
\subfigure[ Energy density parameter vs. the red shift at $\lambda=-2/3$ and $\delta=1$.]{\label{fig:10a}\includegraphics[scale=0.2]{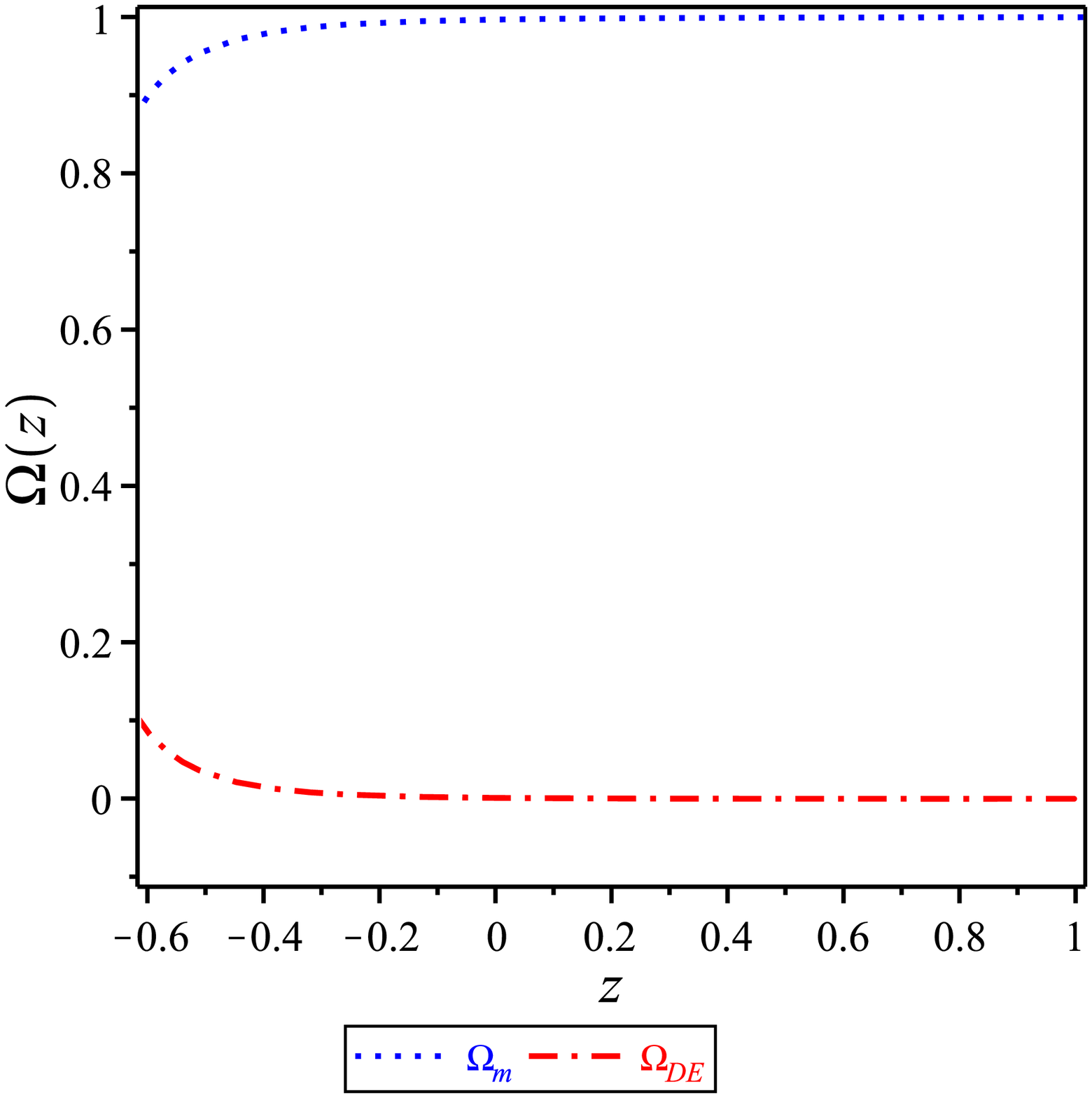}}\hspace{0.2cm}
\subfigure[ Energy density parameter vs. the red shift at $\lambda=-2/3$ and $\delta=1.1$.]{\label{fig:10b}\includegraphics[scale=0.2]{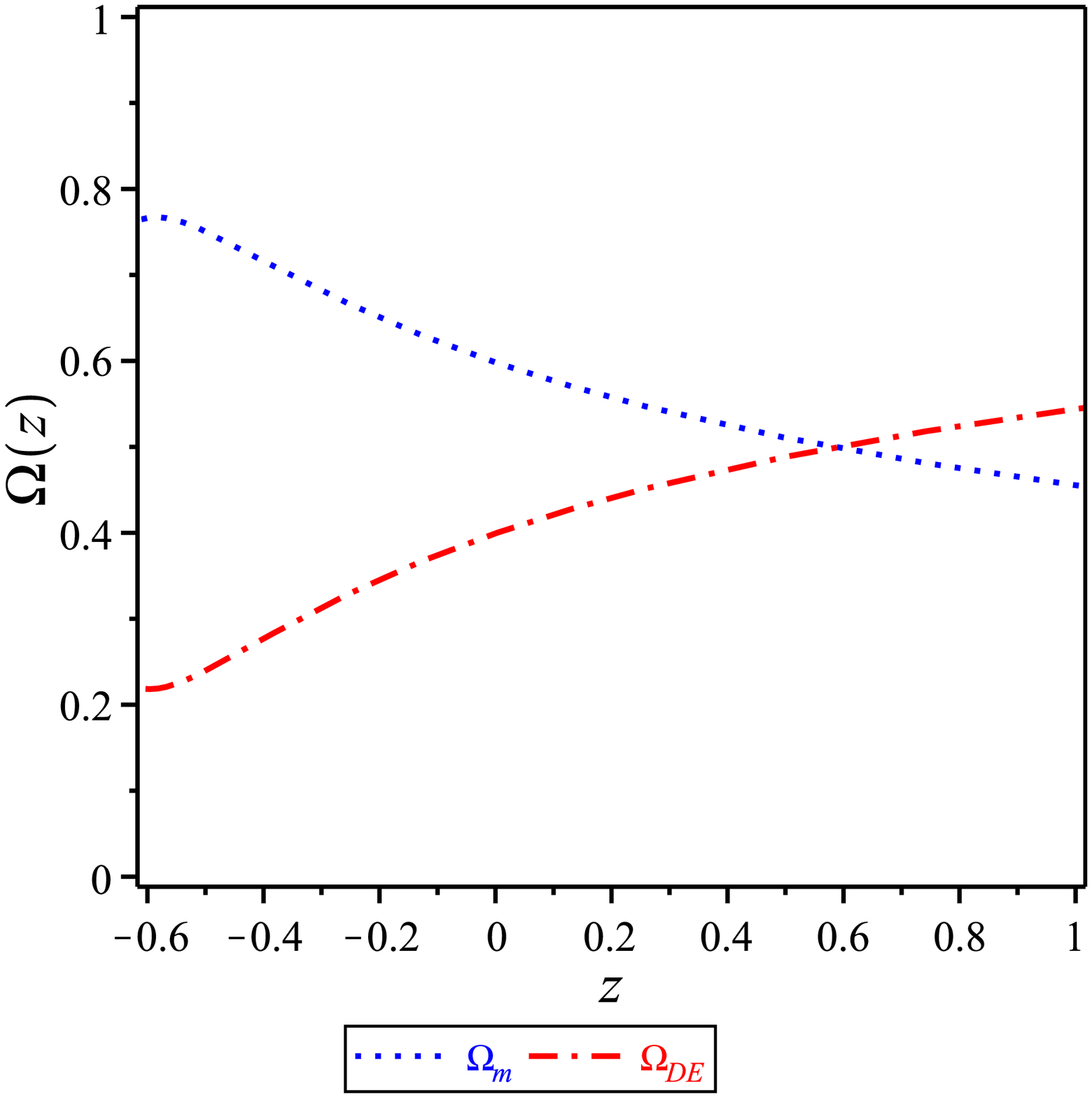}}
\subfigure[ Energy density parameter vs. the red shift at $\lambda=0$ and $\delta=1$.]{\label{fig:10c}\includegraphics[scale=0.2]{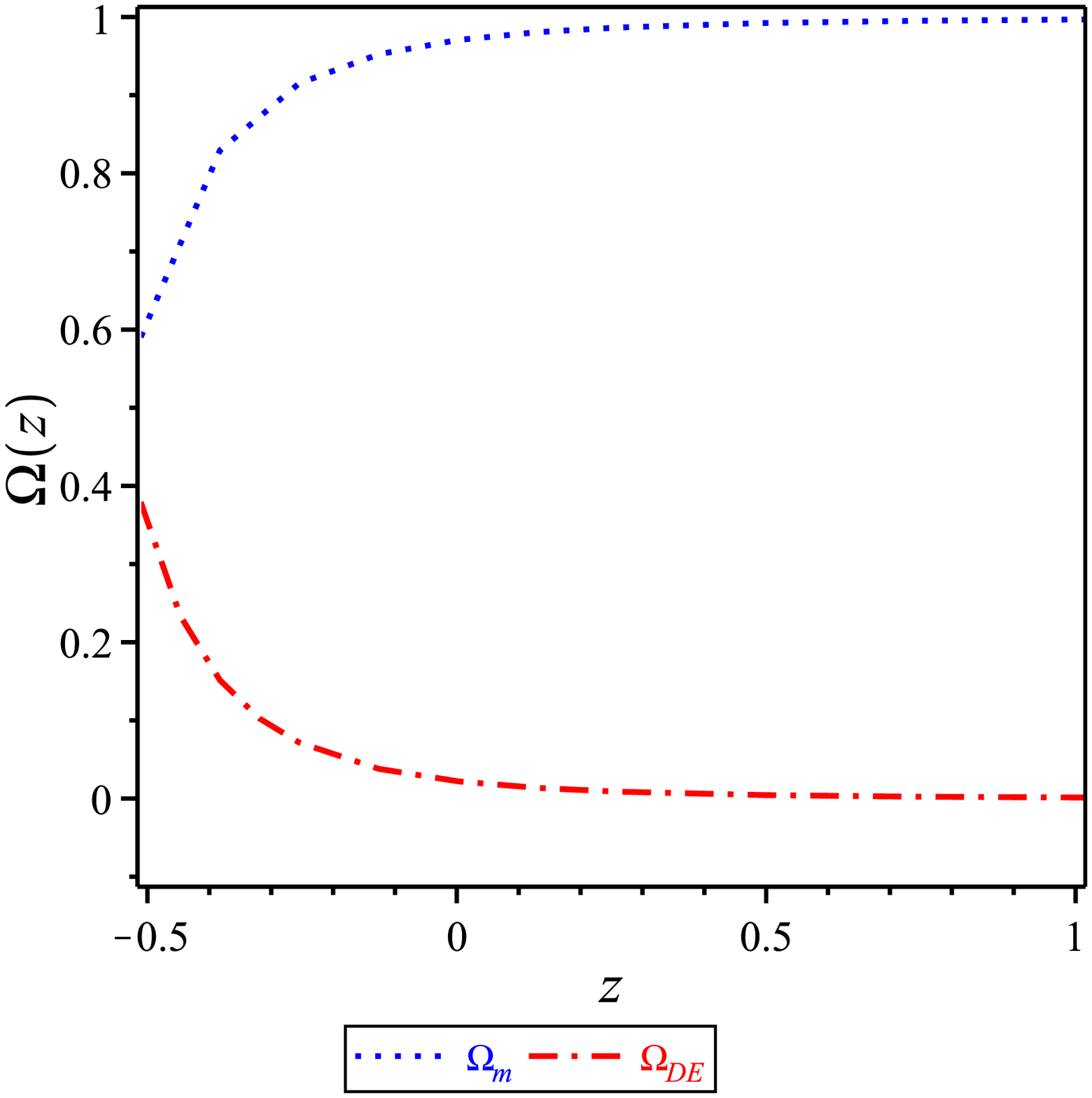}}
\subfigure[ Energy density parameter vs. the red shift at $\lambda=0$ and $\delta=1.1$.]{\label{fig:10d}\includegraphics[scale=0.2]{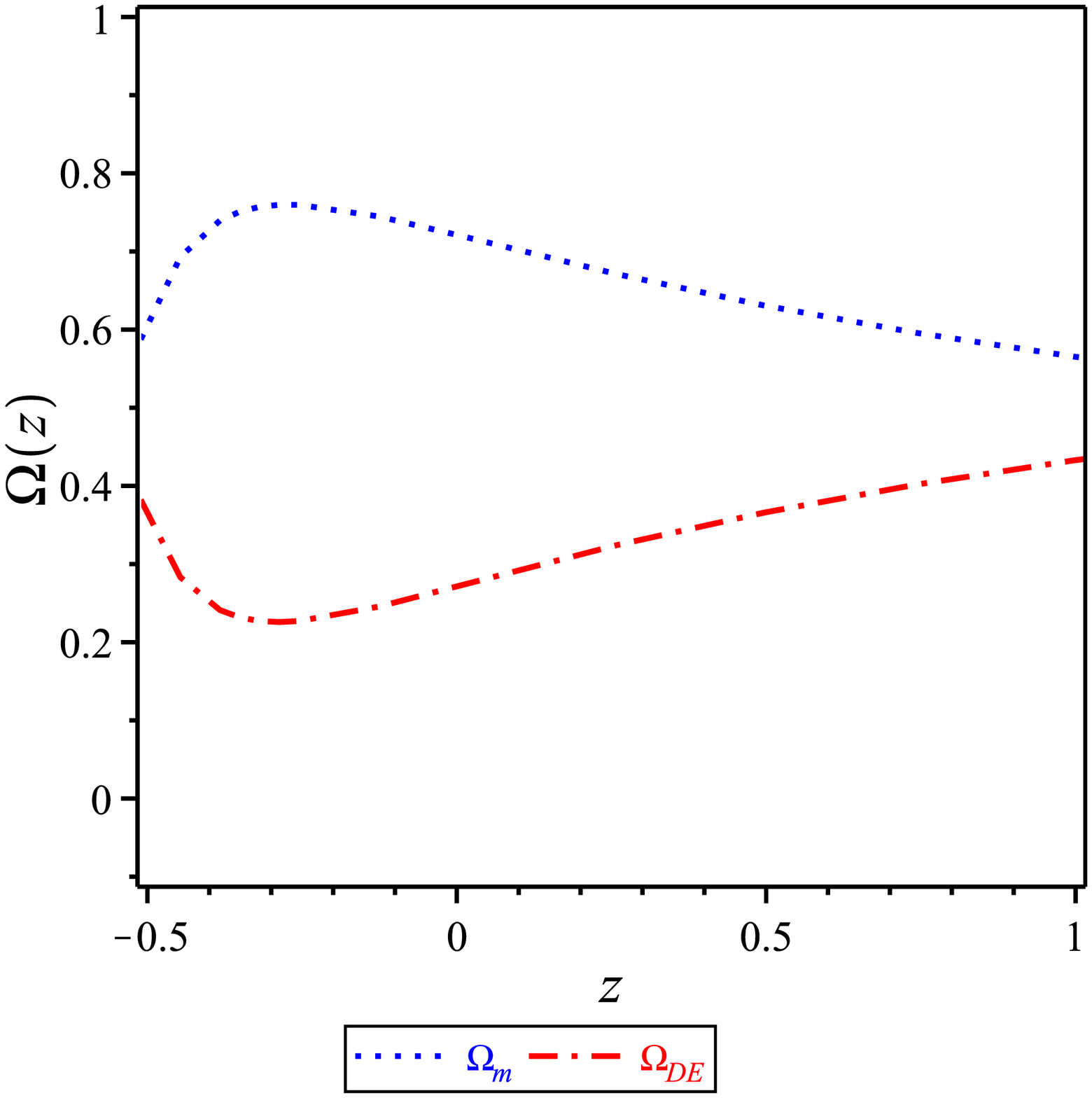}}
\caption{Behaviors of  the energy density parameter versus red shift, for  the radiation case using different values of
model dimensionless parameters $\lambda$  and $\delta$.}
\label{Fig:10}
\end{figure}

{  In Figure \ref{Fig:10}\subref{fig:10a}}, we present the evolution of the matter and dark energy density parameters  for the open universe in the radiation case when $\delta=1$ and $\lambda=-2/3$.  { The Figure} clearly illustrates that $\Omega_{m}$ exerts a dominant effect while $\Omega_{DE}$ has a minimal effect. { In Figure \ref{Fig:10}\subref{fig:10b}},   $\delta$ has a significant effect on the behaviors of $\Omega_{m}$ and  $\Omega_{DE}$.  When $\delta=1$ and $\lambda=0$, the effects of $\Omega_{m}$ and  $\Omega_{DE}$ approximately identical to the previous behavior described  { in Figure \ref{Fig:10}\subref{fig:10c}}, and the parameters have fixed values as the redshift increases in the positive direction. That is $\Omega_{m}$ is dominant and $\Omega_{DE}$ has no effect. Finally, when $\delta=1.1$ and $\lambda=0$, the effects of $\Omega_{m}$  and $\Omega_{DE}$ are approximately identical and these parameters converge toward each other as the redshift increases in the positive direction as shown in { Figure  \ref{Fig:10}\subref{fig:10d}.}  These findings indicate that the behavior of the total matter in a non-flat spatial space differs from that in a flat spatial space.

\newpage
\section{Summary}
  Recently, there is a hot dispute if the spatial curvature of the universe, $k$ has a zero value or not. Especially, there is evidence that if we consider the combined analysis of cosmic
microwave background anisotropy power spectra of the Planck Collaboration with the luminosity
distance data, then a non-flat universe is preferred at 99\% confidence level  \cite{DiValentino:2020hov}. Moreover, the enhanced
lensing amplitude in the cosmic microwave background power spectrum seems to suggest that the curvature index $k$ may be positive.

{ Due to the importance of the spatial curvature as we stated above,   we considered in this work  the behavior of the universe by using MTT of gravitation and a non-flat spatial curvature then discussed the dust,  radiation, and dark energy phases. We explained how the dimensionless parameter $h$, which is a combination between the spatial curvature $k$  and the parameter that characterize MTT $\lambda$,  affects several physical properties, including the cosmological quantities, energy-density, and Hubble parameter,  of the resulting models and make all of them different from their correspondence of GR.} We summarize our results as follows
:\vspace{0.1cm}\\
$\bullet$ The most crucial result of this study is that when the parameter $h$, that characterizes MTT, vanishing then the spatial curvature has a vanishing value, i.e., the resulting models are coincide with the flat universe of GR. \vspace{0.1cm}\\
$\bullet$ In the dust case, i.e., when the EoS parameter vanishes,   $w=0$, the  closed model gives  more satisfactory results than the open model. However, in the  radiation case, $w=1/3$, the  open model gives more satisfactory results than the closed one as shown in Figs. \ref{Fig:2}\subref{fig:2a} (the behavior of the scale factor vs. the cosmic time), \ref{Fig:2}\subref{fig:2b} (the behavior of the Hubble parameter vs. the cosmic time),  and \ref{Fig:2}\subref{fig:2c} (the behavior of the deceleration parameter vs. the cosmic time). Finally, in the case of dark energy ($w=-1$), the AdS/dS model is obtained.\vspace{0.1cm}\\
$\bullet$   To understand the nature of these models and through the use of the first law of thermodynamics, we calculated some quantities, such as apparent horizon, Hawking  temperature, and entropy, for the two cases of dust and radiation.   Despite the fact that these quantities are effected by the M\o ller's parameter, $h$,  their behaviors are generally consistent with observations as shown in Figs.  \ref{Fig:3}\subref{fig:3b}  (the behavior of the Hawking temperature for closed universe),  \ref{Fig:4}\subref{fig:4b}  (the behavior of the Hawking temperature for open universe),  \ref{Fig:6}\subref{fig:6a}  (the behavior of entropy for closed universe) and  \ref{Fig:6}\subref{fig:6b}  (the behavior of entropy for open universe). Moreover, we showed the differences relative to GR that could be observed from our models.\vspace{0.1cm}\\
$\bullet$  Additionally,  we studied the effect of the non-extensive thermodynamics on the two cases, of dust and radiation, and derived the modified Friedmann equations through the use of the first law of thermodynamics.  We derived the dark energy density, pressure, and EoS parameter from these modified equations and observed the effect of the non-extensive parameter $\delta$  on the total matter, i.e., matter and dark energy density parameters. We showed that the total matter is preserved, i.e., the total $\Omega(z)\sim 1$  despite the effect of the M\o ller's parameter as shown in Figs. \ref{Fig:9}\subref{fig:9a} (the behavior of the energy density parameter when $\delta=1$) and  \ref{Fig:9}\subref{fig:9b} (the behavior of the energy density parameter when $\delta=1.1$).\vspace{0.1cm}\\
%$\bullet$ {Finally,  we calculated the luminosity distance, $d_{L}$, for the dust model and showed that our results are consistent with the observational data when $z>0.4$.}

 To conclude,  the incorporation of slightly non-flat spatial geometry to MTT regardless of the agreement with the   analysis of cosmic
microwave background anisotropy power spectra with the luminosity distance data \cite{DiValentino:2020hov},  it improves the phenomenology comparing to the flat case while keeping the
M\o ller's parameter has a negative value.
 \section*{Acknowledgments}
%\begin{acknowledgements}
 The authors would like to thank the anonymous Referees for improving the presentation of the manuscript.

 %%%%%%%%%%%%%%%%%%%%%%%%%%%%%%%%%
%\bibliographystyle{apsrev}
%\bibliography{JRPHSRef}

\begin{thebibliography}{130}%
\makeatletter
\providecommand \@ifxundefined [1]{%
 \@ifx{#1\undefined}
}%
\providecommand \@ifnum [1]{%
 \ifnum #1\expandafter \@firstoftwo
 \else \expandafter \@secondoftwo
 \fi
}%
\providecommand \@ifx [1]{%
 \ifx #1\expandafter \@firstoftwo
 \else \expandafter \@secondoftwo
 \fi
}%
\providecommand \natexlab [1]{#1}%
\providecommand \enquote  [1]{``#1''}%
\providecommand \bibnamefont  [1]{#1}%
\providecommand \bibfnamefont [1]{#1}%
\providecommand \citenamefont [1]{#1}%
\providecommand \href@noop [0]{\@secondoftwo}%
\providecommand \href [0]{\begingroup \@sanitize@url \@href}%
\providecommand \@href[1]{\@@startlink{#1}\@@href}%
\providecommand \@@href[1]{\endgroup#1\@@endlink}%
\providecommand \@sanitize@url [0]{\catcode `\\12\catcode `\$12\catcode
  `\&12\catcode `\#12\catcode `\^12\catcode `\_12\catcode `\%12\relax}%
\providecommand \@@startlink[1]{}%
\providecommand \@@endlink[0]{}%
\providecommand \url  [0]{\begingroup\@sanitize@url \@url }%
\providecommand \@url [1]{\endgroup\@href {#1}{\urlprefix }}%
\providecommand \urlprefix  [0]{URL }%
\providecommand \Eprint [0]{\href }%
\providecommand \doibase [0]{http://dx.doi.org/}%
\providecommand \selectlanguage [0]{\@gobble}%
\providecommand \bibinfo  [0]{\@secondoftwo}%
\providecommand \bibfield  [0]{\@secondoftwo}%
\providecommand \translation [1]{[#1]}%
\providecommand \BibitemOpen [0]{}%
\providecommand \bibitemStop [0]{}%
\providecommand \bibitemNoStop [0]{.\EOS\space}%
\providecommand \EOS [0]{\spacefactor3000\relax}%
\providecommand \BibitemShut  [1]{\csname bibitem#1\endcsname}%
\let\auto@bib@innerbib\@empty
%</preamble>
\bibitem [{\citenamefont {Perlmutter}\ \emph {et~al.}(1998)\citenamefont
  {Perlmutter} \emph {et~al.}}]{Perlmutter:1997zf}%
  \BibitemOpen
  \bibfield  {author} {\bibinfo {author} {\bibfnamefont {S.}~\bibnamefont
  {Perlmutter}} \emph {et~al.} (\bibinfo {collaboration} {Supernova Cosmology
  Project}),\ }\href {\doibase 10.1038/34124} {\bibfield  {journal} {\bibinfo
  {journal} {Nature}\ }\textbf {\bibinfo {volume} {391}},\ \bibinfo {pages}
  {51} (\bibinfo {year} {1998})},\ \Eprint
  {http://arxiv.org/abs/astro-ph/9712212} {arXiv:astro-ph/9712212} \BibitemShut
  {NoStop}%
\bibitem [{\citenamefont {Dalal}\ and\ \citenamefont
  {Griest}(2000)}]{Dalal:2000xw}%
  \BibitemOpen
  \bibfield  {author} {\bibinfo {author} {\bibfnamefont {N.}~\bibnamefont
  {Dalal}}\ and\ \bibinfo {author} {\bibfnamefont {K.}~\bibnamefont {Griest}},\
  }\href {\doibase 10.1016/S0370-2693(00)00987-4} {\bibfield  {journal}
  {\bibinfo  {journal} {Phys. Lett. B}\ }\textbf {\bibinfo {volume} {490}},\
  \bibinfo {pages} {1} (\bibinfo {year} {2000})},\ \Eprint
  {http://arxiv.org/abs/astro-ph/0008260} {arXiv:astro-ph/0008260} \BibitemShut
  {NoStop}%
\bibitem [{\citenamefont {Riess}\ \emph {et~al.}(1998)\citenamefont {Riess}
  \emph {et~al.}}]{Riess:1998cb}%
  \BibitemOpen
  \bibfield  {author} {\bibinfo {author} {\bibfnamefont {A.~G.}\ \bibnamefont
  {Riess}} \emph {et~al.} (\bibinfo {collaboration} {Supernova Search Team}),\
  }\href {\doibase 10.1086/300499} {\bibfield  {journal} {\bibinfo  {journal}
  {Astron. J.}\ }\textbf {\bibinfo {volume} {116}},\ \bibinfo {pages} {1009}
  (\bibinfo {year} {1998})},\ \Eprint {http://arxiv.org/abs/astro-ph/9805201}
  {arXiv:astro-ph/9805201} \BibitemShut {NoStop}%
\bibitem [{\citenamefont {Riess}\ \emph {et~al.}(2004)\citenamefont {Riess}
  \emph {et~al.}}]{Riess:2004nr}%
  \BibitemOpen
  \bibfield  {author} {\bibinfo {author} {\bibfnamefont {A.~G.}\ \bibnamefont
  {Riess}} \emph {et~al.} (\bibinfo {collaboration} {Supernova Search Team}),\
  }\href {\doibase 10.1086/383612} {\bibfield  {journal} {\bibinfo  {journal}
  {Astrophys. J.}\ }\textbf {\bibinfo {volume} {607}},\ \bibinfo {pages} {665}
  (\bibinfo {year} {2004})},\ \Eprint {http://arxiv.org/abs/astro-ph/0402512}
  {arXiv:astro-ph/0402512} \BibitemShut {NoStop}%
\bibitem [{\citenamefont {Sahni}(2002)}]{Sahni:2002kh}%
  \BibitemOpen
  \bibfield  {author} {\bibinfo {author} {\bibfnamefont {V.}~\bibnamefont
  {Sahni}},\ }\href {\doibase 10.1088/0264-9381/19/13/304} {\bibfield
  {journal} {\bibinfo  {journal} {Class. Quant. Grav.}\ }\textbf {\bibinfo
  {volume} {19}},\ \bibinfo {pages} {3435} (\bibinfo {year} {2002})},\ \Eprint
  {http://arxiv.org/abs/astro-ph/0202076} {arXiv:astro-ph/0202076} \BibitemShut
  {NoStop}%
\bibitem [{\citenamefont {Eidelman}\ \emph {et~al.}(2004)\citenamefont
  {Eidelman} \emph {et~al.}}]{Eidelman:2004wy}%
  \BibitemOpen
  \bibfield  {author} {\bibinfo {author} {\bibfnamefont {S.}~\bibnamefont
  {Eidelman}} \emph {et~al.} (\bibinfo {collaboration} {Particle Data Group}),\
  }\href {\doibase 10.1016/j.physletb.2004.06.001} {\bibfield  {journal}
  {\bibinfo  {journal} {Phys. Lett. B}\ }\textbf {\bibinfo {volume} {592}},\
  \bibinfo {pages} {1} (\bibinfo {year} {2004})}\BibitemShut {NoStop}%
\bibitem [{\citenamefont {Kamenshchik}\ \emph {et~al.}(2001)\citenamefont
  {Kamenshchik}, \citenamefont {Moschella},\ and\ \citenamefont
  {Pasquier}}]{Kamenshchik:2001cp}%
  \BibitemOpen
  \bibfield  {author} {\bibinfo {author} {\bibfnamefont {A.~Y.}\ \bibnamefont
  {Kamenshchik}}, \bibinfo {author} {\bibfnamefont {U.}~\bibnamefont
  {Moschella}}, \ and\ \bibinfo {author} {\bibfnamefont {V.}~\bibnamefont
  {Pasquier}},\ }\href {\doibase 10.1016/S0370-2693(01)00571-8} {\bibfield
  {journal} {\bibinfo  {journal} {Phys. Lett. B}\ }\textbf {\bibinfo {volume}
  {511}},\ \bibinfo {pages} {265} (\bibinfo {year} {2001})},\ \Eprint
  {http://arxiv.org/abs/gr-qc/0103004} {arXiv:gr-qc/0103004} \BibitemShut
  {NoStop}%
\bibitem [{\citenamefont {Zhu}\ and\ \citenamefont
  {Fujimoto}(2003)}]{Zhu:2003sq}%
  \BibitemOpen
  \bibfield  {author} {\bibinfo {author} {\bibfnamefont {Z.-H.}\ \bibnamefont
  {Zhu}}\ and\ \bibinfo {author} {\bibfnamefont {M.-K.}\ \bibnamefont
  {Fujimoto}},\ }\href {\doibase 10.1086/346002} {\bibfield  {journal}
  {\bibinfo  {journal} {Astrophys. J.}\ }\textbf {\bibinfo {volume} {585}},\
  \bibinfo {pages} {52} (\bibinfo {year} {2003})},\ \Eprint
  {http://arxiv.org/abs/astro-ph/0303021} {arXiv:astro-ph/0303021} \BibitemShut
  {NoStop}%
\bibitem [{\citenamefont {Sen}\ and\ \citenamefont {Sen}(2003)}]{Sen:2002ss}%
  \BibitemOpen
  \bibfield  {author} {\bibinfo {author} {\bibfnamefont {S.}~\bibnamefont
  {Sen}}\ and\ \bibinfo {author} {\bibfnamefont {A.~A.}\ \bibnamefont {Sen}},\
  }\href {\doibase 10.1086/373900} {\bibfield  {journal} {\bibinfo  {journal}
  {Astrophys. J.}\ }\textbf {\bibinfo {volume} {588}},\ \bibinfo {pages} {1}
  (\bibinfo {year} {2003})},\ \Eprint {http://arxiv.org/abs/astro-ph/0211634}
  {arXiv:astro-ph/0211634} \BibitemShut {NoStop}%
\bibitem [{\citenamefont {Godlowski}\ \emph
  {et~al.}(2004{\natexlab{a}})\citenamefont {Godlowski}, \citenamefont
  {Szydlowski},\ and\ \citenamefont {Krawiec}}]{Godlowski:2003pd}%
  \BibitemOpen
  \bibfield  {author} {\bibinfo {author} {\bibfnamefont {W.}~\bibnamefont
  {Godlowski}}, \bibinfo {author} {\bibfnamefont {M.}~\bibnamefont
  {Szydlowski}}, \ and\ \bibinfo {author} {\bibfnamefont {A.}~\bibnamefont
  {Krawiec}},\ }\href {\doibase 10.1086/382669} {\bibfield  {journal} {\bibinfo
   {journal} {Astrophys. J.}\ }\textbf {\bibinfo {volume} {605}},\ \bibinfo
  {pages} {599} (\bibinfo {year} {2004}{\natexlab{a}})},\ \Eprint
  {http://arxiv.org/abs/astro-ph/0309569} {arXiv:astro-ph/0309569} \BibitemShut
  {NoStop}%
\bibitem [{\citenamefont {Godlowski}\ and\ \citenamefont
  {Szydlowski}(2004)}]{Godlowski:2004pt}%
  \BibitemOpen
  \bibfield  {author} {\bibinfo {author} {\bibfnamefont {W.}~\bibnamefont
  {Godlowski}}\ and\ \bibinfo {author} {\bibfnamefont {M.}~\bibnamefont
  {Szydlowski}},\ }\href {\doibase 10.1023/B:GERG.0000016923.83136.05}
  {\bibfield  {journal} {\bibinfo  {journal} {Gen. Rel. Grav.}\ }\textbf
  {\bibinfo {volume} {36}},\ \bibinfo {pages} {767} (\bibinfo {year} {2004})},\
  \Eprint {http://arxiv.org/abs/astro-ph/0404299} {arXiv:astro-ph/0404299}
  \BibitemShut {NoStop}%
\bibitem [{\citenamefont {Godlowski}\ \emph
  {et~al.}(2004{\natexlab{b}})\citenamefont {Godlowski}, \citenamefont
  {Stelmach},\ and\ \citenamefont {Szydlowski}}]{Godlowski:2004gh}%
  \BibitemOpen
  \bibfield  {author} {\bibinfo {author} {\bibfnamefont {W.}~\bibnamefont
  {Godlowski}}, \bibinfo {author} {\bibfnamefont {J.}~\bibnamefont {Stelmach}},
  \ and\ \bibinfo {author} {\bibfnamefont {M.}~\bibnamefont {Szydlowski}},\
  }\href {\doibase 10.1088/0264-9381/21/16/009} {\bibfield  {journal} {\bibinfo
   {journal} {Class. Quant. Grav.}\ }\textbf {\bibinfo {volume} {21}},\
  \bibinfo {pages} {3953} (\bibinfo {year} {2004}{\natexlab{b}})},\ \Eprint
  {http://arxiv.org/abs/astro-ph/0403534} {arXiv:astro-ph/0403534} \BibitemShut
  {NoStop}%
\bibitem [{\citenamefont {Puetzfeld}\ and\ \citenamefont
  {Chen}(2004)}]{Puetzfeld:2004df}%
  \BibitemOpen
  \bibfield  {author} {\bibinfo {author} {\bibfnamefont {D.}~\bibnamefont
  {Puetzfeld}}\ and\ \bibinfo {author} {\bibfnamefont {X.-l.}\ \bibnamefont
  {Chen}},\ }\href {\doibase 10.1088/0264-9381/21/11/013} {\bibfield  {journal}
  {\bibinfo  {journal} {Class. Quant. Grav.}\ }\textbf {\bibinfo {volume}
  {21}},\ \bibinfo {pages} {2703} (\bibinfo {year} {2004})},\ \Eprint
  {http://arxiv.org/abs/gr-qc/0402026} {arXiv:gr-qc/0402026} \BibitemShut
  {NoStop}%
\bibitem [{\citenamefont {Biesiada}\ \emph {et~al.}(2005)\citenamefont
  {Biesiada}, \citenamefont {Godlowski},\ and\ \citenamefont
  {Szydlowski}}]{Biesiada:2004td}%
  \BibitemOpen
  \bibfield  {author} {\bibinfo {author} {\bibfnamefont {M.}~\bibnamefont
  {Biesiada}}, \bibinfo {author} {\bibfnamefont {W.}~\bibnamefont {Godlowski}},
  \ and\ \bibinfo {author} {\bibfnamefont {M.}~\bibnamefont {Szydlowski}},\
  }\href {\doibase 10.1086/427863} {\bibfield  {journal} {\bibinfo  {journal}
  {Astrophys. J.}\ }\textbf {\bibinfo {volume} {622}},\ \bibinfo {pages} {28}
  (\bibinfo {year} {2005})},\ \Eprint {http://arxiv.org/abs/astro-ph/0403305}
  {arXiv:astro-ph/0403305} \BibitemShut {NoStop}%
\bibitem [{\citenamefont {Carroll}\ \emph {et~al.}(2004)\citenamefont
  {Carroll}, \citenamefont {Duvvuri}, \citenamefont {Trodden},\ and\
  \citenamefont {Turner}}]{Carroll:2003wy}%
  \BibitemOpen
  \bibfield  {author} {\bibinfo {author} {\bibfnamefont {S.~M.}\ \bibnamefont
  {Carroll}}, \bibinfo {author} {\bibfnamefont {V.}~\bibnamefont {Duvvuri}},
  \bibinfo {author} {\bibfnamefont {M.}~\bibnamefont {Trodden}}, \ and\
  \bibinfo {author} {\bibfnamefont {M.~S.}\ \bibnamefont {Turner}},\ }\href
  {\doibase 10.1103/PhysRevD.70.043528} {\bibfield  {journal} {\bibinfo
  {journal} {Phys. Rev. D}\ }\textbf {\bibinfo {volume} {70}},\ \bibinfo
  {pages} {043528} (\bibinfo {year} {2004})},\ \Eprint
  {http://arxiv.org/abs/astro-ph/0306438} {arXiv:astro-ph/0306438} \BibitemShut
  {NoStop}%
\bibitem [{\citenamefont {Capozziello}\ \emph {et~al.}(2003)\citenamefont
  {Capozziello}, \citenamefont {Carloni},\ and\ \citenamefont
  {Troisi}}]{Capozziello:2003tk}%
  \BibitemOpen
  \bibfield  {author} {\bibinfo {author} {\bibfnamefont {S.}~\bibnamefont
  {Capozziello}}, \bibinfo {author} {\bibfnamefont {S.}~\bibnamefont
  {Carloni}}, \ and\ \bibinfo {author} {\bibfnamefont {A.}~\bibnamefont
  {Troisi}},\ }\href@noop {} {\bibfield  {journal} {\bibinfo  {journal} {Recent
  Res. Dev. Astron. Astrophys.}\ }\textbf {\bibinfo {volume} {1}},\ \bibinfo
  {pages} {625} (\bibinfo {year} {2003})},\ \Eprint
  {http://arxiv.org/abs/astro-ph/0303041} {arXiv:astro-ph/0303041} \BibitemShut
  {NoStop}%
\bibitem [{\citenamefont {Deffayet}\ \emph {et~al.}(2002)\citenamefont
  {Deffayet}, \citenamefont {Dvali},\ and\ \citenamefont
  {Gabadadze}}]{Deffayet:2001pu}%
  \BibitemOpen
  \bibfield  {author} {\bibinfo {author} {\bibfnamefont {C.}~\bibnamefont
  {Deffayet}}, \bibinfo {author} {\bibfnamefont {G.}~\bibnamefont {Dvali}}, \
  and\ \bibinfo {author} {\bibfnamefont {G.}~\bibnamefont {Gabadadze}},\ }\href
  {\doibase 10.1103/PhysRevD.65.044023} {\bibfield  {journal} {\bibinfo
  {journal} {Phys. Rev. D}\ }\textbf {\bibinfo {volume} {65}},\ \bibinfo
  {pages} {044023} (\bibinfo {year} {2002})},\ \Eprint
  {http://arxiv.org/abs/astro-ph/0105068} {arXiv:astro-ph/0105068} \BibitemShut
  {NoStop}%
\bibitem [{\citenamefont {Freese}\ and\ \citenamefont
  {Lewis}(2002)}]{freese2002cardassian}%
  \BibitemOpen
  \bibfield  {author} {\bibinfo {author} {\bibfnamefont {K.}~\bibnamefont
  {Freese}}\ and\ \bibinfo {author} {\bibfnamefont {M.}~\bibnamefont {Lewis}},\
  }\href@noop {} {\bibfield  {journal} {\bibinfo  {journal} {Physics Letters
  B}\ }\textbf {\bibinfo {volume} {540}},\ \bibinfo {pages} {1} (\bibinfo
  {year} {2002})}\BibitemShut {NoStop}%
\bibitem [{\citenamefont {Nojiri}\ and\ \citenamefont
  {Odintsov}(2004)}]{Nojiri:2003wx}%
  \BibitemOpen
  \bibfield  {author} {\bibinfo {author} {\bibfnamefont {S.}~\bibnamefont
  {Nojiri}}\ and\ \bibinfo {author} {\bibfnamefont {S.~D.}\ \bibnamefont
  {Odintsov}},\ }\href {\doibase 10.1142/S0217732304013295} {\bibfield
  {journal} {\bibinfo  {journal} {Mod. Phys. Lett. A}\ }\textbf {\bibinfo
  {volume} {19}},\ \bibinfo {pages} {627} (\bibinfo {year} {2004})},\ \Eprint
  {http://arxiv.org/abs/hep-th/0310045} {arXiv:hep-th/0310045} \BibitemShut
  {NoStop}%
\bibitem [{\citenamefont {Arkani-Hamed}\ \emph {et~al.}(2004)\citenamefont
  {Arkani-Hamed}, \citenamefont {Cheng}, \citenamefont {Luty},\ and\
  \citenamefont {Mukohyama}}]{ArkaniHamed:2003uy}%
  \BibitemOpen
  \bibfield  {author} {\bibinfo {author} {\bibfnamefont {N.}~\bibnamefont
  {Arkani-Hamed}}, \bibinfo {author} {\bibfnamefont {H.-C.}\ \bibnamefont
  {Cheng}}, \bibinfo {author} {\bibfnamefont {M.~A.}\ \bibnamefont {Luty}}, \
  and\ \bibinfo {author} {\bibfnamefont {S.}~\bibnamefont {Mukohyama}},\ }\href
  {\doibase 10.1088/1126-6708/2004/05/074} {\bibfield  {journal} {\bibinfo
  {journal} {JHEP}\ }\textbf {\bibinfo {volume} {05}},\ \bibinfo {pages} {074}
  (\bibinfo {year} {2004})},\ \Eprint {http://arxiv.org/abs/hep-th/0312099}
  {arXiv:hep-th/0312099} \BibitemShut {NoStop}%
\bibitem [{\citenamefont {Einstein}(1928)}]{einstein1928riemann}%
  \BibitemOpen
  \bibfield  {author} {\bibinfo {author} {\bibfnamefont {A.}~\bibnamefont
  {Einstein}},\ }\href@noop {} {\bibfield  {journal} {\bibinfo  {journal}
  {Preuss. Akad. Wiss. Phys. Math. Kl}\ }\textbf {\bibinfo {volume} {217}}
  (\bibinfo {year} {1928})}\BibitemShut {NoStop}%
\bibitem [{\citenamefont {M{\o}ller}(1978)}]{moller1978crisis}%
  \BibitemOpen
  \bibfield  {author} {\bibinfo {author} {\bibfnamefont {C.}~\bibnamefont
  {M{\o}ller}},\ }\href@noop {} {\bibfield  {journal} {\bibinfo  {journal} {K.
  Dan. Vidensk. Selsk., Mat.-Fys. Medd}\ }\textbf {\bibinfo {volume} {39}},\
  \bibinfo {pages} {1} (\bibinfo {year} {1978})}\BibitemShut {NoStop}%
\bibitem [{\citenamefont {Saez}(1983)}]{Saez:1983tap}%
  \BibitemOpen
  \bibfield  {author} {\bibinfo {author} {\bibfnamefont {D.}~\bibnamefont
  {Saez}},\ }\href {\doibase 10.1103/PhysRevD.27.2839} {\bibfield  {journal}
  {\bibinfo  {journal} {Phys. Rev. D}\ }\textbf {\bibinfo {volume} {27}},\
  \bibinfo {pages} {2839} (\bibinfo {year} {1983})}\BibitemShut {NoStop}%
\bibitem [{\citenamefont {Meyer}(1982)}]{meyer1982moller}%
  \BibitemOpen
  \bibfield  {author} {\bibinfo {author} {\bibfnamefont {H.}~\bibnamefont
  {Meyer}},\ }\href@noop {} {\bibfield  {journal} {\bibinfo  {journal} {General
  Relativity and Gravitation}\ }\textbf {\bibinfo {volume} {14}},\ \bibinfo
  {pages} {531} (\bibinfo {year} {1982})}\BibitemShut {NoStop}%
\bibitem [{\citenamefont {Hayashi}\ and\ \citenamefont
  {Nakano}(1967{\natexlab{a}})}]{hayashi1967extended}%
  \BibitemOpen
  \bibfield  {author} {\bibinfo {author} {\bibfnamefont {K.}~\bibnamefont
  {Hayashi}}\ and\ \bibinfo {author} {\bibfnamefont {T.}~\bibnamefont
  {Nakano}},\ }\href@noop {} {\bibfield  {journal} {\bibinfo  {journal}
  {Progress of Theoretical Physics}\ }\textbf {\bibinfo {volume} {38}},\
  \bibinfo {pages} {491} (\bibinfo {year} {1967}{\natexlab{a}})}\BibitemShut
  {NoStop}%
\bibitem [{\citenamefont {Hayashi}\ and\ \citenamefont
  {Shirafuji}(1979)}]{Hayashi:1979qx}%
  \BibitemOpen
  \bibfield  {author} {\bibinfo {author} {\bibfnamefont {K.}~\bibnamefont
  {Hayashi}}\ and\ \bibinfo {author} {\bibfnamefont {T.}~\bibnamefont
  {Shirafuji}},\ }\href {\doibase 10.1103/PhysRevD.19.3524} {\bibfield
  {journal} {\bibinfo  {journal} {Phys. Rev. D}\ }\textbf {\bibinfo {volume}
  {19}},\ \bibinfo {pages} {3524} (\bibinfo {year} {1979})},\ \bibinfo {note}
  {[Addendum: Phys.Rev.D 24, 3312--3314 (1982)]}\BibitemShut {NoStop}%
  \bibitem [{\citenamefont {Hohmann}\ and\ \citenamefont
  {Pfeifer}(2021)}]{Hohmann:2020dgy}%
  \BibitemOpen
  \bibfield  {author} {\bibinfo {author} {\bibfnamefont {M.}~\bibnamefont
  {Hohmann}}\ and\ \bibinfo {author} {\bibfnamefont {C.}~\bibnamefont
  {Pfeifer}},\ }\href {\doibase 10.1140/epjc/s10052-021-09165-x} {\bibfield
  {journal} {\bibinfo  {journal} {Eur. Phys. J. C}\ }\textbf {\bibinfo {volume}
  {81}},\ \bibinfo {pages} {376} (\bibinfo {year} {2021})},\ \Eprint
  {http://arxiv.org/abs/2012.14423} {arXiv:2012.14423 [gr-qc]} \BibitemShut
  {NoStop}%
\bibitem [{\citenamefont {Miyamoto}\ and\ \citenamefont
  {Nakano}(1971)}]{miyamoto1971possible}%
  \BibitemOpen
  \bibfield  {author} {\bibinfo {author} {\bibfnamefont {S.}~\bibnamefont
  {Miyamoto}}\ and\ \bibinfo {author} {\bibfnamefont {T.}~\bibnamefont
  {Nakano}},\ }\href@noop {} {\bibfield  {journal} {\bibinfo  {journal}
  {Progress of Theoretical Physics}\ }\textbf {\bibinfo {volume} {45}},\
  \bibinfo {pages} {295} (\bibinfo {year} {1971})}\BibitemShut {NoStop}%
\bibitem [{\citenamefont {S{\'a}ez}\ and\ \citenamefont
  {De~Juan}(1984)}]{saez1984mo11er}%
  \BibitemOpen
  \bibfield  {author} {\bibinfo {author} {\bibfnamefont {D.}~\bibnamefont
  {S{\'a}ez}}\ and\ \bibinfo {author} {\bibfnamefont {T.}~\bibnamefont
  {De~Juan}},\ }\href@noop {} {\bibfield  {journal} {\bibinfo  {journal}
  {General relativity and gravitation}\ }\textbf {\bibinfo {volume} {16}},\
  \bibinfo {pages} {501} (\bibinfo {year} {1984})}\BibitemShut {NoStop}%
\bibitem [{\citenamefont {Schweizer}\ and\ \citenamefont
  {Straumann}(1979)}]{schweizer1979poincare}%
  \BibitemOpen
  \bibfield  {author} {\bibinfo {author} {\bibfnamefont {M.}~\bibnamefont
  {Schweizer}}\ and\ \bibinfo {author} {\bibfnamefont {N.}~\bibnamefont
  {Straumann}},\ }\href@noop {} {\bibfield  {journal} {\bibinfo  {journal}
  {Physics Letters A}\ }\textbf {\bibinfo {volume} {71}},\ \bibinfo {pages}
  {493} (\bibinfo {year} {1979})}\BibitemShut {NoStop}%
\bibitem [{\citenamefont {Schweizer}\ \emph {et~al.}(1980)\citenamefont
  {Schweizer}, \citenamefont {Straumann},\ and\ \citenamefont
  {Wipf}}]{schweizer1980post}%
  \BibitemOpen
  \bibfield  {author} {\bibinfo {author} {\bibfnamefont {M.}~\bibnamefont
  {Schweizer}}, \bibinfo {author} {\bibfnamefont {N.}~\bibnamefont
  {Straumann}}, \ and\ \bibinfo {author} {\bibfnamefont {A.}~\bibnamefont
  {Wipf}},\ }\href@noop {} {\bibfield  {journal} {\bibinfo  {journal} {General
  Relativity and Gravitation}\ }\textbf {\bibinfo {volume} {12}},\ \bibinfo
  {pages} {951} (\bibinfo {year} {1980})}\BibitemShut {NoStop}%
\bibitem [{\citenamefont {Mikhail}\ \emph {et~al.}(1994)\citenamefont
  {Mikhail}, \citenamefont {Wanas}, \citenamefont {Lashin},\ and\ \citenamefont
  {Hindawi}}]{Mikhail:1994bj}%
  \BibitemOpen
  \bibfield  {author} {\bibinfo {author} {\bibfnamefont {F.}~\bibnamefont
  {Mikhail}}, \bibinfo {author} {\bibfnamefont {M.}~\bibnamefont {Wanas}},
  \bibinfo {author} {\bibfnamefont {E.}~\bibnamefont {Lashin}}, \ and\ \bibinfo
  {author} {\bibfnamefont {A.}~\bibnamefont {Hindawi}},\ }\href {\doibase
  10.1007/BF02107145} {\bibfield  {journal} {\bibinfo  {journal} {Gen. Rel.
  Grav.}\ }\textbf {\bibinfo {volume} {26}},\ \bibinfo {pages} {869} (\bibinfo
  {year} {1994})},\ \Eprint {http://arxiv.org/abs/gr-qc/9409039}
  {arXiv:gr-qc/9409039} \BibitemShut {NoStop}%
\bibitem [{\citenamefont {Mikhail}\ \emph {et~al.}(1993)\citenamefont
  {Mikhail}, \citenamefont {Wanas}, \citenamefont {Hindawi},\ and\
  \citenamefont {Lashin}}]{Mikhail:1994rj}%
  \BibitemOpen
  \bibfield  {author} {\bibinfo {author} {\bibfnamefont {F.}~\bibnamefont
  {Mikhail}}, \bibinfo {author} {\bibfnamefont {M.}~\bibnamefont {Wanas}},
  \bibinfo {author} {\bibfnamefont {A.}~\bibnamefont {Hindawi}}, \ and\
  \bibinfo {author} {\bibfnamefont {E.}~\bibnamefont {Lashin}},\ }\href
  {\doibase 10.1007/BF00672861} {\bibfield  {journal} {\bibinfo  {journal}
  {Int. J. Theor. Phys.}\ }\textbf {\bibinfo {volume} {32}},\ \bibinfo {pages}
  {1627} (\bibinfo {year} {1993})},\ \Eprint
  {http://arxiv.org/abs/gr-qc/9406046} {arXiv:gr-qc/9406046} \BibitemShut
  {NoStop}%
\bibitem [{\citenamefont {Shirafuji}\ \emph {et~al.}(1996)\citenamefont
  {Shirafuji}, \citenamefont {Nashed},\ and\ \citenamefont
  {Hayashi}}]{Shirafuji:1995xc}%
  \BibitemOpen
  \bibfield  {author} {\bibinfo {author} {\bibfnamefont {T.}~\bibnamefont
  {Shirafuji}}, \bibinfo {author} {\bibfnamefont {G.~G.}\ \bibnamefont
  {Nashed}}, \ and\ \bibinfo {author} {\bibfnamefont {K.}~\bibnamefont
  {Hayashi}},\ }\href {\doibase 10.1143/PTP.95.665} {\bibfield  {journal}
  {\bibinfo  {journal} {Prog. Theor. Phys.}\ }\textbf {\bibinfo {volume}
  {95}},\ \bibinfo {pages} {665} (\bibinfo {year} {1996})},\ \Eprint
  {http://arxiv.org/abs/gr-qc/9601044} {arXiv:gr-qc/9601044} \BibitemShut
  {NoStop}%
\bibitem [{\citenamefont {S{\'a}ez}(1984)}]{saez1984stationary}%
  \BibitemOpen
  \bibfield  {author} {\bibinfo {author} {\bibfnamefont {D.}~\bibnamefont
  {S{\'a}ez}},\ }\href@noop {} {\bibfield  {journal} {\bibinfo  {journal}
  {Physics Letters A}\ }\textbf {\bibinfo {volume} {106}},\ \bibinfo {pages}
  {293} (\bibinfo {year} {1984})}\BibitemShut {NoStop}%
\bibitem [{\citenamefont {Capozziello}\ and\ \citenamefont
  {De~Laurentis}(2011)}]{Capozziello:2011et}%
  \BibitemOpen
  \bibfield  {author} {\bibinfo {author} {\bibfnamefont {S.}~\bibnamefont
  {Capozziello}}\ and\ \bibinfo {author} {\bibfnamefont {M.}~\bibnamefont
  {De~Laurentis}},\ }\href {\doibase 10.1016/j.physrep.2011.09.003} {\bibfield
  {journal} {\bibinfo  {journal} {Phys. Rept.}\ }\textbf {\bibinfo {volume}
  {509}},\ \bibinfo {pages} {167} (\bibinfo {year} {2011})},\ \Eprint
  {http://arxiv.org/abs/1108.6266} {arXiv:1108.6266 [gr-qc]} \BibitemShut
  {NoStop}%
\bibitem [{\citenamefont {Nojiri}\ and\ \citenamefont
  {Odintsov}(2011)}]{Nojiri:2010wj}%
  \BibitemOpen
  \bibfield  {author} {\bibinfo {author} {\bibfnamefont {S.}~\bibnamefont
  {Nojiri}}\ and\ \bibinfo {author} {\bibfnamefont {S.~D.}\ \bibnamefont
  {Odintsov}},\ }\href {\doibase 10.1016/j.physrep.2011.04.001} {\bibfield
  {journal} {\bibinfo  {journal} {Phys. Rept.}\ }\textbf {\bibinfo {volume}
  {505}},\ \bibinfo {pages} {59} (\bibinfo {year} {2011})},\ \Eprint
  {http://arxiv.org/abs/1011.0544} {arXiv:1011.0544 [gr-qc]} \BibitemShut
  {NoStop}%
%%CITATION = ARXIV:1011.0544;%%
\bibitem [{\citenamefont {Weinberg}(2005)}]{Weinberg:1995mt}%
  \BibitemOpen
  \bibfield  {author} {\bibinfo {author} {\bibfnamefont {S.}~\bibnamefont
  {Weinberg}},\ }\href@noop {} {\emph {\bibinfo {title} {{The Quantum theory of
  fields. Vol. 1: Foundations}}}}\ (\bibinfo  {publisher} {Cambridge University
  Press},\ \bibinfo {year} {2005})\BibitemShut {NoStop}%
\bibitem [{\citenamefont {Clifton}\ \emph {et~al.}(2012)\citenamefont
  {Clifton}, \citenamefont {Ferreira}, \citenamefont {Padilla},\ and\
  \citenamefont {Skordis}}]{Clifton:2011jh}%
  \BibitemOpen
  \bibfield  {author} {\bibinfo {author} {\bibfnamefont {T.}~\bibnamefont
  {Clifton}}, \bibinfo {author} {\bibfnamefont {P.~G.}\ \bibnamefont
  {Ferreira}}, \bibinfo {author} {\bibfnamefont {A.}~\bibnamefont {Padilla}}, \
  and\ \bibinfo {author} {\bibfnamefont {C.}~\bibnamefont {Skordis}},\ }\href
  {\doibase 10.1016/j.physrep.2012.01.001} {\bibfield  {journal} {\bibinfo
  {journal} {Phys. Rept.}\ }\textbf {\bibinfo {volume} {513}},\ \bibinfo
  {pages} {1} (\bibinfo {year} {2012})},\ \Eprint
  {http://arxiv.org/abs/1106.2476} {arXiv:1106.2476 [astro-ph.CO]} \BibitemShut
  {NoStop}%
%%CITATION = ARXIV:1106.2476;%%
\bibitem [{\citenamefont {De~Felice}\ and\ \citenamefont
  {Tsujikawa}(2010)}]{DeFelice:2010aj}%
  \BibitemOpen
  \bibfield  {author} {\bibinfo {author} {\bibfnamefont {A.}~\bibnamefont
  {De~Felice}}\ and\ \bibinfo {author} {\bibfnamefont {S.}~\bibnamefont
  {Tsujikawa}},\ }\href {\doibase 10.12942/lrr-2010-3} {\bibfield  {journal}
  {\bibinfo  {journal} {Living Rev. Rel.}\ }\textbf {\bibinfo {volume} {13}},\
  \bibinfo {pages} {3} (\bibinfo {year} {2010})},\ \Eprint
  {http://arxiv.org/abs/1002.4928} {arXiv:1002.4928 [gr-qc]} \BibitemShut
  {NoStop}%
%%CITATION = ARXIV:1002.4928;%%
\bibitem [{\citenamefont {Unzicker}\ and\ \citenamefont
  {Case}(2005)}]{Unzicker:2005in}%
  \BibitemOpen
  \bibfield  {author} {\bibinfo {author} {\bibfnamefont {A.}~\bibnamefont
  {Unzicker}}\ and\ \bibinfo {author} {\bibfnamefont {T.}~\bibnamefont
  {Case}},\ }\href@noop {} {\  (\bibinfo {year} {2005})},\ \Eprint
  {http://arxiv.org/abs/physics/0503046} {arXiv:physics/0503046} \BibitemShut
  {NoStop}%
\bibitem [{\citenamefont {Aldrovandi}\ and\ \citenamefont
  {Pereira}(2013)}]{Aldrovandi:2013wha}%
  \BibitemOpen
  \bibfield  {author} {\bibinfo {author} {\bibfnamefont {R.}~\bibnamefont
  {Aldrovandi}}\ and\ \bibinfo {author} {\bibfnamefont {J.~G.}\ \bibnamefont
  {Pereira}},\ }\href {\doibase 10.1007/978-94-007-5143-9} {\emph {\bibinfo
  {title} {{Teleparallel Gravity}: {An Introduction}}}}\ (\bibinfo  {publisher}
  {Springer},\ \bibinfo {year} {2013})\BibitemShut {NoStop}%
\bibitem [{\citenamefont {Maluf}(2013)}]{Maluf:2013gaa}%
  \BibitemOpen
  \bibfield  {author} {\bibinfo {author} {\bibfnamefont {J.~W.}\ \bibnamefont
  {Maluf}},\ }\href {\doibase 10.1002/andp.201200272} {\bibfield  {journal}
  {\bibinfo  {journal} {Annalen Phys.}\ }\textbf {\bibinfo {volume} {525}},\
  \bibinfo {pages} {339} (\bibinfo {year} {2013})},\ \Eprint
  {http://arxiv.org/abs/1303.3897} {arXiv:1303.3897 [gr-qc]} \BibitemShut
  {NoStop}%
\bibitem [{\citenamefont {Cai}\ \emph {et~al.}(2016)\citenamefont {Cai},
  \citenamefont {Capozziello}, \citenamefont {De~Laurentis},\ and\
  \citenamefont {Saridakis}}]{Cai:2015emx}%
  \BibitemOpen
  \bibfield  {author} {\bibinfo {author} {\bibfnamefont {Y.-F.}\ \bibnamefont
  {Cai}}, \bibinfo {author} {\bibfnamefont {S.}~\bibnamefont {Capozziello}},
  \bibinfo {author} {\bibfnamefont {M.}~\bibnamefont {De~Laurentis}}, \ and\
  \bibinfo {author} {\bibfnamefont {E.~N.}\ \bibnamefont {Saridakis}},\ }\href
  {\doibase 10.1088/0034-4885/79/10/106901} {\bibfield  {journal} {\bibinfo
  {journal} {Rept. Prog. Phys.}\ }\textbf {\bibinfo {volume} {79}},\ \bibinfo
  {pages} {106901} (\bibinfo {year} {2016})},\ \Eprint
  {http://arxiv.org/abs/1511.07586} {arXiv:1511.07586 [gr-qc]} \BibitemShut
  {NoStop}%
\bibitem [{\citenamefont {Bengochea}\ and\ \citenamefont
  {Ferraro}(2009)}]{Bengochea:2008gz}%
  \BibitemOpen
  \bibfield  {author} {\bibinfo {author} {\bibfnamefont {G.~R.}\ \bibnamefont
  {Bengochea}}\ and\ \bibinfo {author} {\bibfnamefont {R.}~\bibnamefont
  {Ferraro}},\ }\href {\doibase 10.1103/PhysRevD.79.124019} {\bibfield
  {journal} {\bibinfo  {journal} {Phys. Rev. D}\ }\textbf {\bibinfo {volume}
  {79}},\ \bibinfo {pages} {124019} (\bibinfo {year} {2009})},\ \Eprint
  {http://arxiv.org/abs/0812.1205} {arXiv:0812.1205 [astro-ph]} \BibitemShut
  {NoStop}%
\bibitem [{\citenamefont {Linder}(2010)}]{Linder:2010py}%
  \BibitemOpen
  \bibfield  {author} {\bibinfo {author} {\bibfnamefont {E.~V.}\ \bibnamefont
  {Linder}},\ }\href {\doibase 10.1103/PhysRevD.81.127301,
  10.1103/PhysRevD.82.109902} {\bibfield  {journal} {\bibinfo  {journal} {Phys.
  Rev.}\ }\textbf {\bibinfo {volume} {D81}},\ \bibinfo {pages} {127301}
  (\bibinfo {year} {2010})},\ \bibinfo {note} {[Erratum: Phys.
  Rev.D82,109902(2010)]},\ \Eprint {http://arxiv.org/abs/1005.3039}
  {arXiv:1005.3039 [astro-ph.CO]} \BibitemShut {NoStop}%
\bibitem [{\citenamefont {Kofinas}\ and\ \citenamefont
  {Saridakis}(2014{\natexlab{a}})}]{Kofinas:2014owa}%
  \BibitemOpen
  \bibfield  {author} {\bibinfo {author} {\bibfnamefont {G.}~\bibnamefont
  {Kofinas}}\ and\ \bibinfo {author} {\bibfnamefont {E.~N.}\ \bibnamefont
  {Saridakis}},\ }\href {\doibase 10.1103/PhysRevD.90.084044} {\bibfield
  {journal} {\bibinfo  {journal} {Phys. Rev. D}\ }\textbf {\bibinfo {volume}
  {90}},\ \bibinfo {pages} {084044} (\bibinfo {year} {2014}{\natexlab{a}})},\
  \Eprint {http://arxiv.org/abs/1404.2249} {arXiv:1404.2249 [gr-qc]}
  \BibitemShut {NoStop}%
\bibitem [{\citenamefont {Kofinas}\ and\ \citenamefont
  {Saridakis}(2014{\natexlab{b}})}]{Kofinas:2014daa}%
  \BibitemOpen
  \bibfield  {author} {\bibinfo {author} {\bibfnamefont {G.}~\bibnamefont
  {Kofinas}}\ and\ \bibinfo {author} {\bibfnamefont {E.~N.}\ \bibnamefont
  {Saridakis}},\ }\href {\doibase 10.1103/PhysRevD.90.084045} {\bibfield
  {journal} {\bibinfo  {journal} {Phys. Rev.}\ }\textbf {\bibinfo {volume}
  {D90}},\ \bibinfo {pages} {084045} (\bibinfo {year} {2014}{\natexlab{b}})},\
  \Eprint {http://arxiv.org/abs/1408.0107} {arXiv:1408.0107 [gr-qc]}
  \BibitemShut {NoStop}%
\bibitem [{\citenamefont {Bahamonde}\ \emph {et~al.}(2015)\citenamefont
  {Bahamonde}, \citenamefont {B\"ohmer},\ and\ \citenamefont
  {Wright}}]{Bahamonde:2015zma}%
  \BibitemOpen
  \bibfield  {author} {\bibinfo {author} {\bibfnamefont {S.}~\bibnamefont
  {Bahamonde}}, \bibinfo {author} {\bibfnamefont {C.~G.}\ \bibnamefont
  {B\"ohmer}}, \ and\ \bibinfo {author} {\bibfnamefont {M.}~\bibnamefont
  {Wright}},\ }\href {\doibase 10.1103/PhysRevD.92.104042} {\bibfield
  {journal} {\bibinfo  {journal} {Phys. Rev. D}\ }\textbf {\bibinfo {volume}
  {92}},\ \bibinfo {pages} {104042} (\bibinfo {year} {2015})},\ \Eprint
  {http://arxiv.org/abs/1508.05120} {arXiv:1508.05120 [gr-qc]} \BibitemShut
  {NoStop}%
  \bibitem [{\citenamefont {Awad}\ and\ \citenamefont
  {Nashed}(2017)}]{Awad:2017sau}%
  \BibitemOpen
  \bibfield  {author} {\bibinfo {author} {\bibfnamefont {A.}~\bibnamefont
  {Awad}}\ and\ \bibinfo {author} {\bibfnamefont {G.}~\bibnamefont {Nashed}},\
  }\href {\doibase 10.1088/1475-7516/2017/02/046} {\bibfield  {journal}
  {\bibinfo  {journal} {JCAP}\ }\textbf {\bibinfo {volume} {02}},\ \bibinfo
  {pages} {046} (\bibinfo {year} {2017})},\ \Eprint
  {http://arxiv.org/abs/1701.06899} {arXiv:1701.06899 [gr-qc]} \BibitemShut
  {NoStop}%
  \bibitem [{\citenamefont {Nashed}(2014)}]{Nashed:2014sea}%
  \BibitemOpen
  \bibfield  {author} {\bibinfo {author} {\bibfnamefont {G.~G.~L.}\
  \bibnamefont {Nashed}},\ }\href {\doibase 10.1209/0295-5075/105/10001}
  {\bibfield  {journal} {\bibinfo  {journal} {EPL}\ }\textbf {\bibinfo {volume}
  {105}},\ \bibinfo {pages} {10001} (\bibinfo {year} {2014})},\ \Eprint
  {http://arxiv.org/abs/1501.00974} {arXiv:1501.00974 [gr-qc]} \BibitemShut
  {NoStop}%
  \bibitem [{\citenamefont {Elizalde}\ \emph {et~al.}(2020)\citenamefont
 {Elizalde}, \citenamefont {Nashed}, \citenamefont {Nojiri},\ and\
 \citenamefont {Odintsov}}]{Elizalde:2020icc}%
 \BibitemOpen
 \bibfield {author} {\bibinfo {author} {\bibfnamefont {E.}~\bibnamefont
 {Elizalde}}, \bibinfo {author} {\bibfnamefont {G.~G.~L.}\ \bibnamefont
 {Nashed}}, \bibinfo {author} {\bibfnamefont {S.}~\bibnamefont {Nojiri}}, \
 and\ \bibinfo {author} {\bibfnamefont {S.~D.}\ \bibnamefont {Odintsov}},\
 }\href {\doibase 10.1140/epjc/s10052-020-7686-3} {\bibfield {journal}
 {\bibinfo {journal} {Eur. Phys. J.}\ }\textbf {\bibinfo {volume} {C80}},\
 \bibinfo {pages} {109} (\bibinfo {year} {2020})},\ \Eprint
 {http://arxiv.org/abs/2001.11357} {arXiv:2001.11357 [gr-qc]} \BibitemShut
 {NoStop}%
\bibitem [{\citenamefont {Karpathopoulos}\ \emph {et~al.}(2018)\citenamefont
  {Karpathopoulos}, \citenamefont {Basilakos}, \citenamefont {Leon},
  \citenamefont {Paliathanasis},\ and\ \citenamefont
  {Tsamparlis}}]{Karpathopoulos:2017arc}%
  \BibitemOpen
  \bibfield  {author} {\bibinfo {author} {\bibfnamefont {L.}~\bibnamefont
  {Karpathopoulos}}, \bibinfo {author} {\bibfnamefont {S.}~\bibnamefont
  {Basilakos}}, \bibinfo {author} {\bibfnamefont {G.}~\bibnamefont {Leon}},
  \bibinfo {author} {\bibfnamefont {A.}~\bibnamefont {Paliathanasis}}, \ and\
  \bibinfo {author} {\bibfnamefont {M.}~\bibnamefont {Tsamparlis}},\ }\href
  {\doibase 10.1007/s10714-018-2400-6} {\bibfield  {journal} {\bibinfo
  {journal} {Gen. Rel. Grav.}\ }\textbf {\bibinfo {volume} {50}},\ \bibinfo
  {pages} {79} (\bibinfo {year} {2018})},\ \Eprint
  {http://arxiv.org/abs/1709.02197} {arXiv:1709.02197 [gr-qc]} \BibitemShut
  {NoStop}%
\bibitem [{\citenamefont {Ren}\ \emph {et~al.}(2021{\natexlab{a}})\citenamefont
  {Ren}, \citenamefont {Zhao}, \citenamefont {Saridakis},\ and\ \citenamefont
  {Cai}}]{Ren:2021uqb}%
  \BibitemOpen
  \bibfield  {author} {\bibinfo {author} {\bibfnamefont {X.}~\bibnamefont
  {Ren}}, \bibinfo {author} {\bibfnamefont {Y.}~\bibnamefont {Zhao}}, \bibinfo
  {author} {\bibfnamefont {E.~N.}\ \bibnamefont {Saridakis}}, \ and\ \bibinfo
  {author} {\bibfnamefont {Y.-F.}\ \bibnamefont {Cai}},\ }\href@noop {} {\
  (\bibinfo {year} {2021}{\natexlab{a}})},\ \Eprint
  {http://arxiv.org/abs/2105.04578} {arXiv:2105.04578 [astro-ph.CO]}
  \BibitemShut {NoStop}%
\bibitem [{\citenamefont {B\"ohmer}\ and\ \citenamefont
  {Jensko}(2021)}]{Bohmer:2021eoo}%
  \BibitemOpen
  \bibfield  {author} {\bibinfo {author} {\bibfnamefont {C.~G.}\ \bibnamefont
  {B\"ohmer}}\ and\ \bibinfo {author} {\bibfnamefont {E.}~\bibnamefont
  {Jensko}},\ }\href@noop {} {\  (\bibinfo {year} {2021})},\ \Eprint
  {http://arxiv.org/abs/2103.15906} {arXiv:2103.15906 [gr-qc]} \BibitemShut
  {NoStop}%
\bibitem [{\citenamefont {Geng}\ \emph {et~al.}(2011)\citenamefont {Geng},
  \citenamefont {Lee}, \citenamefont {Saridakis},\ and\ \citenamefont
  {Wu}}]{Geng:2011aj}%
  \BibitemOpen
  \bibfield  {author} {\bibinfo {author} {\bibfnamefont {C.-Q.}\ \bibnamefont
  {Geng}}, \bibinfo {author} {\bibfnamefont {C.-C.}\ \bibnamefont {Lee}},
  \bibinfo {author} {\bibfnamefont {E.~N.}\ \bibnamefont {Saridakis}}, \ and\
  \bibinfo {author} {\bibfnamefont {Y.-P.}\ \bibnamefont {Wu}},\ }\href
  {\doibase 10.1016/j.physletb.2011.09.082} {\bibfield  {journal} {\bibinfo
  {journal} {Phys. Lett. B}\ }\textbf {\bibinfo {volume} {704}},\ \bibinfo
  {pages} {384} (\bibinfo {year} {2011})},\ \Eprint
  {http://arxiv.org/abs/1109.1092} {arXiv:1109.1092 [hep-th]} \BibitemShut
  {NoStop}%
\bibitem [{\citenamefont {Hohmann}\ \emph {et~al.}(2018)\citenamefont
  {Hohmann}, \citenamefont {JÃ¤rv},\ and\ \citenamefont
  {Ualikhanova}}]{Hohmann:2018rwf}%
  \BibitemOpen
  \bibfield  {author} {\bibinfo {author} {\bibfnamefont {M.}~\bibnamefont
  {Hohmann}}, \bibinfo {author} {\bibfnamefont {L.}~\bibnamefont {JÃ¤rv}}, \
  and\ \bibinfo {author} {\bibfnamefont {U.}~\bibnamefont {Ualikhanova}},\
  }\href {\doibase 10.1103/PhysRevD.97.104011} {\bibfield  {journal} {\bibinfo
  {journal} {Phys. Rev.}\ }\textbf {\bibinfo {volume} {D97}},\ \bibinfo {pages}
  {104011} (\bibinfo {year} {2018})},\ \Eprint
  {http://arxiv.org/abs/1801.05786} {arXiv:1801.05786 [gr-qc]} \BibitemShut
  {NoStop}%
\bibitem [{\citenamefont {Bahamonde}\ \emph {et~al.}(2019)\citenamefont
  {Bahamonde}, \citenamefont {Dialektopoulos},\ and\ \citenamefont
  {Levi~Said}}]{Bahamonde:2019shr}%
  \BibitemOpen
  \bibfield  {author} {\bibinfo {author} {\bibfnamefont {S.}~\bibnamefont
  {Bahamonde}}, \bibinfo {author} {\bibfnamefont {K.~F.}\ \bibnamefont
  {Dialektopoulos}}, \ and\ \bibinfo {author} {\bibfnamefont {J.}~\bibnamefont
  {Levi~Said}},\ }\href {\doibase 10.1103/PhysRevD.100.064018} {\bibfield
  {journal} {\bibinfo  {journal} {Phys. Rev. D}\ }\textbf {\bibinfo {volume}
  {100}},\ \bibinfo {pages} {064018} (\bibinfo {year} {2019})},\ \Eprint
  {http://arxiv.org/abs/1904.10791} {arXiv:1904.10791 [gr-qc]} \BibitemShut
  {NoStop}%
\bibitem [{\citenamefont {Chen}\ \emph {et~al.}(2011)\citenamefont {Chen},
  \citenamefont {Dent}, \citenamefont {Dutta},\ and\ \citenamefont
  {Saridakis}}]{Chen:2010va}%
  \BibitemOpen
  \bibfield  {author} {\bibinfo {author} {\bibfnamefont {S.-H.}\ \bibnamefont
  {Chen}}, \bibinfo {author} {\bibfnamefont {J.~B.}\ \bibnamefont {Dent}},
  \bibinfo {author} {\bibfnamefont {S.}~\bibnamefont {Dutta}}, \ and\ \bibinfo
  {author} {\bibfnamefont {E.~N.}\ \bibnamefont {Saridakis}},\ }\href {\doibase
  10.1103/PhysRevD.83.023508} {\bibfield  {journal} {\bibinfo  {journal} {Phys.
  Rev.}\ }\textbf {\bibinfo {volume} {D83}},\ \bibinfo {pages} {023508}
  (\bibinfo {year} {2011})},\ \Eprint {http://arxiv.org/abs/1008.1250}
  {arXiv:1008.1250 [astro-ph.CO]} \BibitemShut {NoStop}%
\bibitem [{\citenamefont {Zheng}\ and\ \citenamefont
  {Huang}(2011)}]{Zheng:2010am}%
  \BibitemOpen
  \bibfield  {author} {\bibinfo {author} {\bibfnamefont {R.}~\bibnamefont
  {Zheng}}\ and\ \bibinfo {author} {\bibfnamefont {Q.-G.}\ \bibnamefont
  {Huang}},\ }\href {\doibase 10.1088/1475-7516/2011/03/002} {\bibfield
  {journal} {\bibinfo  {journal} {JCAP}\ }\textbf {\bibinfo {volume} {1103}},\
  \bibinfo {pages} {002} (\bibinfo {year} {2011})},\ \Eprint
  {http://arxiv.org/abs/1010.3512} {arXiv:1010.3512 [gr-qc]} \BibitemShut
  {NoStop}%
\bibitem [{\citenamefont {Bamba}\ \emph {et~al.}(2011)\citenamefont {Bamba},
  \citenamefont {Geng}, \citenamefont {Lee},\ and\ \citenamefont
  {Luo}}]{Bamba:2010wb}%
  \BibitemOpen
  \bibfield  {author} {\bibinfo {author} {\bibfnamefont {K.}~\bibnamefont
  {Bamba}}, \bibinfo {author} {\bibfnamefont {C.-Q.}\ \bibnamefont {Geng}},
  \bibinfo {author} {\bibfnamefont {C.-C.}\ \bibnamefont {Lee}}, \ and\
  \bibinfo {author} {\bibfnamefont {L.-W.}\ \bibnamefont {Luo}},\ }\href
  {\doibase 10.1088/1475-7516/2011/01/021} {\bibfield  {journal} {\bibinfo
  {journal} {JCAP}\ }\textbf {\bibinfo {volume} {1101}},\ \bibinfo {pages}
  {021} (\bibinfo {year} {2011})},\ \Eprint {http://arxiv.org/abs/1011.0508}
  {arXiv:1011.0508 [astro-ph.CO]} \BibitemShut {NoStop}%
\bibitem [{\citenamefont {Cai}\ \emph {et~al.}(2011)\citenamefont {Cai},
  \citenamefont {Chen}, \citenamefont {Dent}, \citenamefont {Dutta},\ and\
  \citenamefont {Saridakis}}]{Cai:2011tc}%
  \BibitemOpen
  \bibfield  {author} {\bibinfo {author} {\bibfnamefont {Y.-F.}\ \bibnamefont
  {Cai}}, \bibinfo {author} {\bibfnamefont {S.-H.}\ \bibnamefont {Chen}},
  \bibinfo {author} {\bibfnamefont {J.~B.}\ \bibnamefont {Dent}}, \bibinfo
  {author} {\bibfnamefont {S.}~\bibnamefont {Dutta}}, \ and\ \bibinfo {author}
  {\bibfnamefont {E.~N.}\ \bibnamefont {Saridakis}},\ }\href {\doibase
  10.1088/0264-9381/28/21/215011} {\bibfield  {journal} {\bibinfo  {journal}
  {Class. Quant. Grav.}\ }\textbf {\bibinfo {volume} {28}},\ \bibinfo {pages}
  {215011} (\bibinfo {year} {2011})},\ \Eprint {http://arxiv.org/abs/1104.4349}
  {arXiv:1104.4349 [astro-ph.CO]} \BibitemShut {NoStop}%
\bibitem [{\citenamefont {Capozziello}\ \emph {et~al.}(2011)\citenamefont
  {Capozziello}, \citenamefont {Cardone}, \citenamefont {Farajollahi},\ and\
  \citenamefont {Ravanpak}}]{Capozziello:2011hj}%
  \BibitemOpen
  \bibfield  {author} {\bibinfo {author} {\bibfnamefont {S.}~\bibnamefont
  {Capozziello}}, \bibinfo {author} {\bibfnamefont {V.~F.}\ \bibnamefont
  {Cardone}}, \bibinfo {author} {\bibfnamefont {H.}~\bibnamefont
  {Farajollahi}}, \ and\ \bibinfo {author} {\bibfnamefont {A.}~\bibnamefont
  {Ravanpak}},\ }\href {\doibase 10.1103/PhysRevD.84.043527} {\bibfield
  {journal} {\bibinfo  {journal} {Phys. Rev.}\ }\textbf {\bibinfo {volume}
  {D84}},\ \bibinfo {pages} {043527} (\bibinfo {year} {2011})},\ \Eprint
  {http://arxiv.org/abs/1108.2789} {arXiv:1108.2789 [astro-ph.CO]} \BibitemShut
  {NoStop}%
\bibitem [{\citenamefont {Wei}\ \emph {et~al.}(2012)\citenamefont {Wei},
  \citenamefont {Guo},\ and\ \citenamefont {Wang}}]{Wei:2011aa}%
  \BibitemOpen
  \bibfield  {author} {\bibinfo {author} {\bibfnamefont {H.}~\bibnamefont
  {Wei}}, \bibinfo {author} {\bibfnamefont {X.-J.}\ \bibnamefont {Guo}}, \ and\
  \bibinfo {author} {\bibfnamefont {L.-F.}\ \bibnamefont {Wang}},\ }\href
  {\doibase 10.1016/j.physletb.2011.12.039} {\bibfield  {journal} {\bibinfo
  {journal} {Phys. Lett.}\ }\textbf {\bibinfo {volume} {B707}},\ \bibinfo
  {pages} {298} (\bibinfo {year} {2012})},\ \Eprint
  {http://arxiv.org/abs/1112.2270} {arXiv:1112.2270 [gr-qc]} \BibitemShut
  {NoStop}%
\bibitem [{\citenamefont {AmorÃ³s}\ \emph {et~al.}(2013)\citenamefont
  {AmorÃ³s}, \citenamefont {de~Haro},\ and\ \citenamefont
  {Odintsov}}]{Amoros:2013nxa}%
  \BibitemOpen
  \bibfield  {author} {\bibinfo {author} {\bibfnamefont {J.}~\bibnamefont
  {AmorÃ³s}}, \bibinfo {author} {\bibfnamefont {J.}~\bibnamefont {de~Haro}}, \
  and\ \bibinfo {author} {\bibfnamefont {S.~D.}\ \bibnamefont {Odintsov}},\
  }\href {\doibase 10.1103/PhysRevD.87.104037} {\bibfield  {journal} {\bibinfo
  {journal} {Phys. Rev.}\ }\textbf {\bibinfo {volume} {D87}},\ \bibinfo {pages}
  {104037} (\bibinfo {year} {2013})},\ \Eprint {http://arxiv.org/abs/1305.2344}
  {arXiv:1305.2344 [gr-qc]} \BibitemShut {NoStop}%
\bibitem [{\citenamefont {Otalora}(2013)}]{Otalora:2013dsa}%
  \BibitemOpen
  \bibfield  {author} {\bibinfo {author} {\bibfnamefont {G.}~\bibnamefont
  {Otalora}},\ }\href {\doibase 10.1103/PhysRevD.88.063505} {\bibfield
  {journal} {\bibinfo  {journal} {Phys. Rev.}\ }\textbf {\bibinfo {volume}
  {D88}},\ \bibinfo {pages} {063505} (\bibinfo {year} {2013})},\ \Eprint
  {http://arxiv.org/abs/1305.5896} {arXiv:1305.5896 [gr-qc]} \BibitemShut
  {NoStop}%
\bibitem [{\citenamefont {Bamba}\ \emph {et~al.}(2013)\citenamefont {Bamba},
  \citenamefont {Odintsov},\ and\ \citenamefont
  {SÃ¡ez-GÃ³mez}}]{Bamba:2013jqa}%
  \BibitemOpen
  \bibfield  {author} {\bibinfo {author} {\bibfnamefont {K.}~\bibnamefont
  {Bamba}}, \bibinfo {author} {\bibfnamefont {S.~D.}\ \bibnamefont {Odintsov}},
  \ and\ \bibinfo {author} {\bibfnamefont {D.}~\bibnamefont {SÃ¡ez-GÃ³mez}},\
  }\href {\doibase 10.1103/PhysRevD.88.084042} {\bibfield  {journal} {\bibinfo
  {journal} {Phys. Rev.}\ }\textbf {\bibinfo {volume} {D88}},\ \bibinfo {pages}
  {084042} (\bibinfo {year} {2013})},\ \Eprint {http://arxiv.org/abs/1308.5789}
  {arXiv:1308.5789 [gr-qc]} \BibitemShut {NoStop}%
\bibitem [{\citenamefont {Li}\ \emph {et~al.}(2013)\citenamefont {Li},
  \citenamefont {Lee},\ and\ \citenamefont {Geng}}]{Li:2013xea}%
  \BibitemOpen
  \bibfield  {author} {\bibinfo {author} {\bibfnamefont {J.-T.}\ \bibnamefont
  {Li}}, \bibinfo {author} {\bibfnamefont {C.-C.}\ \bibnamefont {Lee}}, \ and\
  \bibinfo {author} {\bibfnamefont {C.-Q.}\ \bibnamefont {Geng}},\ }\href
  {\doibase 10.1140/epjc/s10052-013-2315-z} {\bibfield  {journal} {\bibinfo
  {journal} {Eur. Phys. J.}\ }\textbf {\bibinfo {volume} {C73}},\ \bibinfo
  {pages} {2315} (\bibinfo {year} {2013})},\ \Eprint
  {http://arxiv.org/abs/1302.2688} {arXiv:1302.2688 [gr-qc]} \BibitemShut
  {NoStop}%
\bibitem [{\citenamefont {Nashed}(2007)}]{Nashed20071047}%
  \BibitemOpen
  \bibfield  {author} {\bibinfo {author} {\bibfnamefont {G.}~\bibnamefont
  {Nashed}},\ }\href {\doibase 10.1142/S021773230702141X} {\bibfield  {journal}
  {\bibinfo  {journal} {Modern Physics Letters A}\ }\textbf {\bibinfo {volume}
  {22}},\ \bibinfo {pages} {1047} (\bibinfo {year} {2007})},\ \bibinfo {note}
  {cited By 25}\BibitemShut {NoStop}%
\bibitem [{\citenamefont {Paliathanasis}\ \emph {et~al.}(2014)\citenamefont
  {Paliathanasis}, \citenamefont {Basilakos}, \citenamefont {Saridakis},
  \citenamefont {Capozziello}, \citenamefont {Atazadeh}, \citenamefont
  {Darabi},\ and\ \citenamefont {Tsamparlis}}]{Paliathanasis:2014iva}%
  \BibitemOpen
  \bibfield  {author} {\bibinfo {author} {\bibfnamefont {A.}~\bibnamefont
  {Paliathanasis}}, \bibinfo {author} {\bibfnamefont {S.}~\bibnamefont
  {Basilakos}}, \bibinfo {author} {\bibfnamefont {E.~N.}\ \bibnamefont
  {Saridakis}}, \bibinfo {author} {\bibfnamefont {S.}~\bibnamefont
  {Capozziello}}, \bibinfo {author} {\bibfnamefont {K.}~\bibnamefont
  {Atazadeh}}, \bibinfo {author} {\bibfnamefont {F.}~\bibnamefont {Darabi}}, \
  and\ \bibinfo {author} {\bibfnamefont {M.}~\bibnamefont {Tsamparlis}},\
  }\href {\doibase 10.1103/PhysRevD.89.104042} {\bibfield  {journal} {\bibinfo
  {journal} {Phys. Rev.}\ ,\ \bibinfo {pages} {104042}} (\bibinfo {year}
  {2014})},\ \Eprint {http://arxiv.org/abs/1402.5935} {arXiv:1402.5935 [gr-qc]}
  \BibitemShut {NoStop}%
\bibitem [{\citenamefont {Malekjani}\ \emph {et~al.}(2017)\citenamefont
  {Malekjani}, \citenamefont {Haidari},\ and\ \citenamefont
  {Basilakos}}]{Malekjani:2016mtm}%
  \BibitemOpen
  \bibfield  {author} {\bibinfo {author} {\bibfnamefont {M.}~\bibnamefont
  {Malekjani}}, \bibinfo {author} {\bibfnamefont {N.}~\bibnamefont {Haidari}},
  \ and\ \bibinfo {author} {\bibfnamefont {S.}~\bibnamefont {Basilakos}},\
  }\href {\doibase 10.1093/mnras/stw3367} {\bibfield  {journal} {\bibinfo
  {journal} {Mon. Not. Roy. Astron. Soc.}\ }\textbf {\bibinfo {volume} {466}},\
  \bibinfo {pages} {3488} (\bibinfo {year} {2017})},\ \Eprint
  {http://arxiv.org/abs/1609.01964} {arXiv:1609.01964 [gr-qc]} \BibitemShut
  {NoStop}%
\bibitem [{\citenamefont {Farrugia}\ and\ \citenamefont
  {Said}(2016)}]{Farrugia:2016qqe}%
  \BibitemOpen
  \bibfield  {author} {\bibinfo {author} {\bibfnamefont {G.}~\bibnamefont
  {Farrugia}}\ and\ \bibinfo {author} {\bibfnamefont {J.~L.}\ \bibnamefont
  {Said}},\ }\href {\doibase 10.1103/PhysRevD.94.124054} {\bibfield  {journal}
  {\bibinfo  {journal} {Phys. Rev.}\ }\textbf {\bibinfo {volume} {D94}},\
  \bibinfo {pages} {124054} (\bibinfo {year} {2016})},\ \Eprint
  {http://arxiv.org/abs/1701.00134} {arXiv:1701.00134 [gr-qc]} \BibitemShut
  {NoStop}%
\bibitem [{\citenamefont {Qi}\ \emph {et~al.}(2017)\citenamefont {Qi},
  \citenamefont {Cao}, \citenamefont {Biesiada}, \citenamefont {Zheng},\ and\
  \citenamefont {Zhu}}]{Qi:2017xzl}%
  \BibitemOpen
  \bibfield  {author} {\bibinfo {author} {\bibfnamefont {J.-Z.}\ \bibnamefont
  {Qi}}, \bibinfo {author} {\bibfnamefont {S.}~\bibnamefont {Cao}}, \bibinfo
  {author} {\bibfnamefont {M.}~\bibnamefont {Biesiada}}, \bibinfo {author}
  {\bibfnamefont {X.}~\bibnamefont {Zheng}}, \ and\ \bibinfo {author}
  {\bibfnamefont {H.}~\bibnamefont {Zhu}},\ }\href {\doibase
  10.1140/epjc/s10052-017-5069-1} {\bibfield  {journal} {\bibinfo  {journal}
  {Eur. Phys. J.}\ }\textbf {\bibinfo {volume} {C77}},\ \bibinfo {pages} {502}
  (\bibinfo {year} {2017})},\ \Eprint {http://arxiv.org/abs/1708.08603}
  {arXiv:1708.08603 [astro-ph.CO]} \BibitemShut {NoStop}%
\bibitem [{\citenamefont {Cai}\ \emph {et~al.}(2018)\citenamefont {Cai},
  \citenamefont {Li}, \citenamefont {Saridakis},\ and\ \citenamefont
  {Xue}}]{Cai:2018rzd}%
  \BibitemOpen
  \bibfield  {author} {\bibinfo {author} {\bibfnamefont {Y.-F.}\ \bibnamefont
  {Cai}}, \bibinfo {author} {\bibfnamefont {C.}~\bibnamefont {Li}}, \bibinfo
  {author} {\bibfnamefont {E.~N.}\ \bibnamefont {Saridakis}}, \ and\ \bibinfo
  {author} {\bibfnamefont {L.}~\bibnamefont {Xue}},\ }\href {\doibase
  10.1103/PhysRevD.97.103513} {\bibfield  {journal} {\bibinfo  {journal} {Phys.
  Rev.}\ }\textbf {\bibinfo {volume} {D97}},\ \bibinfo {pages} {103513}
  (\bibinfo {year} {2018})},\ \Eprint {http://arxiv.org/abs/1801.05827}
  {arXiv:1801.05827 [gr-qc]} \BibitemShut {NoStop}%
\bibitem [{\citenamefont {Nashed}(2006)}]{Nashed20062241}%
  \BibitemOpen
  \bibfield  {author} {\bibinfo {author} {\bibfnamefont {G.}~\bibnamefont
  {Nashed}},\ }\href {\doibase 10.1142/S0217732306020445} {\bibfield  {journal}
  {\bibinfo  {journal} {Modern Physics Letters A}\ }\textbf {\bibinfo {volume}
  {21}},\ \bibinfo {pages} {2241} (\bibinfo {year} {2006})},\ \bibinfo {note}
  {cited By 24}\BibitemShut {NoStop}%
\bibitem [{\citenamefont {Anagnostopoulos}\ \emph {et~al.}(2019)\citenamefont
  {Anagnostopoulos}, \citenamefont {Basilakos},\ and\ \citenamefont
  {Saridakis}}]{Anagnostopoulos:2019miu}%
  \BibitemOpen
  \bibfield  {author} {\bibinfo {author} {\bibfnamefont {F.~K.}\ \bibnamefont
  {Anagnostopoulos}}, \bibinfo {author} {\bibfnamefont {S.}~\bibnamefont
  {Basilakos}}, \ and\ \bibinfo {author} {\bibfnamefont {E.~N.}\ \bibnamefont
  {Saridakis}},\ }\href {\doibase 10.1103/PhysRevD.100.083517} {\bibfield
  {journal} {\bibinfo  {journal} {Phys. Rev. D}\ }\textbf {\bibinfo {volume}
  {100}},\ \bibinfo {pages} {083517} (\bibinfo {year} {2019})},\ \Eprint
  {http://arxiv.org/abs/1907.07533} {arXiv:1907.07533 [astro-ph.CO]}
  \BibitemShut {NoStop}%
\bibitem [{\citenamefont {Nunes}\ \emph {et~al.}(2019)\citenamefont {Nunes},
  \citenamefont {Alves},\ and\ \citenamefont {de~Araujo}}]{Nunes:2019bjq}%
  \BibitemOpen
  \bibfield  {author} {\bibinfo {author} {\bibfnamefont {R.~C.}\ \bibnamefont
  {Nunes}}, \bibinfo {author} {\bibfnamefont {M.~E.~S.}\ \bibnamefont {Alves}},
  \ and\ \bibinfo {author} {\bibfnamefont {J.~C.~N.}\ \bibnamefont
  {de~Araujo}},\ }\href {\doibase 10.1103/PhysRevD.100.064012} {\bibfield
  {journal} {\bibinfo  {journal} {Phys. Rev. D}\ }\textbf {\bibinfo {volume}
  {100}},\ \bibinfo {pages} {064012} (\bibinfo {year} {2019})},\ \Eprint
  {http://arxiv.org/abs/1905.03237} {arXiv:1905.03237 [gr-qc]} \BibitemShut
  {NoStop}%
\bibitem [{\citenamefont {Cai}\ \emph {et~al.}(2020)\citenamefont {Cai},
  \citenamefont {Khurshudyan},\ and\ \citenamefont {Saridakis}}]{Cai:2019bdh}%
  \BibitemOpen
  \bibfield  {author} {\bibinfo {author} {\bibfnamefont {Y.-F.}\ \bibnamefont
  {Cai}}, \bibinfo {author} {\bibfnamefont {M.}~\bibnamefont {Khurshudyan}}, \
  and\ \bibinfo {author} {\bibfnamefont {E.~N.}\ \bibnamefont {Saridakis}},\
  }\href {\doibase 10.3847/1538-4357/ab5a7f} {\bibfield  {journal} {\bibinfo
  {journal} {Astrophys. J.}\ }\textbf {\bibinfo {volume} {888}},\ \bibinfo
  {pages} {62} (\bibinfo {year} {2020})},\ \Eprint
  {http://arxiv.org/abs/1907.10813} {arXiv:1907.10813 [astro-ph.CO]}
  \BibitemShut {NoStop}%
\bibitem [{\citenamefont {Yan}\ \emph {et~al.}(2020)\citenamefont {Yan},
  \citenamefont {Zhang}, \citenamefont {Chen}, \citenamefont {Zhang},
  \citenamefont {Cai},\ and\ \citenamefont {Saridakis}}]{Yan:2019gbw}%
  \BibitemOpen
  \bibfield  {author} {\bibinfo {author} {\bibfnamefont {S.-F.}\ \bibnamefont
  {Yan}}, \bibinfo {author} {\bibfnamefont {P.}~\bibnamefont {Zhang}}, \bibinfo
  {author} {\bibfnamefont {J.-W.}\ \bibnamefont {Chen}}, \bibinfo {author}
  {\bibfnamefont {X.-Z.}\ \bibnamefont {Zhang}}, \bibinfo {author}
  {\bibfnamefont {Y.-F.}\ \bibnamefont {Cai}}, \ and\ \bibinfo {author}
  {\bibfnamefont {E.~N.}\ \bibnamefont {Saridakis}},\ }\href {\doibase
  10.1103/PhysRevD.101.121301} {\bibfield  {journal} {\bibinfo  {journal}
  {Phys. Rev. D}\ }\textbf {\bibinfo {volume} {101}},\ \bibinfo {pages}
  {121301} (\bibinfo {year} {2020})},\ \Eprint
  {http://arxiv.org/abs/1909.06388} {arXiv:1909.06388 [astro-ph.CO]}
  \BibitemShut {NoStop}%
\bibitem [{\citenamefont {El~Hanafy}\ and\ \citenamefont
  {Nashed}(2019)}]{ElHanafy:2019zhr}%
  \BibitemOpen
  \bibfield  {author} {\bibinfo {author} {\bibfnamefont {W.}~\bibnamefont
  {El~Hanafy}}\ and\ \bibinfo {author} {\bibfnamefont {G.~G.~L.}\ \bibnamefont
  {Nashed}},\ }\href {\doibase 10.1103/PhysRevD.100.083535} {\bibfield
  {journal} {\bibinfo  {journal} {Phys. Rev. D}\ }\textbf {\bibinfo {volume}
  {100}},\ \bibinfo {pages} {083535} (\bibinfo {year} {2019})},\ \Eprint
  {http://arxiv.org/abs/1910.04160} {arXiv:1910.04160 [gr-qc]} \BibitemShut
  {NoStop}%
\bibitem [{\citenamefont {Saridakis}\ \emph {et~al.}(2020)\citenamefont
  {Saridakis}, \citenamefont {Myrzakul}, \citenamefont {Myrzakulov},\ and\
  \citenamefont {Yerzhanov}}]{Saridakis:2019qwt}%
  \BibitemOpen
  \bibfield  {author} {\bibinfo {author} {\bibfnamefont {E.~N.}\ \bibnamefont
  {Saridakis}}, \bibinfo {author} {\bibfnamefont {S.}~\bibnamefont {Myrzakul}},
  \bibinfo {author} {\bibfnamefont {K.}~\bibnamefont {Myrzakulov}}, \ and\
  \bibinfo {author} {\bibfnamefont {K.}~\bibnamefont {Yerzhanov}},\ }\href
  {\doibase 10.1103/PhysRevD.102.023525} {\bibfield  {journal} {\bibinfo
  {journal} {Phys. Rev. D}\ }\textbf {\bibinfo {volume} {102}},\ \bibinfo
  {pages} {023525} (\bibinfo {year} {2020})},\ \Eprint
  {http://arxiv.org/abs/1912.03882} {arXiv:1912.03882 [gr-qc]} \BibitemShut
  {NoStop}%
\bibitem [{\citenamefont {Awad}\ and\ \citenamefont {Nashed}(2017)}]{Awad2017}%
  \BibitemOpen
  \bibfield  {author} {\bibinfo {author} {\bibfnamefont {A.}~\bibnamefont
  {Awad}}\ and\ \bibinfo {author} {\bibfnamefont {G.}~\bibnamefont {Nashed}},\
  }\href {\doibase 10.1088/1475-7516/2017/02/046} {\bibfield  {journal}
  {\bibinfo  {journal} {Journal of Cosmology and Astroparticle Physics}\
  }\textbf {\bibinfo {volume} {2017}} (\bibinfo {year} {2017}),\
  10.1088/1475-7516/2017/02/046},\ \bibinfo {note} {cited By 20}\BibitemShut
  {NoStop}%
\bibitem [{\citenamefont {Wang}\ and\ \citenamefont
  {Mota}(2020)}]{Wang:2020zfv}%
  \BibitemOpen
  \bibfield  {author} {\bibinfo {author} {\bibfnamefont {D.}~\bibnamefont
  {Wang}}\ and\ \bibinfo {author} {\bibfnamefont {D.}~\bibnamefont {Mota}},\
  }\href {\doibase 10.1103/PhysRevD.102.063530} {\bibfield  {journal} {\bibinfo
   {journal} {Phys. Rev. D}\ }\textbf {\bibinfo {volume} {102}},\ \bibinfo
  {pages} {063530} (\bibinfo {year} {2020})},\ \Eprint
  {http://arxiv.org/abs/2003.10095} {arXiv:2003.10095 [astro-ph.CO]}
  \BibitemShut {NoStop}%
\bibitem [{\citenamefont {Ren}\ \emph {et~al.}(2021{\natexlab{b}})\citenamefont
  {Ren}, \citenamefont {Wong}, \citenamefont {Cai},\ and\ \citenamefont
  {Saridakis}}]{Ren:2021tfi}%
  \BibitemOpen
  \bibfield  {author} {\bibinfo {author} {\bibfnamefont {X.}~\bibnamefont
  {Ren}}, \bibinfo {author} {\bibfnamefont {T.~H.~T.}\ \bibnamefont {Wong}},
  \bibinfo {author} {\bibfnamefont {Y.-F.}\ \bibnamefont {Cai}}, \ and\
  \bibinfo {author} {\bibfnamefont {E.~N.}\ \bibnamefont {Saridakis}},\ }\href
  {\doibase 10.1016/j.dark.2021.100812} {\bibfield  {journal} {\bibinfo
  {journal} {Phys. Dark Univ.}\ }\textbf {\bibinfo {volume} {32}},\ \bibinfo
  {pages} {100812} (\bibinfo {year} {2021}{\natexlab{b}})},\ \Eprint
  {http://arxiv.org/abs/2103.01260} {arXiv:2103.01260 [astro-ph.CO]}
  \BibitemShut {NoStop}%
\bibitem [{\citenamefont {Boehmer}\ \emph {et~al.}(2011)\citenamefont
  {Boehmer}, \citenamefont {Mussa},\ and\ \citenamefont
  {Tamanini}}]{Boehmer:2011gw}%
  \BibitemOpen
  \bibfield  {author} {\bibinfo {author} {\bibfnamefont {C.~G.}\ \bibnamefont
  {Boehmer}}, \bibinfo {author} {\bibfnamefont {A.}~\bibnamefont {Mussa}}, \
  and\ \bibinfo {author} {\bibfnamefont {N.}~\bibnamefont {Tamanini}},\ }\href
  {\doibase 10.1088/0264-9381/28/24/245020} {\bibfield  {journal} {\bibinfo
  {journal} {Class. Quant. Grav.}\ }\textbf {\bibinfo {volume} {28}},\ \bibinfo
  {pages} {245020} (\bibinfo {year} {2011})},\ \Eprint
  {http://arxiv.org/abs/1107.4455} {arXiv:1107.4455 [gr-qc]} \BibitemShut
  {NoStop}%
\bibitem [{\citenamefont {Gonzalez}\ \emph {et~al.}(2012)\citenamefont
  {Gonzalez}, \citenamefont {Saridakis},\ and\ \citenamefont
  {Vasquez}}]{Gonzalez:2011dr}%
  \BibitemOpen
  \bibfield  {author} {\bibinfo {author} {\bibfnamefont {P.~A.}\ \bibnamefont
  {Gonzalez}}, \bibinfo {author} {\bibfnamefont {E.~N.}\ \bibnamefont
  {Saridakis}}, \ and\ \bibinfo {author} {\bibfnamefont {Y.}~\bibnamefont
  {Vasquez}},\ }\href {\doibase 10.1007/JHEP07(2012)053} {\bibfield  {journal}
  {\bibinfo  {journal} {JHEP}\ }\textbf {\bibinfo {volume} {07}},\ \bibinfo
  {pages} {053} (\bibinfo {year} {2012})},\ \Eprint
  {http://arxiv.org/abs/1110.4024} {arXiv:1110.4024 [gr-qc]} \BibitemShut
  {NoStop}%
\bibitem [{\citenamefont {Ferraro}\ and\ \citenamefont
  {Fiorini}(2011)}]{Ferraro:2011ks}%
  \BibitemOpen
  \bibfield  {author} {\bibinfo {author} {\bibfnamefont {R.}~\bibnamefont
  {Ferraro}}\ and\ \bibinfo {author} {\bibfnamefont {F.}~\bibnamefont
  {Fiorini}},\ }\href {\doibase 10.1103/PhysRevD.84.083518} {\bibfield
  {journal} {\bibinfo  {journal} {Phys. Rev. D}\ }\textbf {\bibinfo {volume}
  {84}},\ \bibinfo {pages} {083518} (\bibinfo {year} {2011})},\ \Eprint
  {http://arxiv.org/abs/1109.4209} {arXiv:1109.4209 [gr-qc]} \BibitemShut
  {NoStop}%
\bibitem [{\citenamefont {Nashed}(2010)}]{Nashed2010}%
  \BibitemOpen
  \bibfield  {author} {\bibinfo {author} {\bibfnamefont {G.}~\bibnamefont
  {Nashed}},\ }\href {\doibase 10.1088/1674-1056/19/2/020401} {\bibfield
  {journal} {\bibinfo  {journal} {Chinese Physics B}\ }\textbf {\bibinfo
  {volume} {19}} (\bibinfo {year} {2010}),\ 10.1088/1674-1056/19/2/020401},\
  \bibinfo {note} {cited By 24}\BibitemShut {NoStop}%
\bibitem [{\citenamefont {Wang}(2011)}]{Wang:2011xf}%
  \BibitemOpen
  \bibfield  {author} {\bibinfo {author} {\bibfnamefont {T.}~\bibnamefont
  {Wang}},\ }\href {\doibase 10.1103/PhysRevD.84.024042} {\bibfield  {journal}
  {\bibinfo  {journal} {Phys. Rev. D}\ }\textbf {\bibinfo {volume} {84}},\
  \bibinfo {pages} {024042} (\bibinfo {year} {2011})},\ \Eprint
  {http://arxiv.org/abs/1102.4410} {arXiv:1102.4410 [gr-qc]} \BibitemShut
  {NoStop}%
\bibitem [{\citenamefont {Atazadeh}\ and\ \citenamefont
  {Mousavi}(2013)}]{Atazadeh:2012am}%
  \BibitemOpen
  \bibfield  {author} {\bibinfo {author} {\bibfnamefont {K.}~\bibnamefont
  {Atazadeh}}\ and\ \bibinfo {author} {\bibfnamefont {M.}~\bibnamefont
  {Mousavi}},\ }\href {\doibase 10.1140/epjc/s10052-012-2272-y} {\bibfield
  {journal} {\bibinfo  {journal} {Eur. Phys. J. C}\ }\textbf {\bibinfo {volume}
  {73}},\ \bibinfo {pages} {2272} (\bibinfo {year} {2013})},\ \Eprint
  {http://arxiv.org/abs/1212.3764} {arXiv:1212.3764 [gr-qc]} \BibitemShut
  {NoStop}%
\bibitem [{\citenamefont {Rodrigues}\ \emph {et~al.}(2013)\citenamefont
  {Rodrigues}, \citenamefont {Houndjo}, \citenamefont {Tossa}, \citenamefont
  {Momeni},\ and\ \citenamefont {Myrzakulov}}]{Rodrigues:2013ifa}%
  \BibitemOpen
  \bibfield  {author} {\bibinfo {author} {\bibfnamefont {M.~E.}\ \bibnamefont
  {Rodrigues}}, \bibinfo {author} {\bibfnamefont {M.~J.~S.}\ \bibnamefont
  {Houndjo}}, \bibinfo {author} {\bibfnamefont {J.}~\bibnamefont {Tossa}},
  \bibinfo {author} {\bibfnamefont {D.}~\bibnamefont {Momeni}}, \ and\ \bibinfo
  {author} {\bibfnamefont {R.}~\bibnamefont {Myrzakulov}},\ }\href {\doibase
  10.1088/1475-7516/2013/11/024} {\bibfield  {journal} {\bibinfo  {journal}
  {JCAP}\ }\textbf {\bibinfo {volume} {11}},\ \bibinfo {pages} {024} (\bibinfo
  {year} {2013})},\ \Eprint {http://arxiv.org/abs/1306.2280} {arXiv:1306.2280
  [gr-qc]} \BibitemShut {NoStop}%
\bibitem [{\citenamefont {Nashed}(2013)}]{Nashed:2013bfa}%
  \BibitemOpen
  \bibfield  {author} {\bibinfo {author} {\bibfnamefont {G.~G.~L.}\
  \bibnamefont {Nashed}},\ }\href {\doibase 10.1103/PhysRevD.88.104034}
  {\bibfield  {journal} {\bibinfo  {journal} {Phys. Rev.}\ }\textbf {\bibinfo
  {volume} {D88}},\ \bibinfo {pages} {104034} (\bibinfo {year} {2013})},\
  \Eprint {http://arxiv.org/abs/1311.3131} {arXiv:1311.3131 [gr-qc]}
  \BibitemShut {NoStop}%
\bibitem [{\citenamefont {Nashed}(2014)}]{Nashed:2014iua}%
  \BibitemOpen
  \bibfield  {author} {\bibinfo {author} {\bibfnamefont {G.~G.~L.}\
  \bibnamefont {Nashed}},\ }\href {\doibase 10.1155/2014/830109} {\bibfield
  {journal} {\bibinfo  {journal} {Adv. High Energy Phys.}\ }\textbf {\bibinfo
  {volume} {2014}},\ \bibinfo {pages} {830109} (\bibinfo {year}
  {2014})}\BibitemShut {NoStop}%
\bibitem [{\citenamefont {Junior}\ \emph {et~al.}(2015)\citenamefont {Junior},
  \citenamefont {Rodrigues},\ and\ \citenamefont {Houndjo}}]{Junior:2015fya}%
  \BibitemOpen
  \bibfield  {author} {\bibinfo {author} {\bibfnamefont {E.~L.~B.}\
  \bibnamefont {Junior}}, \bibinfo {author} {\bibfnamefont {M.~E.}\
  \bibnamefont {Rodrigues}}, \ and\ \bibinfo {author} {\bibfnamefont
  {M.~J.~S.}\ \bibnamefont {Houndjo}},\ }\href {\doibase
  10.1088/1475-7516/2015/10/060} {\bibfield  {journal} {\bibinfo  {journal}
  {JCAP}\ }\textbf {\bibinfo {volume} {1510}},\ \bibinfo {pages} {060}
  (\bibinfo {year} {2015})},\ \Eprint {http://arxiv.org/abs/1503.07857}
  {arXiv:1503.07857 [gr-qc]} \BibitemShut {NoStop}%
\bibitem [{\citenamefont {Shirafuji}\ and\ \citenamefont
  {Nashed}(1997)}]{Shirafuji19971355}%
  \BibitemOpen
  \bibfield  {author} {\bibinfo {author} {\bibfnamefont {T.}~\bibnamefont
  {Shirafuji}}\ and\ \bibinfo {author} {\bibfnamefont {G.}~\bibnamefont
  {Nashed}},\ }\href {\doibase 10.1143/PTP.98.1355} {\bibfield  {journal}
  {\bibinfo  {journal} {Progress of Theoretical Physics}\ }\textbf {\bibinfo
  {volume} {98}},\ \bibinfo {pages} {1355} (\bibinfo {year} {1997})},\ \bibinfo
  {note} {cited By 25}\BibitemShut {NoStop}%
\bibitem [{\citenamefont {Kofinas}\ \emph {et~al.}(2015)\citenamefont
  {Kofinas}, \citenamefont {Papantonopoulos},\ and\ \citenamefont
  {Saridakis}}]{Kofinas:2015hla}%
  \BibitemOpen
  \bibfield  {author} {\bibinfo {author} {\bibfnamefont {G.}~\bibnamefont
  {Kofinas}}, \bibinfo {author} {\bibfnamefont {E.}~\bibnamefont
  {Papantonopoulos}}, \ and\ \bibinfo {author} {\bibfnamefont {E.~N.}\
  \bibnamefont {Saridakis}},\ }\href {\doibase 10.1103/PhysRevD.91.104034}
  {\bibfield  {journal} {\bibinfo  {journal} {Phys. Rev.}\ }\textbf {\bibinfo
  {volume} {D91}},\ \bibinfo {pages} {104034} (\bibinfo {year} {2015})},\
  \Eprint {http://arxiv.org/abs/1501.00365} {arXiv:1501.00365 [gr-qc]}
  \BibitemShut {NoStop}%
  \bibitem [{\citenamefont {Shirafuji}\ and\ \citenamefont
  {Nashed}(1997)}]{Shirafuji:1997wy}%
  \BibitemOpen
  \bibfield  {author} {\bibinfo {author} {\bibfnamefont {T.}~\bibnamefont
  {Shirafuji}}\ and\ \bibinfo {author} {\bibfnamefont {G.~G.~L.}\ \bibnamefont
  {Nashed}},\ }\href {\doibase 10.1143/PTP.98.1355} {\bibfield  {journal}
  {\bibinfo  {journal} {Prog. Theor. Phys.}\ }\textbf {\bibinfo {volume}
  {98}},\ \bibinfo {pages} {1355} (\bibinfo {year} {1997})},\ \Eprint
  {http://arxiv.org/abs/gr-qc/9711010} {arXiv:gr-qc/9711010} \BibitemShut
  {NoStop}%
\bibitem [{\citenamefont {Das}\ \emph {et~al.}(2015)\citenamefont {Das},
  \citenamefont {Rahaman}, \citenamefont {Guha},\ and\ \citenamefont
  {Ray}}]{Das:2015gwa}%
  \BibitemOpen
  \bibfield  {author} {\bibinfo {author} {\bibfnamefont {A.}~\bibnamefont
  {Das}}, \bibinfo {author} {\bibfnamefont {F.}~\bibnamefont {Rahaman}},
  \bibinfo {author} {\bibfnamefont {B.~K.}\ \bibnamefont {Guha}}, \ and\
  \bibinfo {author} {\bibfnamefont {S.}~\bibnamefont {Ray}},\ }\href {\doibase
  10.1007/s10509-015-2441-1} {\bibfield  {journal} {\bibinfo  {journal}
  {Astrophys. Space Sci.}\ }\textbf {\bibinfo {volume} {358}},\ \bibinfo
  {pages} {36} (\bibinfo {year} {2015})},\ \Eprint
  {http://arxiv.org/abs/1507.04959} {arXiv:1507.04959 [gr-qc]} \BibitemShut
  {NoStop}%
\bibitem [{\citenamefont {Nashed}(2002)}]{Nashed2002521}%
  \BibitemOpen
  \bibfield  {author} {\bibinfo {author} {\bibfnamefont {G.}~\bibnamefont
  {Nashed}},\ }\href
  {https://www.scopus.com/inward/record.uri?eid=2-s2.0-0141726732&partnerID=40&md5=020641db2d2584aa35ed4a6c59bef75e}
  {\bibfield  {journal} {\bibinfo  {journal} {Nuovo Cimento della Societa
  Italiana di Fisica B}\ }\textbf {\bibinfo {volume} {117}},\ \bibinfo {pages}
  {521} (\bibinfo {year} {2002})},\ \bibinfo {note} {cited By 25}\BibitemShut
  {NoStop}%
\bibitem [{\citenamefont {Rani}\ \emph {et~al.}(2016)\citenamefont {Rani},
  \citenamefont {Jawad},\ and\ \citenamefont {Amin}}]{Rani:2016gnl}%
  \BibitemOpen
  \bibfield  {author} {\bibinfo {author} {\bibfnamefont {S.}~\bibnamefont
  {Rani}}, \bibinfo {author} {\bibfnamefont {A.}~\bibnamefont {Jawad}}, \ and\
  \bibinfo {author} {\bibfnamefont {M.~B.}\ \bibnamefont {Amin}},\ }\href
  {\doibase 10.1088/0253-6102/66/4/411} {\bibfield  {journal} {\bibinfo
  {journal} {Commun. Theor. Phys.}\ }\textbf {\bibinfo {volume} {66}},\
  \bibinfo {pages} {411} (\bibinfo {year} {2016})}\BibitemShut {NoStop}%
\bibitem [{\citenamefont {Rodrigues}\ and\ \citenamefont
  {Junior}(2018)}]{Rodrigues:2016uor}%
  \BibitemOpen
  \bibfield  {author} {\bibinfo {author} {\bibfnamefont {M.~E.}\ \bibnamefont
  {Rodrigues}}\ and\ \bibinfo {author} {\bibfnamefont {E.~L.~B.}\ \bibnamefont
  {Junior}},\ }\href {\doibase 10.1007/s10509-018-3262-9} {\bibfield  {journal}
  {\bibinfo  {journal} {Astrophys. Space Sci.}\ }\textbf {\bibinfo {volume}
  {363}},\ \bibinfo {pages} {43} (\bibinfo {year} {2018})},\ \Eprint
  {http://arxiv.org/abs/1606.04918} {arXiv:1606.04918 [gr-qc]} \BibitemShut
  {NoStop}%
\bibitem [{\citenamefont {Mai}\ and\ \citenamefont {Lu}(2017)}]{Mai:2017riq}%
  \BibitemOpen
  \bibfield  {author} {\bibinfo {author} {\bibfnamefont {Z.-F.}\ \bibnamefont
  {Mai}}\ and\ \bibinfo {author} {\bibfnamefont {H.}~\bibnamefont {Lu}},\
  }\href {\doibase 10.1103/PhysRevD.95.124024} {\bibfield  {journal} {\bibinfo
  {journal} {Phys. Rev.}\ }\textbf {\bibinfo {volume} {D95}},\ \bibinfo {pages}
  {124024} (\bibinfo {year} {2017})},\ \Eprint
  {http://arxiv.org/abs/1704.05919} {arXiv:1704.05919 [hep-th]} \BibitemShut
  {NoStop}%
\bibitem [{\citenamefont {Newton~Singh}\ \emph {et~al.}(2019)\citenamefont
  {Newton~Singh}, \citenamefont {Rahaman},\ and\ \citenamefont
  {Banerjee}}]{Singh:2019ykp}%
  \BibitemOpen
  \bibfield  {author} {\bibinfo {author} {\bibfnamefont {K.}~\bibnamefont
  {Newton~Singh}}, \bibinfo {author} {\bibfnamefont {F.}~\bibnamefont
  {Rahaman}}, \ and\ \bibinfo {author} {\bibfnamefont {A.}~\bibnamefont
  {Banerjee}},\ }\href {\doibase 10.1103/PhysRevD.100.084023} {\bibfield
  {journal} {\bibinfo  {journal} {Phys. Rev. D}\ }\textbf {\bibinfo {volume}
  {100}},\ \bibinfo {pages} {084023} (\bibinfo {year} {2019})},\ \Eprint
  {http://arxiv.org/abs/1909.10882} {arXiv:1909.10882 [gr-qc]} \BibitemShut
  {NoStop}%
\bibitem [{\citenamefont {Nashed}\ and\ \citenamefont
  {Capozziello}(2020)}]{Nashed:2020kjh}%
  \BibitemOpen
  \bibfield  {author} {\bibinfo {author} {\bibfnamefont {G.~G.~L.}\
  \bibnamefont {Nashed}}\ and\ \bibinfo {author} {\bibfnamefont
  {S.}~\bibnamefont {Capozziello}},\ }\href {\doibase
  10.1140/epjc/s10052-020-08551-1} {\bibfield  {journal} {\bibinfo  {journal}
  {Eur. Phys. J. C}\ }\textbf {\bibinfo {volume} {80}},\ \bibinfo {pages} {969}
  (\bibinfo {year} {2020})},\ \Eprint {http://arxiv.org/abs/2010.06355}
  {arXiv:2010.06355 [gr-qc]} \BibitemShut {NoStop}%
\bibitem [{\citenamefont {Bhatti}(2018)}]{Bhatti:2018fsc}%
  \BibitemOpen
  \bibfield  {author} {\bibinfo {author} {\bibfnamefont {M.~Z.}\ \bibnamefont
  {Bhatti}},\ }\href {\doibase 10.1140/epjp/i2018-12214-8} {\bibfield
  {journal} {\bibinfo  {journal} {Eur. Phys. J. Plus}\ }\textbf {\bibinfo
  {volume} {133}},\ \bibinfo {pages} {431} (\bibinfo {year}
  {2018})}\BibitemShut {NoStop}%
\bibitem [{\citenamefont {Ashraf}\ \emph {et~al.}(2020)\citenamefont {Ashraf},
  \citenamefont {Zhang}, \citenamefont {Ditta},\ and\ \citenamefont
  {Mustafa}}]{Ashraf:2020yyo}%
  \BibitemOpen
  \bibfield  {author} {\bibinfo {author} {\bibfnamefont {A.}~\bibnamefont
  {Ashraf}}, \bibinfo {author} {\bibfnamefont {Z.}~\bibnamefont {Zhang}},
  \bibinfo {author} {\bibfnamefont {A.}~\bibnamefont {Ditta}}, \ and\ \bibinfo
  {author} {\bibfnamefont {G.}~\bibnamefont {Mustafa}},\ }\href {\doibase
  10.1016/j.aop.2020.168322} {\bibfield  {journal} {\bibinfo  {journal} {Annals
  Phys.}\ }\textbf {\bibinfo {volume} {422}},\ \bibinfo {pages} {168322}
  (\bibinfo {year} {2020})}\BibitemShut {NoStop}%
  \bibitem[{\citenamefont{El~Hanafy and Nashed}(2016)}]{ElHanafy:2015egm}
\bibinfo{author}{\bibfnamefont{W.}~\bibnamefont{El~Hanafy}} \bibnamefont{and}
  \bibinfo{author}{\bibfnamefont{G.~G.~L.} \bibnamefont{Nashed}},
  \bibinfo{journal}{Astrophys. Space Sci.} \textbf{\bibinfo{volume}{361}},
  \bibinfo{pages}{68} (\bibinfo{year}{2016}), \eprint{1507.07377}.
\bibitem [{\citenamefont {Ditta}\ \emph {et~al.}(2021)\citenamefont {Ditta},
  \citenamefont {Ahmad}, \citenamefont {Hussain},\ and\ \citenamefont
  {Mustafa}}]{Ditta:2021wfl}%
  \BibitemOpen
  \bibfield  {author} {\bibinfo {author} {\bibfnamefont {A.}~\bibnamefont
  {Ditta}}, \bibinfo {author} {\bibfnamefont {M.}~\bibnamefont {Ahmad}},
  \bibinfo {author} {\bibfnamefont {I.}~\bibnamefont {Hussain}}, \ and\
  \bibinfo {author} {\bibfnamefont {G.}~\bibnamefont {Mustafa}},\ }\href
  {\doibase 10.1088/1674-1137/abdfbd} {\bibfield  {journal} {\bibinfo
  {journal} {Chin. Phys. C}\ }\textbf {\bibinfo {volume} {45}},\ \bibinfo
  {pages} {045102} (\bibinfo {year} {2021})}\BibitemShut {NoStop}%
\bibitem [{\citenamefont {Hayashi}\ and\ \citenamefont
  {Nakano}(1967{\natexlab{b}})}]{Hayashi:1967se}%
  \BibitemOpen
  \bibfield  {author} {\bibinfo {author} {\bibfnamefont {K.}~\bibnamefont
  {Hayashi}}\ and\ \bibinfo {author} {\bibfnamefont {T.}~\bibnamefont
  {Nakano}},\ }\href {\doibase 10.1143/PTP.38.491} {\bibfield  {journal}
  {\bibinfo  {journal} {Prog. Theor. Phys.}\ }\textbf {\bibinfo {volume}
  {38}},\ \bibinfo {pages} {491} (\bibinfo {year}
  {1967}{\natexlab{b}})}\BibitemShut {NoStop}%
\bibitem [{\citenamefont {{Saez}}\ and\ \citenamefont {{de
  Juan}}(1984)}]{1984GReGr..16..501S}%
  \BibitemOpen
  \bibfield  {author} {\bibinfo {author} {\bibfnamefont {D.}~\bibnamefont
  {{Saez}}}\ and\ \bibinfo {author} {\bibfnamefont {T.}~\bibnamefont {{de
  Juan}}},\ }\href {\doibase 10.1007/BF00762343} {\bibfield  {journal}
  {\bibinfo  {journal} {General Relativity and Gravitation}\ }\textbf {\bibinfo
  {volume} {16}},\ \bibinfo {pages} {501} (\bibinfo {year} {1984})}\BibitemShut
  {NoStop}%
\bibitem [{\citenamefont {Robertson}(1932)}]{robertson1932groups}%
  \BibitemOpen
  \bibfield  {author} {\bibinfo {author} {\bibfnamefont {H.}~\bibnamefont
  {Robertson}},\ }\href@noop {} {\bibfield  {journal} {\bibinfo  {journal}
  {Annals of Mathematics}\ ,\ \bibinfo {pages} {496}} (\bibinfo {year}
  {1932})}\BibitemShut {NoStop}%
\bibitem [{\citenamefont {Weinberg}(1972)}]{Weinberg:1972kfs}%
  \BibitemOpen
  \bibfield  {author} {\bibinfo {author} {\bibfnamefont {S.}~\bibnamefont
  {Weinberg}},\ }\href@noop {} {\emph {\bibinfo {title} {{Gravitation and
  Cosmology}: {Principles and Applications of the General Theory of
  Relativity}}}}\ (\bibinfo  {publisher} {John Wiley and Sons},\ \bibinfo
  {address} {New York},\ \bibinfo {year} {1972})\BibitemShut {NoStop}%
\bibitem [{\citenamefont {Capozziello}\ \emph {et~al.}(2014)\citenamefont
  {Capozziello}, \citenamefont {Farooq}, \citenamefont {Luongo},\ and\
  \citenamefont {Ratra}}]{Capozziello:2014zda}%
  \BibitemOpen
  \bibfield  {author} {\bibinfo {author} {\bibfnamefont {S.}~\bibnamefont
  {Capozziello}}, \bibinfo {author} {\bibfnamefont {O.}~\bibnamefont {Farooq}},
  \bibinfo {author} {\bibfnamefont {O.}~\bibnamefont {Luongo}}, \ and\ \bibinfo
  {author} {\bibfnamefont {B.}~\bibnamefont {Ratra}},\ }\href {\doibase
  10.1103/PhysRevD.90.044016} {\bibfield  {journal} {\bibinfo  {journal} {Phys.
  Rev. D}\ }\textbf {\bibinfo {volume} {90}},\ \bibinfo {pages} {044016}
  (\bibinfo {year} {2014})},\ \Eprint {http://arxiv.org/abs/1403.1421}
  {arXiv:1403.1421 [gr-qc]} \BibitemShut {NoStop}%
\bibitem [{\citenamefont {Awad}\ \emph {et~al.}(2018)\citenamefont {Awad},
  \citenamefont {El~Hanafy}, \citenamefont {Nashed},\ and\ \citenamefont
  {Saridakis}}]{Awad:2017yod}%
  \BibitemOpen
  \bibfield  {author} {\bibinfo {author} {\bibfnamefont {A.}~\bibnamefont
  {Awad}}, \bibinfo {author} {\bibfnamefont {W.}~\bibnamefont {El~Hanafy}},
  \bibinfo {author} {\bibfnamefont {G.~G.~L.}\ \bibnamefont {Nashed}}, \ and\
  \bibinfo {author} {\bibfnamefont {E.~N.}\ \bibnamefont {Saridakis}},\ }\href
  {\doibase 10.1088/1475-7516/2018/02/052} {\bibfield  {journal} {\bibinfo
  {journal} {JCAP}\ }\textbf {\bibinfo {volume} {1802}},\ \bibinfo {pages}
  {052} (\bibinfo {year} {2018})},\ \Eprint {http://arxiv.org/abs/1710.10194}
  {arXiv:1710.10194 [gr-qc]} \BibitemShut {NoStop}%
\bibitem [{\citenamefont {Akarsu}\ and\ \citenamefont
  {Dereli}(2012)}]{Akarsu:2011zd}%
  \BibitemOpen
  \bibfield  {author} {\bibinfo {author} {\bibfnamefont {O.}~\bibnamefont
  {Akarsu}}\ and\ \bibinfo {author} {\bibfnamefont {T.}~\bibnamefont
  {Dereli}},\ }\href {\doibase 10.1007/s10773-011-0941-5} {\bibfield  {journal}
  {\bibinfo  {journal} {Int. J. Theor. Phys.}\ }\textbf {\bibinfo {volume}
  {51}},\ \bibinfo {pages} {612} (\bibinfo {year} {2012})},\ \Eprint
  {http://arxiv.org/abs/1102.0915} {arXiv:1102.0915 [gr-qc]} \BibitemShut
  {NoStop}%
\bibitem [{\citenamefont {Wetterich}(2013)}]{Wetterich:2013aca}%
  \BibitemOpen
  \bibfield  {author} {\bibinfo {author} {\bibfnamefont {C.}~\bibnamefont
  {Wetterich}},\ }\href {\doibase 10.1016/j.dark.2013.10.002} {\bibfield
  {journal} {\bibinfo  {journal} {Phys. Dark Univ.}\ }\textbf {\bibinfo
  {volume} {2}},\ \bibinfo {pages} {184} (\bibinfo {year} {2013})},\ \Eprint
  {http://arxiv.org/abs/1303.6878} {arXiv:1303.6878 [astro-ph.CO]} \BibitemShut
  {NoStop}%
\bibitem [{\citenamefont {Mamon}\ and\ \citenamefont
  {Das}(2017)}]{Mamon:2016dlv}%
  \BibitemOpen
  \bibfield  {author} {\bibinfo {author} {\bibfnamefont {A.~A.}\ \bibnamefont
  {Mamon}}\ and\ \bibinfo {author} {\bibfnamefont {S.}~\bibnamefont {Das}},\
  }\href {\doibase 10.1140/epjc/s10052-017-5066-4} {\bibfield  {journal}
  {\bibinfo  {journal} {Eur. Phys. J. C}\ }\textbf {\bibinfo {volume} {77}},\
  \bibinfo {pages} {495} (\bibinfo {year} {2017})},\ \Eprint
  {http://arxiv.org/abs/1610.07337} {arXiv:1610.07337 [gr-qc]} \BibitemShut
  {NoStop}%
\bibitem [{\citenamefont {Akbar}\ and\ \citenamefont
  {Cai}(2007{\natexlab{a}})}]{Akbar:2006kj}%
  \BibitemOpen
  \bibfield  {author} {\bibinfo {author} {\bibfnamefont {M.}~\bibnamefont
  {Akbar}}\ and\ \bibinfo {author} {\bibfnamefont {R.-G.}\ \bibnamefont
  {Cai}},\ }\href {\doibase 10.1103/PhysRevD.75.084003} {\bibfield  {journal}
  {\bibinfo  {journal} {Phys. Rev. D}\ }\textbf {\bibinfo {volume} {75}},\
  \bibinfo {pages} {084003} (\bibinfo {year} {2007}{\natexlab{a}})},\ \Eprint
  {http://arxiv.org/abs/hep-th/0609128} {arXiv:hep-th/0609128} \BibitemShut
  {NoStop}%
\bibitem [{\citenamefont {Cai}\ \emph {et~al.}(2009)\citenamefont {Cai},
  \citenamefont {Cao},\ and\ \citenamefont {Hu}}]{Cai:2008gw}%
  \BibitemOpen
  \bibfield  {author} {\bibinfo {author} {\bibfnamefont {R.-G.}\ \bibnamefont
  {Cai}}, \bibinfo {author} {\bibfnamefont {L.-M.}\ \bibnamefont {Cao}}, \ and\
  \bibinfo {author} {\bibfnamefont {Y.-P.}\ \bibnamefont {Hu}},\ }\href
  {\doibase 10.1088/0264-9381/26/15/155018} {\bibfield  {journal} {\bibinfo
  {journal} {Class. Quant. Grav.}\ }\textbf {\bibinfo {volume} {26}},\ \bibinfo
  {pages} {155018} (\bibinfo {year} {2009})},\ \Eprint
  {http://arxiv.org/abs/0809.1554} {arXiv:0809.1554 [hep-th]} \BibitemShut
  {NoStop}%
\bibitem [{\citenamefont {Akbar}\ and\ \citenamefont
  {Cai}(2007{\natexlab{b}})}]{Akbar:2006mq}%
  \BibitemOpen
  \bibfield  {author} {\bibinfo {author} {\bibfnamefont {M.}~\bibnamefont
  {Akbar}}\ and\ \bibinfo {author} {\bibfnamefont {R.-G.}\ \bibnamefont
  {Cai}},\ }\href {\doibase 10.1016/j.physletb.2007.03.005} {\bibfield
  {journal} {\bibinfo  {journal} {Phys. Lett. B}\ }\textbf {\bibinfo {volume}
  {648}},\ \bibinfo {pages} {243} (\bibinfo {year} {2007}{\natexlab{b}})},\
  \Eprint {http://arxiv.org/abs/gr-qc/0612089} {arXiv:gr-qc/0612089}
  \BibitemShut {NoStop}%
\bibitem [{\citenamefont {Tsallis}\ and\ \citenamefont
  {Cirto}(2013)}]{Tsallis:2012js}%
  \BibitemOpen
  \bibfield  {author} {\bibinfo {author} {\bibfnamefont {C.}~\bibnamefont
  {Tsallis}}\ and\ \bibinfo {author} {\bibfnamefont {L.~J.~L.}\ \bibnamefont
  {Cirto}},\ }\href {\doibase 10.1140/epjc/s10052-013-2487-6} {\bibfield
  {journal} {\bibinfo  {journal} {Eur. Phys. J. C}\ }\textbf {\bibinfo {volume}
  {73}},\ \bibinfo {pages} {2487} (\bibinfo {year} {2013})},\ \Eprint
  {http://arxiv.org/abs/1202.2154} {arXiv:1202.2154 [cond-mat.stat-mech]}
  \BibitemShut {NoStop}%
\bibitem [{\citenamefont {Cai}\ and\ \citenamefont {Kim}(2005)}]{Cai:2005ra}%
  \BibitemOpen
  \bibfield  {author} {\bibinfo {author} {\bibfnamefont {R.-G.}\ \bibnamefont
  {Cai}}\ and\ \bibinfo {author} {\bibfnamefont {S.~P.}\ \bibnamefont {Kim}},\
  }\href {\doibase 10.1088/1126-6708/2005/02/050} {\bibfield  {journal}
  {\bibinfo  {journal} {JHEP}\ }\textbf {\bibinfo {volume} {02}},\ \bibinfo
  {pages} {050} (\bibinfo {year} {2005})},\ \Eprint
  {http://arxiv.org/abs/hep-th/0501055} {arXiv:hep-th/0501055} \BibitemShut
  {NoStop}%
\bibitem [{\citenamefont {Lymperis}\ and\ \citenamefont
  {Saridakis}(2018)}]{Lymperis:2018iuz}%
  \BibitemOpen
  \bibfield  {author} {\bibinfo {author} {\bibfnamefont {A.}~\bibnamefont
  {Lymperis}}\ and\ \bibinfo {author} {\bibfnamefont {E.~N.}\ \bibnamefont
  {Saridakis}},\ }\href {\doibase 10.1140/epjc/s10052-018-6480-y} {\bibfield
  {journal} {\bibinfo  {journal} {Eur. Phys. J. C}\ }\textbf {\bibinfo {volume}
  {78}},\ \bibinfo {pages} {993} (\bibinfo {year} {2018})},\ \Eprint
  {http://arxiv.org/abs/1806.04614} {arXiv:1806.04614 [gr-qc]} \BibitemShut
  {NoStop}%
\bibitem [{\citenamefont {Azmi}\ and\ \citenamefont
  {Cleymans}(2015)}]{Azmi:2015xqa}%
  \BibitemOpen
  \bibfield  {author} {\bibinfo {author} {\bibfnamefont {M.~D.}\ \bibnamefont
  {Azmi}}\ and\ \bibinfo {author} {\bibfnamefont {J.}~\bibnamefont
  {Cleymans}},\ }\href {\doibase 10.1140/epjc/s10052-015-3629-9} {\bibfield
  {journal} {\bibinfo  {journal} {Eur. Phys. J. C}\ }\textbf {\bibinfo {volume}
  {75}},\ \bibinfo {pages} {430} (\bibinfo {year} {2015})},\ \Eprint
  {http://arxiv.org/abs/1501.07127} {arXiv:1501.07127 [hep-ph]} \BibitemShut
  {NoStop}%
\bibitem [{\citenamefont {Cleymans}\ \emph {et~al.}(2013)\citenamefont
  {Cleymans}, \citenamefont {Lykasov}, \citenamefont {Parvan}, \citenamefont
  {Sorin}, \citenamefont {Teryaev},\ and\ \citenamefont
  {Worku}}]{Cleymans:2013rfq}%
  \BibitemOpen
  \bibfield  {author} {\bibinfo {author} {\bibfnamefont {J.}~\bibnamefont
  {Cleymans}}, \bibinfo {author} {\bibfnamefont {G.~I.}\ \bibnamefont
  {Lykasov}}, \bibinfo {author} {\bibfnamefont {A.~S.}\ \bibnamefont {Parvan}},
  \bibinfo {author} {\bibfnamefont {A.~S.}\ \bibnamefont {Sorin}}, \bibinfo
  {author} {\bibfnamefont {O.~V.}\ \bibnamefont {Teryaev}}, \ and\ \bibinfo
  {author} {\bibfnamefont {D.}~\bibnamefont {Worku}},\ }\href {\doibase
  10.1016/j.physletb.2013.05.029} {\bibfield  {journal} {\bibinfo  {journal}
  {Phys. Lett. B}\ }\textbf {\bibinfo {volume} {723}},\ \bibinfo {pages} {351}
  (\bibinfo {year} {2013})},\ \Eprint {http://arxiv.org/abs/1302.1970}
  {arXiv:1302.1970 [hep-ph]} \BibitemShut {NoStop}%
\bibitem [{\citenamefont {{Berman}}(1983)}]{1983NCimB..74..182B}%
  \BibitemOpen
  \bibfield  {author} {\bibinfo {author} {\bibfnamefont {M.~S.}\ \bibnamefont
  {{Berman}}},\ }\href {\doibase 10.1007/BF02721676} {\bibfield  {journal}
  {\bibinfo  {journal} {Nuovo Cimento B Serie}\ }\textbf {\bibinfo {volume}
  {74}},\ \bibinfo {pages} {182} (\bibinfo {year} {1983})}\BibitemShut
  {NoStop}%
\bibitem [{\citenamefont {Hobson}\ \emph {et~al.}(2006)\citenamefont {Hobson},
  \citenamefont {Efstathiou},\ and\ \citenamefont {Lasenby}}]{Hobson:2006se}%
  \BibitemOpen
  \bibfield  {author} {\bibinfo {author} {\bibfnamefont {M.~P.}\ \bibnamefont
  {Hobson}}, \bibinfo {author} {\bibfnamefont {G.~P.}\ \bibnamefont
  {Efstathiou}}, \ and\ \bibinfo {author} {\bibfnamefont {A.~N.}\ \bibnamefont
  {Lasenby}},\ }\href@noop {} {\emph {\bibinfo {title} {{General relativity: An
  introduction for physicists}}}}\ (\bibinfo {year} {2006})\BibitemShut
  {NoStop}%
\bibitem [{\citenamefont {Garriga}\ \emph {et~al.}(1999)\citenamefont
  {Garriga}, \citenamefont {Tanaka},\ and\ \citenamefont
  {Vilenkin}}]{Garriga:1998px}%
  \BibitemOpen
  \bibfield  {author} {\bibinfo {author} {\bibfnamefont {J.}~\bibnamefont
  {Garriga}}, \bibinfo {author} {\bibfnamefont {T.}~\bibnamefont {Tanaka}}, \
  and\ \bibinfo {author} {\bibfnamefont {A.}~\bibnamefont {Vilenkin}},\ }\href
  {\doibase 10.1103/PhysRevD.60.023501} {\bibfield  {journal} {\bibinfo
  {journal} {Phys. Rev. D}\ }\textbf {\bibinfo {volume} {60}},\ \bibinfo
  {pages} {023501} (\bibinfo {year} {1999})},\ \Eprint
  {http://arxiv.org/abs/astro-ph/9803268} {arXiv:astro-ph/9803268} \BibitemShut
  {NoStop}%
\bibitem [{\citenamefont {Chiba}\ and\ \citenamefont
  {Nakamura}(1998)}]{Chiba:1998tc}%
  \BibitemOpen
  \bibfield  {author} {\bibinfo {author} {\bibfnamefont {T.}~\bibnamefont
  {Chiba}}\ and\ \bibinfo {author} {\bibfnamefont {T.}~\bibnamefont
  {Nakamura}},\ }\href {\doibase 10.1143/PTP.100.1077} {\bibfield  {journal}
  {\bibinfo  {journal} {Prog. Theor. Phys.}\ }\textbf {\bibinfo {volume}
  {100}},\ \bibinfo {pages} {1077} (\bibinfo {year} {1998})},\ \Eprint
  {http://arxiv.org/abs/astro-ph/9808022} {arXiv:astro-ph/9808022} \BibitemShut
  {NoStop}%
\bibitem [{\citenamefont {{Suzuki}}\ \emph {et~al.}(2012)\citenamefont
  {{Suzuki}}, \citenamefont {{Rubin}}, \citenamefont {{Lidman}}, \citenamefont
  {{Aldering}}, \citenamefont {{Amanullah}}, \citenamefont {{Barbary}},
  \citenamefont {{Barrientos}}, \citenamefont {{Botyanszki}}, \citenamefont
  {{Brodwin}}, \citenamefont {{Connolly}}, \citenamefont {{Dawson}},
  \citenamefont {{Dey}}, \citenamefont {{Doi}}, \citenamefont {{Donahue}},
  \citenamefont {{Deustua}}, \citenamefont {{Eisenhardt}}, \citenamefont
  {{Ellingson}}, \citenamefont {{Faccioli}}, \citenamefont {{Fadeyev}},
  \citenamefont {{Fakhouri}}, \citenamefont {{Fruchter}}, \citenamefont
  {{Gilbank}}, \citenamefont {{Gladders}}, \citenamefont {{Goldhaber}},
  \citenamefont {{Gonzalez}}, \citenamefont {{Goobar}}, \citenamefont {{Gude}},
  \citenamefont {{Hattori}}, \citenamefont {{Hoekstra}}, \citenamefont
  {{Hsiao}}, \citenamefont {{Huang}}, \citenamefont {{Ihara}}, \citenamefont
  {{Jee}}, \citenamefont {{Johnston}}, \citenamefont {{Kashikawa}},
  \citenamefont {{Koester}}, \citenamefont {{Konishi}}, \citenamefont
  {{Kowalski}}, \citenamefont {{Linder}}, \citenamefont {{Lubin}},
  \citenamefont {{Melbourne}}, \citenamefont {{Meyers}}, \citenamefont
  {{Morokuma}}, \citenamefont {{Munshi}}, \citenamefont {{Mullis}},
  \citenamefont {{Oda}}, \citenamefont {{Panagia}}, \citenamefont
  {{Perlmutter}}, \citenamefont {{Postman}}, \citenamefont {{Pritchard}},
  \citenamefont {{Rhodes}}, \citenamefont {{Ripoche}}, \citenamefont
  {{Rosati}}, \citenamefont {{Schlegel}}, \citenamefont {{Spadafora}},
  \citenamefont {{Stanford}}, \citenamefont {{Stanishev}}, \citenamefont
  {{Stern}}, \citenamefont {{Strovink}}, \citenamefont {{Takanashi}},
  \citenamefont {{Tokita}}, \citenamefont {{Wagner}}, \citenamefont {{Wang}},
  \citenamefont {{Yasuda}}, \citenamefont {{Yee}},\ and\ \citenamefont
  {{Supernova Cosmology Project}}}]{2012ApJ...746...85S}%
  \BibitemOpen
  \bibfield  {author} {\bibinfo {author} {\bibfnamefont {N.}~\bibnamefont
  {{Suzuki}}}, \bibinfo {author} {\bibfnamefont {D.}~\bibnamefont {{Rubin}}},
  \bibinfo {author} {\bibfnamefont {C.}~\bibnamefont {{Lidman}}}, \bibinfo
  {author} {\bibfnamefont {G.}~\bibnamefont {{Aldering}}}, \bibinfo {author}
  {\bibfnamefont {R.}~\bibnamefont {{Amanullah}}}, \bibinfo {author}
  {\bibfnamefont {K.}~\bibnamefont {{Barbary}}}, \bibinfo {author}
  {\bibfnamefont {L.~F.}\ \bibnamefont {{Barrientos}}}, \bibinfo {author}
  {\bibfnamefont {J.}~\bibnamefont {{Botyanszki}}}, \bibinfo {author}
  {\bibfnamefont {M.}~\bibnamefont {{Brodwin}}}, \bibinfo {author}
  {\bibfnamefont {N.}~\bibnamefont {{Connolly}}}, \bibinfo {author}
  {\bibfnamefont {K.~S.}\ \bibnamefont {{Dawson}}}, \bibinfo {author}
  {\bibfnamefont {A.}~\bibnamefont {{Dey}}}, \bibinfo {author} {\bibfnamefont
  {M.}~\bibnamefont {{Doi}}}, \bibinfo {author} {\bibfnamefont
  {M.}~\bibnamefont {{Donahue}}}, \bibinfo {author} {\bibfnamefont
  {S.}~\bibnamefont {{Deustua}}}, \bibinfo {author} {\bibfnamefont
  {P.}~\bibnamefont {{Eisenhardt}}}, \bibinfo {author} {\bibfnamefont
  {E.}~\bibnamefont {{Ellingson}}}, \bibinfo {author} {\bibfnamefont
  {L.}~\bibnamefont {{Faccioli}}}, \bibinfo {author} {\bibfnamefont
  {V.}~\bibnamefont {{Fadeyev}}}, \bibinfo {author} {\bibfnamefont {H.~K.}\
  \bibnamefont {{Fakhouri}}}, \bibinfo {author} {\bibfnamefont {A.~S.}\
  \bibnamefont {{Fruchter}}}, \bibinfo {author} {\bibfnamefont {D.~G.}\
  \bibnamefont {{Gilbank}}}, \bibinfo {author} {\bibfnamefont {M.~D.}\
  \bibnamefont {{Gladders}}}, \bibinfo {author} {\bibfnamefont
  {G.}~\bibnamefont {{Goldhaber}}}, \bibinfo {author} {\bibfnamefont {A.~H.}\
  \bibnamefont {{Gonzalez}}}, \bibinfo {author} {\bibfnamefont
  {A.}~\bibnamefont {{Goobar}}}, \bibinfo {author} {\bibfnamefont
  {A.}~\bibnamefont {{Gude}}}, \bibinfo {author} {\bibfnamefont
  {T.}~\bibnamefont {{Hattori}}}, \bibinfo {author} {\bibfnamefont
  {H.}~\bibnamefont {{Hoekstra}}}, \bibinfo {author} {\bibfnamefont
  {E.}~\bibnamefont {{Hsiao}}}, \bibinfo {author} {\bibfnamefont
  {X.}~\bibnamefont {{Huang}}}, \bibinfo {author} {\bibfnamefont
  {Y.}~\bibnamefont {{Ihara}}}, \bibinfo {author} {\bibfnamefont {M.~J.}\
  \bibnamefont {{Jee}}}, \bibinfo {author} {\bibfnamefont {D.}~\bibnamefont
  {{Johnston}}}, \bibinfo {author} {\bibfnamefont {N.}~\bibnamefont
  {{Kashikawa}}}, \bibinfo {author} {\bibfnamefont {B.}~\bibnamefont
  {{Koester}}}, \bibinfo {author} {\bibfnamefont {K.}~\bibnamefont
  {{Konishi}}}, \bibinfo {author} {\bibfnamefont {M.}~\bibnamefont
  {{Kowalski}}}, \bibinfo {author} {\bibfnamefont {E.~V.}\ \bibnamefont
  {{Linder}}}, \bibinfo {author} {\bibfnamefont {L.}~\bibnamefont {{Lubin}}},
  \bibinfo {author} {\bibfnamefont {J.}~\bibnamefont {{Melbourne}}}, \bibinfo
  {author} {\bibfnamefont {J.}~\bibnamefont {{Meyers}}}, \bibinfo {author}
  {\bibfnamefont {T.}~\bibnamefont {{Morokuma}}}, \bibinfo {author}
  {\bibfnamefont {F.}~\bibnamefont {{Munshi}}}, \bibinfo {author}
  {\bibfnamefont {C.}~\bibnamefont {{Mullis}}}, \bibinfo {author}
  {\bibfnamefont {T.}~\bibnamefont {{Oda}}}, \bibinfo {author} {\bibfnamefont
  {N.}~\bibnamefont {{Panagia}}}, \bibinfo {author} {\bibfnamefont
  {S.}~\bibnamefont {{Perlmutter}}}, \bibinfo {author} {\bibfnamefont
  {M.}~\bibnamefont {{Postman}}}, \bibinfo {author} {\bibfnamefont
  {T.}~\bibnamefont {{Pritchard}}}, \bibinfo {author} {\bibfnamefont
  {J.}~\bibnamefont {{Rhodes}}}, \bibinfo {author} {\bibfnamefont
  {P.}~\bibnamefont {{Ripoche}}}, \bibinfo {author} {\bibfnamefont
  {P.}~\bibnamefont {{Rosati}}}, \bibinfo {author} {\bibfnamefont {D.~J.}\
  \bibnamefont {{Schlegel}}}, \bibinfo {author} {\bibfnamefont
  {A.}~\bibnamefont {{Spadafora}}}, \bibinfo {author} {\bibfnamefont {S.~A.}\
  \bibnamefont {{Stanford}}}, \bibinfo {author} {\bibfnamefont
  {V.}~\bibnamefont {{Stanishev}}}, \bibinfo {author} {\bibfnamefont
  {D.}~\bibnamefont {{Stern}}}, \bibinfo {author} {\bibfnamefont
  {M.}~\bibnamefont {{Strovink}}}, \bibinfo {author} {\bibfnamefont
  {N.}~\bibnamefont {{Takanashi}}}, \bibinfo {author} {\bibfnamefont
  {K.}~\bibnamefont {{Tokita}}}, \bibinfo {author} {\bibfnamefont
  {M.}~\bibnamefont {{Wagner}}}, \bibinfo {author} {\bibfnamefont
  {L.}~\bibnamefont {{Wang}}}, \bibinfo {author} {\bibfnamefont
  {N.}~\bibnamefont {{Yasuda}}}, \bibinfo {author} {\bibfnamefont {H.~K.~C.}\
  \bibnamefont {{Yee}}}, \ and\ \bibinfo {author} {\bibfnamefont
  {T.}~\bibnamefont {{Supernova Cosmology Project}}},\ }\href {\doibase
  10.1088/0004-637X/746/1/85} {\bibfield  {journal} {\bibinfo  {journal}
  {\apj}\ }\textbf {\bibinfo {volume} {746}},\ \bibinfo {eid} {85} (\bibinfo
  {year} {2012})},\ \Eprint {http://arxiv.org/abs/1105.3470} {arXiv:1105.3470
  [astro-ph.CO]} \BibitemShut {NoStop}%
\bibitem [{\citenamefont {Suzuki}\ \emph {et~al.}(2012)\citenamefont {Suzuki}
  \emph {et~al.}}]{Suzuki:2011hu}%
  \BibitemOpen
  \bibfield  {author} {\bibinfo {author} {\bibfnamefont {N.}~\bibnamefont
  {Suzuki}} \emph {et~al.} (\bibinfo {collaboration} {Supernova Cosmology
  Project}),\ }\href {\doibase 10.1088/0004-637X/746/1/85} {\bibfield
  {journal} {\bibinfo  {journal} {Astrophys. J.}\ }\textbf {\bibinfo {volume}
  {746}},\ \bibinfo {pages} {85} (\bibinfo {year} {2012})},\ \Eprint
  {http://arxiv.org/abs/1105.3470} {arXiv:1105.3470 [astro-ph.CO]} \BibitemShut
  {NoStop}%
\bibitem [{\citenamefont {Amanullah}\ \emph {et~al.}(2010)\citenamefont
  {Amanullah} \emph {et~al.}}]{Amanullah:2010vv}%
  \BibitemOpen
  \bibfield  {author} {\bibinfo {author} {\bibfnamefont {R.}~\bibnamefont
  {Amanullah}} \emph {et~al.},\ }\href {\doibase 10.1088/0004-637X/716/1/712}
  {\bibfield  {journal} {\bibinfo  {journal} {Astrophys. J.}\ }\textbf
  {\bibinfo {volume} {716}},\ \bibinfo {pages} {712} (\bibinfo {year}
  {2010})},\ \Eprint {http://arxiv.org/abs/1004.1711} {arXiv:1004.1711
  [astro-ph.CO]} \BibitemShut {NoStop}%
\bibitem [{\citenamefont {Astier}\ \emph {et~al.}(2006)\citenamefont {Astier}
  \emph {et~al.}}]{Astier:2005qq}%
  \BibitemOpen
  \bibfield  {author} {\bibinfo {author} {\bibfnamefont {P.}~\bibnamefont
  {Astier}} \emph {et~al.} (\bibinfo {collaboration} {SNLS}),\ }\href {\doibase
  10.1051/0004-6361:20054185} {\bibfield  {journal} {\bibinfo  {journal}
  {Astron. Astrophys.}\ }\textbf {\bibinfo {volume} {447}},\ \bibinfo {pages}
  {31} (\bibinfo {year} {2006})},\ \Eprint
  {http://arxiv.org/abs/astro-ph/0510447} {arXiv:astro-ph/0510447} \BibitemShut
  {NoStop}%
\bibitem [{\citenamefont {Scolnic}\ \emph {et~al.}(2018)\citenamefont {Scolnic}
  \emph {et~al.}}]{Scolnic:2017caz}%
  \BibitemOpen
  \bibfield  {author} {\bibinfo {author} {\bibfnamefont {D.~M.}\ \bibnamefont
  {Scolnic}} \emph {et~al.},\ }\href {\doibase 10.3847/1538-4357/aab9bb}
  {\bibfield  {journal} {\bibinfo  {journal} {Astrophys. J.}\ }\textbf
  {\bibinfo {volume} {859}},\ \bibinfo {pages} {101} (\bibinfo {year}
  {2018})},\ \Eprint {http://arxiv.org/abs/1710.00845} {arXiv:1710.00845
  [astro-ph.CO]} \BibitemShut {NoStop}%
\bibitem [{\citenamefont {Riess}\ \emph {et~al.}(2007)\citenamefont {Riess}
  \emph {et~al.}}]{Riess:2006fw}%
  \BibitemOpen
  \bibfield  {author} {\bibinfo {author} {\bibfnamefont {A.~G.}\ \bibnamefont
  {Riess}} \emph {et~al.},\ }\href {\doibase 10.1086/510378} {\bibfield
  {journal} {\bibinfo  {journal} {Astrophys. J.}\ }\textbf {\bibinfo {volume}
  {659}},\ \bibinfo {pages} {98} (\bibinfo {year} {2007})},\ \Eprint
  {http://arxiv.org/abs/astro-ph/0611572} {arXiv:astro-ph/0611572} \BibitemShut
  {NoStop}%
\bibitem [{\citenamefont {Di~Valentino}\ \emph {et~al.}(2021)\citenamefont
  {Di~Valentino}, \citenamefont {Melchiorri},\ and\ \citenamefont
  {Silk}}]{DiValentino:2020hov}%
  \BibitemOpen
  \bibfield  {author} {\bibinfo {author} {\bibfnamefont {E.}~\bibnamefont
  {Di~Valentino}}, \bibinfo {author} {\bibfnamefont {A.}~\bibnamefont
  {Melchiorri}}, \ and\ \bibinfo {author} {\bibfnamefont {J.}~\bibnamefont
  {Silk}},\ }\href {\doibase 10.3847/2041-8213/abe1c4} {\bibfield  {journal}
  {\bibinfo  {journal} {Astrophys. J. Lett.}\ }\textbf {\bibinfo {volume}
  {908}},\ \bibinfo {pages} {L9} (\bibinfo {year} {2021})},\ \Eprint
  {http://arxiv.org/abs/2003.04935} {arXiv:2003.04935 [astro-ph.CO]}
  \BibitemShut {NoStop}%

\end{thebibliography}

\end{document}